\documentclass[usenatbib]{mn2e}
\usepackage{graphicx}
\usepackage{epstopdf}
\usepackage{amssymb,amsmath,afterpage}
\usepackage{float,flafter}
\title[Baryons and the spins and shapes of DM haloes]{The impact of baryons on the spins and shapes of dark matter haloes}
\author[S. E. Bryan et al.]
{
{\parbox{\textwidth}{S. E. Bryan,$^{1}$\thanks{E-mail:sarah.bryan@gmail.com}
S. T. Kay,$^{1}$
A. R. Duffy, $^{2,3}$
J. Schaye,$^{4}$
C. Dalla Vecchia $^{4,5}$ and \\
C. M. Booth$^{6,7,4}$}}\vspace{0.4cm}\\
\parbox{\textwidth}{$^{1}$Jodrell Bank Centre for Astrophysics, School of
  Physics and Astronomy, The University of Manchester, Manchester M13 9PL\\
$^{2}$School of Physics, University of Melbourne, Parkville, VIC 3010, Australia.\\
$^{3}$ICRAR, University of Western Australia, WA 6009, Australia.\\
$^{4}$Leiden Observatory, Leiden University, Postbus 9513, 2300 RA Leiden, The
Netherlands.\\
$^{5}$Max Planck Institute for Extraterrestrial Physics, Giessenbachstra\ss{}e 1, 85748 Garching, Germany.\\
$^{6}$Department of Astronomy \& Astrophysics, The University of Chicago, Chicago, IL 60637, USA. \\
$^{7}$Kavli Institute for Cosmological Physics and Enrico Fermi Institute, The University of Chicago, Chicago, IL, 60637, USA. \\
}
}
\begin{document}
% bibliography and bibfile journal definitions (taken from aa.cls)

% Astronomy and Astrophysics
\def\aj{AJ}					% Astronomical Journal
\def\araa{ARA\&A}				% Annual Reviews of Astronomy and Astrophysics
\def\apj{ApJ}					% Astrophysical Journal
\def\apjl{ApJL}					% Astrophysical Journal, Letters
\def\apjs{ApJS}					% Astrophysical Journal, Supplement Series
\def\apss{Astrophysics and Space Science}
\def\capsp{Comments on Astrophysics and Space Physics}
\def\aap{A\&A}					% Astronomy and Astrophysics
\def\aapr{A\&A~Rev.}				% Astronomy and Astrophysics Reviews
\def\aaps{A\&AS}				% Astronomy and Astrophysics, Supplement
\def\azh{AZh}					% Astronomicheskii Zhurnal
\def\baas{BAAS}					% Bulletin of the AAS
\def\jrasc{JRASC}				% Journal of the RAS of Canada
\def\memras{MmRAS}				% Memoirs of the RAS
\def\mnras{MNRAS}					% Monthly Notices of the Royal Astronomical Society
\def\pasp{PASP}					% Publications of the ASP
\def\pasj{PASJ}					% Publications of the ASJ
\def\qjras{QJRAS}				% Quarterly Journal of the RAS
\def\skytel{S\&T}				% Sky and Telescope
\def\solphys{Sol.~Phys.}			% Solar Physics
\def\sovast{Soviet~Ast.}			% Soviet Astronomy
\def\ssr{Space~Sci.~Rev.}			% Space Science Reviews
\def\zap{ZAp}					% Zeitschrift fuer Astrophysik
\def\na{New Astronomy}				% New Astronomy
\def\iaucirc{IAU~Circ.}				% IAU Cirulars
\def\aplett{Astrophys.~Lett.}			% Astrophysics Letters
\def\apspr{Astrophys.~Space~Phys.~Res.}		% Astrophysics Space Physics Research
\def\bain{Bull.~Astron.~Inst.~Netherlands}	% Bulletin Astronomical Institute of the Netherlands
\def\memsai{Mem.~Soc.~Astron.~Italiana}		% Mem. Societa Astronomica Italiana

% Optics
\def\ao{Appl.~Opt.}				% Applied Optics

% General physics
\def\pra{Phys.~Rev.~A}				% Physical Review A: General Physics
\def\prb{Phys.~Rev.~B}				% Physical Review B: Solid State
\def\prc{Phys.~Rev.~C}				% Physical Review C
\def\prd{Phys.~Rev.~D}				% Physical Review D
\def\pre{Phys.~Rev.~E}				% Physical Review E
\def\prl{Phys.~Rev.~Lett.}			% Physical Review Letters
\def\nat{Nature}				% Nature
\def\fcp{Fund.~Cosmic~Phys.}			% Fundamental Cosmic Physics
\def\gca{Geochim.~Cosmochim.~Acta}		% Geochimica Cosmochimica Acta
\def\grl{Geophys.~Res.~Lett.}			% Geophysics Research Letters
\def\jcp{J.~Chem.~Phys.}			% Journal of Chemical Physics
\def\jgr{J.~Geophys.~Res.}			% Journal of Geophysics Research
\def\jqsrt{J.~Quant.~Spec.~Radiat.~Transf.}	% Journal of Quantitiative Spectroscopy and Radiative Trasfer
\def\nphysa{Nucl.~Phys.~A}			% Nuclear Physics A
\def\physrep{Phys.~Rep.}			% Physics Reports
\def\physscr{Phys.~Scr}				% Physica Scripta
\def\planss{Planet.~Space~Sci.}			% Planetary Space Science
\def\procspie{Proc.~SPIE}			% Proceedings of the SPIE
\def\rpp{Rep.~Prog.~Phys.}			% Rep. Prog. Phys.
\let\astap=\aap
\let\apjlett=\apjl
\let\apjsupp=\apjs
\let\applopt=\ao
\let\prep=\physrep

% end of file

\date{Accepted ...... Received ...... ; in original form......   }

\pagerange{\pageref{firstpage}--\pageref{lastpage}} \pubyear{2012}
\maketitle
\label{firstpage}
\begin{abstract}
We use numerical simulations to investigate how the statistical properties of dark matter (DM) 
haloes are affected by the baryonic processes associated with galaxy formation. We focus 
on how these processes influence the spin and shape of a large number of DM haloes 
covering a wide range of mass scales, from galaxies to clusters at redshifts zero and one, extending to dwarf galaxies at redshift two. 
The haloes are extracted from the OverWhelmingly Large
Simulations (OWLS), a suite of state-of-the-art high-resolution cosmological simulations run with a range
of feedback prescriptions. We find that the median spin parameter
in DM-only simulations is independent of mass, redshift and cosmology.  At $z = 0$ baryons increase
the spin of the DM in the central region ($\le 0.25 \, r_{200}$) by up to 30 per cent when feedback is weak or absent. This increase can be attributed to the transfer of angular momentum
from baryons to the DM, but is no longer present at $z = 2$. We also present fits to the mass dependence of the DM halo shape at both low
and high redshift. At $z=0$ the sphericity (triaxiality) is negatively (positively) correlated with halo mass
and both results are independent of cosmology. Interestingly, these mass-dependent trends are markedly 
weaker at $z=2$. While the cooling of  baryons acts to make the overall DM halo more spherical, stronger
feedback prescriptions (e.g. from active galactic nuclei) tend to reduce the impact of baryons by reducing
the central halo mass concentration. More generally, we demonstrate a strongly positive (negative) 
correlation between
halo sphericity (triaxiality) and galaxy formation efficiency, with the latter measured using 
the central halo baryon fraction. In 
conclusion, our results suggest that the effects of baryons on the DM halo spin and shape are minor when 
the effects of cooling are mitigated, as required by realistic models of galaxy formation, although they remain significant for the inner halo.
\end{abstract}


\section{Introduction}
\label{shapesintro}

A natural consequence of the standard hierarchical structure formation
paradigm is that the shapes of dark matter haloes are triaxial, a property that is 
inherited from their progenitor density perturbations \citep{bib:Bardeen86}. This
additionally leads to aspherical growth as the halo accretes matter from
preferential directions, associated with the surrounding sheets and filaments.
The anisotropic accretion history of a halo also affects its angular
momentum distribution, through the presence of non-zero torques. It is therefore clear 
that both the shape and spin of a dark matter halo are important diagnostics for 
an accurate determination of their structure and formation history.

While the spin and shape of a dark matter halo are not directly observable, they 
have important consequences for the structure and dynamics of galaxies. For example, halo spin is 
an important parameter in galaxy formation models as it affects the size of the embedded
galactic disc (e.g. \citealt{bib:Mo98}, but see \citealt{bib:Sales12}). Deviations from axisymmetry in elliptical 
galaxies are likely to influence the gas kinematics of the system \citep{bib:deZeeuw89}, 
and may be responsible for exciting or sustaining warps and stabilising or deforming polar rings
\citep{bib:Steiman92}. Axisymmetry may also influence the fuelling efficiency of the central black
hole \citep{bib:Franx91}. Misalignment of the angular momentum of the halo and the galaxy may be
responsible for the anisotropic distribution of subhaloes and satellite
galaxies (\citealt{bib:Holmberg74, bib:Knebe04, bib:Kang05, bib:Libeskind05, bib:Zentner05, bib:Libeskind07, bib:Knebe10}) and could cause
galactic warps (\citealt{bib:Ostriker89, bib:Debattista99, bib:Bailin04}).

On cluster scales, asphericity in the dark matter halo will naturally correspond to 
asphericity in the gas density and will impact the shape of X-ray isophotes and the 
Sunyaev-Zel'dovich signal. Understanding the intrinsic shape of dark matter
haloes is also important for weak lensing analysis (see, for example, the
discussions in \citealt{bib:Becker11, bib:Bett12}) and it is well known that intrinsic ellipticity
can contribute significantly to a lensing halo's ability to form arcs \citep{bib:Oguri03}.

Several methods are used to constrain galaxy and halo shapes observationally (see, for example, \citealt{bib:Sackett99, bib:Merrifield04}).  Unfortunately studies performed to date do not yet reveal a consistent picture (see the discussion in \citealt{bib:Obrien10}).  The observations cover a large range of systems and vary in the extent to which the halo is probed, making a direct comparison somewhat difficult.  Whether the discrepancies result from halo-to-halo scatter or from systematic errors in the observed estimates is unclear.  However, given the rapidly accumulating number of data sets, ever increasing sophistication of the data analysis tools and the development of more realistic mock observations from simulations, one can soon expect the situation to change substantially. 

Theoretically, the predictions for the distribution of angular momentum and halo 
shapes have been studied extensively, primarily using cosmological $N$-body simulations 
(\citealt{bib:Frenk88}; \citealt{bib:Dubinski91}; \citealt{bib:Warren92}; 
 \citealt{bib:Cole96}; \citealt{bib:Bullock02}; \citealt{bib:Jing02}; 
 \citealt{bib:Bailin05}; \citealt{bib:Allgood06}; \citealt{bib:Bett07}; 
 \citealt{bib:Maccio08}; \citealt{bib:JeesonDaniel11}; \citealt{bib:Vera-Ciro11}; 
 \citealt{bib:Bett12}; \citealt{bib:Zemp12}).
 There is a general consensus that cold dark matter haloes have approximately log-normal spin 
distributions and are triaxial, with sphericities, $(c/a) \simeq 0.5-0.8$
\footnote{Where $a > b > c$ are the eigenvalues of the halo's inertia tensor.} 
and elongations, $(b/a)\simeq 0.4-1$. Furthermore, haloes are generally found to be 
highly flattened and show a tendency toward prolate shapes $(c/b > b/a)$, especially 
in the inner regions. There is also general agreement that the sphericity decreases with 
increasing halo mass and that the spin is independent of mass
(\citealt{bib:Bullock02, bib:Jing02, bib:Springel04, bib:Hopkins05, bib:Bett07, bib:Maccio08, bib:JeesonDaniel11}).  In addition, \cite{bib:JeesonDaniel11} have shown that sphericity is strongly correlated with concentration, while both triaxiality and spin are anti-correlated with concentration.

While these results are interesting, a significant uncertainty is how the dark matter
halo is affected by the additional, non-gravitational processes acting on the baryons.
Recent work has clearly established that the condensation of baryons to the centre of 
dark matter haloes tends to increase the central angular momentum of 
the halo (see for example, \citealt{bib:Sharma05, bib:Tonini06a, bib:Kaufmann07, bib:Abadi10, bib:Bett10}) and 
to make the halo more spherical or axisymmetric (see, for example, \citealt{bib:Katz91,
bib:Dubinski94, bib:Evrard94, bib:Barnes96,   bib:Tissera98,bib:Springel04,bib:Kazantzidis04, 
bib:Debattista08, bib:Pedrosa10, bib:Tissera10, bib:Bryan11,bib:Zemp12}).  This result has been 
used to explain the discrepancy between the strongly-prolate triaxial shape found in 
$N$-body simulations and the more spherical central regions of observed systems.

Incorporating baryonic physics in cosmological simulations is a
non-trivial task and the computational cost of this process has placed limits
on both the parameter space and the size of the sample of haloes explored to
date. The detailed nature of the baryonic processes involved in galaxy formation 
and the precise influence of these processes on galaxies therefore remains largely uncertain.
In this paper, we attempt to make progress on both fronts, by studying the spin and shape
distributions for a large ($>1000$) sample of dark matter haloes, spanning a range of mass
(from dwarf galaxies to clusters) and redshift ($z=0,1$ and $2$). We do this using 
the OverWhelmingly Large Simulations (OWLS; \citealt{bib:Schaye10}) -- a suite of cosmological hydrodynamical 
simulations run with many different physical prescriptions for the baryons.
By providing identical simulations run with different implementations of the subgrid physics, OWLS
offers the opportunity to explore the effects of baryons under a range of physical conditions,
for the same population of haloes. In particular, we use the OWLS data to quantify, in a statistically 
meaningful way, the influence of feedback processes (from no feedback, to feedback from stars and 
black holes) on the spin and shape distributions of dark matter haloes.

The paper is organised as follows. The simulations used in this 
analysis, and the methods used to identify haloes and to estimate their spins and 
shapes are outlined in section \ref{simshapes}. Our results are presented in 
section \ref{results}, including fitting formulae for the predicted correlations 
between halo shape and mass. The robustness of our results is demonstrated via a resolution
study, given in the appendix. Finally, we summarise our main results in section \ref{summaryshapes}.

\section{Methodology}
\label{simshapes}

\subsection{Simulation details}

\begin{table*}
\caption[]{A list of the OWLS runs used in this analysis.  We use the same identifier for each run as in \citet{bib:Schaye10} and comment 
on the subgrid physics implemented in each case. Key global properties of the simulations are also listed, namely 
the number of dark matter and baryonic particles; 
the comoving box length; 
and the dark matter particle mass. 
Values are presented for all runs analysed at $z = 0$. The $100 \, h^{-1} {\rm Mpc}$ boxes are also used for results 
at $z=1$ and $z = 2$, while $25 \, h^{-1} {\rm Mpc}$ boxes are used only at $z=2$ (values for these runs are given in brackets).  The maximum force softening was 0.5, 2.0 and 8.0 $h^{-1} {\rm kpc}$ for the $25$, $100$ and $400 {\, h^{-1} \rm Mpc}$ boxes respectively. 
}
\centering 
\begin{tabular}{ l l l l l l} 
\hline
Name & Description  & $N_{\rm DM}$ & $N_{\rm baryons}$ & Box length & $m_{\rm DM}$ \\
&& &  & ($h^{-1}$ Mpc)  &  $(h^{-1}\, \mbox{M}_{\odot})$  \\
\hline 
DMONLY &  Dark matter only runs  & & & &\\
 & {\it WMAP}\,1 & $216^3$ & - & $50$ & $8.6 \times 10^{8}$\\
 & {\it WMAP}\,3  & $512^3$ & -  & $100$  ($25$) & $4.9\times 10^{8}$ $(7.7 \times 10^6)$ \\
 & {\it WMAP}\,3 & $512^3$ & - & $400$  & $3.1 \times 10^{10} $\\
 & {\it WMAP}\,5 & $512^3$ & -  & $100$  ($25$) & $5.3\times 10^{8}$ $(8.3 \times 10^6)$ \\
 & {\it WMAP}\,5 & $512^3$ & - & $400$  & $3.4 \times 10^{10} $\\
NOSN\_NOZCOOL & No feedback, primordial abundances for cooling & $512^3$ & $512^3$& $100$ ($25$) & $4.1 \times 10^{8} $ $(6.3 \times 10^6)$\\
REF & Weak stellar feedback, metal cooling &$512^3$ & $512^3$ & $100$  ($25$) &$4.1 \times 10^{8} $ $(6.3 \times 10^6)$\\
WDENS & Strong stellar feedback, metal cooling &$512^3$  & $512^3$&  $100$ ($25$)&$4.1 \times 10^{8} $ $(6.3 \times 10^6)$\\
AGN &  Weak stellar \& AGN feedback, metal cooling & $512^3$ & $512^3$&  $100$ ($25$) &$4.1 \times 10^{8} $ $(6.3 \times 10^6)$\\
\hline 
\end{tabular}
\label{table:simtypes} 
\end{table*}

The haloes used for this analysis were extracted from a subset of the OWLS runs. For detailed information about these 
simulations the reader is referred to \cite{bib:Schaye10}; here, the most relevant aspects are briefly reviewed for convenience. 

For all runs, cosmological initial conditions were set up using a transfer function generated with {\sc{cmbfast}} \citep{bib:Seljak96}.  Initial ($z=127$) positions and 
velocities were computed using the \cite{bib:Zeldovich70} approximation from an initial glass-like state \citep{bib:White96}.  All simulations were run using a modified 
version of {\small GADGET-3} \citep{bib:Gadget2}.  For a given box size the same initial conditions were used in each run, allowing us to directly
 compare the same haloes evolved using different prescriptions for the sub-grid physics.  A summary of the simulations used in this analysis, including values for 
key numerical parameters, can be found in Table \ref{table:simtypes}.

We begin by exploring dark matter only (DMONLY) simulations, as this allows us to validate our results by comparing 
to the existing literature, as well as to set the scene for exploring the impact of baryons.  We consider three sets 
of dark matter only simulations, each run with different values for the cosmological parameters. Our main results (including baryons) assume values taken from the 3rd year  {\it Wilkinson Microwave Anisotropy Probe} data ({\it WMAP}\,3; \citealt{bib:WMAP3}) with $[\Omega_m$,$\Omega_\Lambda,\Omega_b,n,\sigma_8]=[0.238,0.762,0.0418,0.95,0.74]$. This model was run with
dark matter only, using $512^3$ particles in three box sizes. The two larger boxes ($100$ and $400 \, h^{-1} {\rm Mpc}$) were run to $z=0$ while a smaller, high-resolution box ($25 \, h^{-1} {\rm Mpc}$) was run to $z=2$; we use the latter to study the properties of haloes at high redshift.  The comoving softening length was set to $1/25$ of the initial mean interparticle spacing until $z = 2.91$; below this redshift the softening was held fixed in proper units.  The maximum physical softening length in the $25 \, h^{-1} {\rm Mpc}$ box was 0.5 $ \, h^{-1} {\rm kpc}$, while in the $100$ and $400 \, h^{-1} {\rm Mpc}$ it was 4 and 16 times larger respectively. We also present results for the newer
 {\it WMAP}\,5 cosmology (\citealt{bib:Komatsu09}) with $[\Omega_m$,$\Omega_\Lambda,\Omega_b,n,\sigma_8]=[0.258,0.742,0.0441,0.963,0.796]$ using the same boxes 
 as before. Finally, we consider a run with the {\it WMAP}\,1 cosmology \citep{bib:wmap1_03} with 
 $[\Omega_m$,$\Omega_\Lambda,\Omega_b,n,\sigma_8]=[0.25,0.75,0.045,1,0.9]$. 
This simulation was run using $216^3$ particles in a 50 $h^{-1}$ Mpc box, matching the resolution of the 
Millennium Simulation (\citealt{bib:Springel05}), which was also run with the {\it WMAP}\,1 parameters.

To study the effect of varying levels of feedback on the spin and shape parameters of haloes extracted from $\Lambda$CDM 
simulations, four baryon runs from the OWLS simulations are considered, all run with the {\it WMAP}\,3 cosmology. These 
runs model the gaseous component using smoothed particle hydrodynamics (SPH).

All of the simulations include radiative cooling (\citealt{bib:Wiersma09}), star formation (\citealt{bib:Schaye08}) and metal enrichment (\citealt{bib:Wiersma09b}), but differ in their feedback prescriptions (for more details see \citealt{bib:DallaVecchia08}).  All models followed the timed release of H, He and 9 different heavy elements produced by massive stars, AGB stars and supernovae of type I and II (\citealt{bib:Wiersma09b})  The first run (NOSN\_NOZCOOL) did not include any feedback processes\footnote{This simulation also neglects enhanced cooling through metal-lines.}; 
the second (REF) includes weak feedback from stellar winds and supernovae; the third (WDENS) includes stronger stellar feedback with a wind-speed that depends on the local gas density; and the final run (AGN), feedback from both stars and active galactic nuclei (\citealt{bib:Booth09}). In the feedback models,
enhanced cooling from metal lines was also accounted for. All four models were run using 
the $100 \, h^{-1} {\rm Mpc}$ box (to $z=0$) and the $25 \, h^{-1} {\rm Mpc}$ box (to $z=2$). The same number of gas 
particles as dark matter particles ($512^3$) was adopted for each box.

\subsection{Halo sample}

Haloes were extracted from each simulation using the Friends-of-Friends (FoF) algorithm \citep{bib:Davis85} which links particles together within a fixed 
comoving separation. This
separation, known as the linking length, was set to 0.2 times the mean inter-particle distance. The FoF groups were then decomposed into self-bound
sub-haloes using the {\sc subfind} algorithm \citep{bib:Springel01, bib:Dolag08}. Finally, a sphere was grown around the most bound particle of the most massive sub-halo until the mean
total mass density was equal to 200 times the critical density.  We choose to define our haloes as including all particles contained within this sphere. The mass and radius of the halo are referred to as $M_{200}$ and $r_{200}$ 
respectively. 

Only haloes that contain at least 1000 particles are considered in this analysis, as this ensures that the results are fully converged (see the appendix for a 
discussion of the effects of resolution).  While estimates of the spin parameter are found to be well resolved for haloes with 
more than 300 particles (as in \citealt{bib:Bett07})  a larger number of particles is required to resolve the halo shape (in particular the 
triaxiality of the halo), in agreement with \cite{bib:Maccio08}.  This cut imposes a minimum halo mass of 
$4.9 \times 10^{11}\,\, h^{-1}\, \mbox{M}_{\odot}$ at $z = 0$ and $7.7 \times 10^{9}\,\, h^{-1}\, \mbox{M}_{\odot}$ at $z = 2$, for our {\it WMAP}\,3 runs.  

We have investigated whether our results are affected by restricting our sample to dynamically relaxed haloes. We estimated the dynamical state 
using the centroid shift, defined to be the distance between the minimum potential position (the halo centre) and the centre of mass of the halo. In accordance
with \citet{bib:Neto07}, a halo is defined to be relaxed if the centroid shift is less than $0.07\, r_{200}$. We find that while relaxed haloes are typically more spherical and less triaxial, this restriction does not affect the trends found in this 
work and so we chose to present our analysis for the total halo sample (except where explicitly stated otherwise). Not only does this result in larger numbers
of objects, our choice is also motivated by the fact that an observational cut based on the relaxation state of a halo is not a straightforward task.  

\begin{table*}
\caption[]{Number of haloes (containing at least 1000 particles) and median halo mass $M_{200}$ (in  $10^{12} h^{-1} {\rm M}_{\odot}$) for each simulation at  $z = 0$, $1$ and $2$.} 
\centering 
\begin{tabular}{l  l l l l l l}
\hline 
& \multicolumn{2}{c} {$ z = 0 $} &  \multicolumn{2}{c}{$z = 1 $} &\multicolumn{2}{c} {$ z = 2 $ } \\

Simulation &  $N_{\mbox{haloes}}$ & $M_{200}$ & $N_{\mbox{haloes}}$ & $M_{200}$ & $N_{\mbox{haloes}}$ & $M_{200}$ \\
\hline
DMONLY  {\it WMAP}\,1 & 431  & 1.7 & -  & - & - & -\\
DMONLY  {\it WMAP}\,3 & 7094 & 2.3 & 3765 &  0.77 & 5334 & 0.030\\
DMONLY  {\it WMAP}\,5 & 8188 & 3.6 & - & - & - & - \\
&&&\\
AGN           & 3878 & 0.88 & 3375 & 0.83 & 4743 & 0.033\\
WDENS         & 3884 & 0.91 & 3371 & 0.84 & 4699 & 0.035\\
REF           & 3990 & 0.99 & 3580 & 0.93 &  4823 & 0.044\\
NOSN\_NOZCOOL          & 4634 & 1.0 & 4041  & 0.91 & 5683 & 0.040\\
\hline 
\end{tabular}
\label{numhaloes} 
\end{table*}

The numbers of haloes in our final samples are presented in Table \ref{numhaloes}. Note that the number of haloes in DMONLY {\it WMAP}\,3 at $z=0$
is almost twice as large as in the baryon runs; this is due to the inclusion of haloes from the $400 \, h^{-1} {\rm Mpc}$ box in the former case. Overall,
our DMONLY sample spans around three orders of magnitude in mass (from galaxy to cluster scales, i.e. $M_{200} \sim 10^{12}-10^{15} \, h^{-1} {\rm M}_{\odot}$) 
at $z=0$. Our $z=2$ sample covers a similar dynamic range but at lower mass (dwarf galaxy to group scales, $M_{200} \sim 10^{10}-10^{13} \, h^{-1} {\rm M}_{\odot}$).

\subsection{Defining halo spin and shape}
\label{methodshapes}

The dimensionless spin parameter $\lambda$ provides a useful measure of the amount of rotational support present within a dark matter halo. 
We estimate this property for the dark matter particles using the modified expression given by \cite{bib:Bullock01} 
\begin{equation}
\lambda'(r) = \frac{J_{{\rm DM}}}{\sqrt{2} M_{{\rm DM}} \, v_c \, r},
\label{lambdadef}
\end{equation}
where $J_{\rm DM}$ is the angular momentum of the DM within a sphere of radius $r$, containing mass $M_{\rm DM}$ and $v_{\rm c}=\sqrt{GM_{\rm {tot}}\left( < r \right)/r}$ is the halo circular velocity at this radius. This expression reduces to the standard spin parameter (\citealt{bib:Peebles69}) when measured at the virial radius of a truncated singular isothermal halo.  For a comparison of the different definitions for the spin parameter, see \cite{bib:Maccio07}; hereafter we drop the prime and refer to the modified spin parameter as 
$\lambda$.  

To characterise the halo shape, we use the mass distribution tensor
$\boldsymbol{\mathcal{M}}$ which has been used extensively in the halo shape
literature  (e.g. \citealt{bib:Cole96, bib:Bailin05}). The components of the tensor (a square matrix) are 
\begin{equation}
\mathcal{M}_{ij} = \sum_k{{m_k r_{k,i} r_{k,j}}},
\end{equation}
where the index $k$ runs over all dark matter particles within a given radius, 
$m_{k}$ is the mass of the $k$th particle and $r_{k,i}$ the $i$th component of its position vector from the halo centre.
The square roots of the eigenvalues of the mass distribution tensor, obtained using Jacobi transformations, are defined as $a,b,c$ (where $a > b > c$) and are used to measure the shape of the simulated haloes.  Note that the shapes obtained using the inertia tensor and the mass distribution tensor $\boldsymbol{\mathcal{M}}$ are equivalent (\citealt{bib:Bett07}). 

Our shape results are presented using the following parameters: $s=c/a$ is used as a measure of halo sphericity; $e=b/a$ as a measure of elongation; 
and $T = (a^2-b^2)/(a^2 - c^2)$ as a measure of the triaxiality of the halo. A purely spherical halo will have $s=e=1$ with $T$ being undefined. Low values of
$T$ ($T \rightarrow 0$) correspond to oblate haloes while high values ($T \rightarrow 1$) correspond to prolate haloes.

We note that computation of the mass tensor in a spherical region biases the shape towards higher sphericity; this is corrected for, as suggested in
 \citealt{bib:Bailin05}, by re-scaling the axis ratios $s \rightarrow s^{\sqrt{3}}$ and $e \rightarrow e^{\sqrt{3}}$.
While our adopted method described above may not be the most robust way of describing the physical halo shape (e.g. see the discussion in \citealt{bib:Zemp11}), we follow \cite{bib:Bett12} and use this simple approach as it is most directly comparable with observations and is adequate for the comparison we wish to present here.  We checked that using the reduced mass distribution tensor, and also using an iterative technique to determine the halo shapes, do not result in 
systematically different results.

Our main results for spin and shape parameters are presented for the dark matter within $r_{200}$, but we also consider results within the central
region ($0.25\, r_{200}$) where baryonic physics plays a more significant role. The convergence radius (see \citealt{bib:Power03}) is less than 
$0.25\, r_{200}$ 
for almost all of the haloes considered here. (The haloes that are not converged within $0.25\, r_{200} $ are excluded from the studies of the central region. 
In all cases this is less than one per cent of the sample).  Resolution tests for this region are discussed in the appendix. 

\section{Results}
\label{results}

We begin by presenting our results for the halo spin parameter, then go on to explore the shapes of a large number of haloes selected from our cosmological sample.  In both cases, we first discuss the results from dark matter only simulations before studying the impact of the baryons.  Quantities are measured within the central region ($0.25\, r_{200}$) and at $r_{200}$.  
 
\subsection{Halo Spin}
\label{spinresults}

\begin{figure*}
%made using SpinPlots
\begin{tabular}{l l}
\includegraphics[width=8.4cm,height=8.1cm,angle=-90,keepaspectratio]{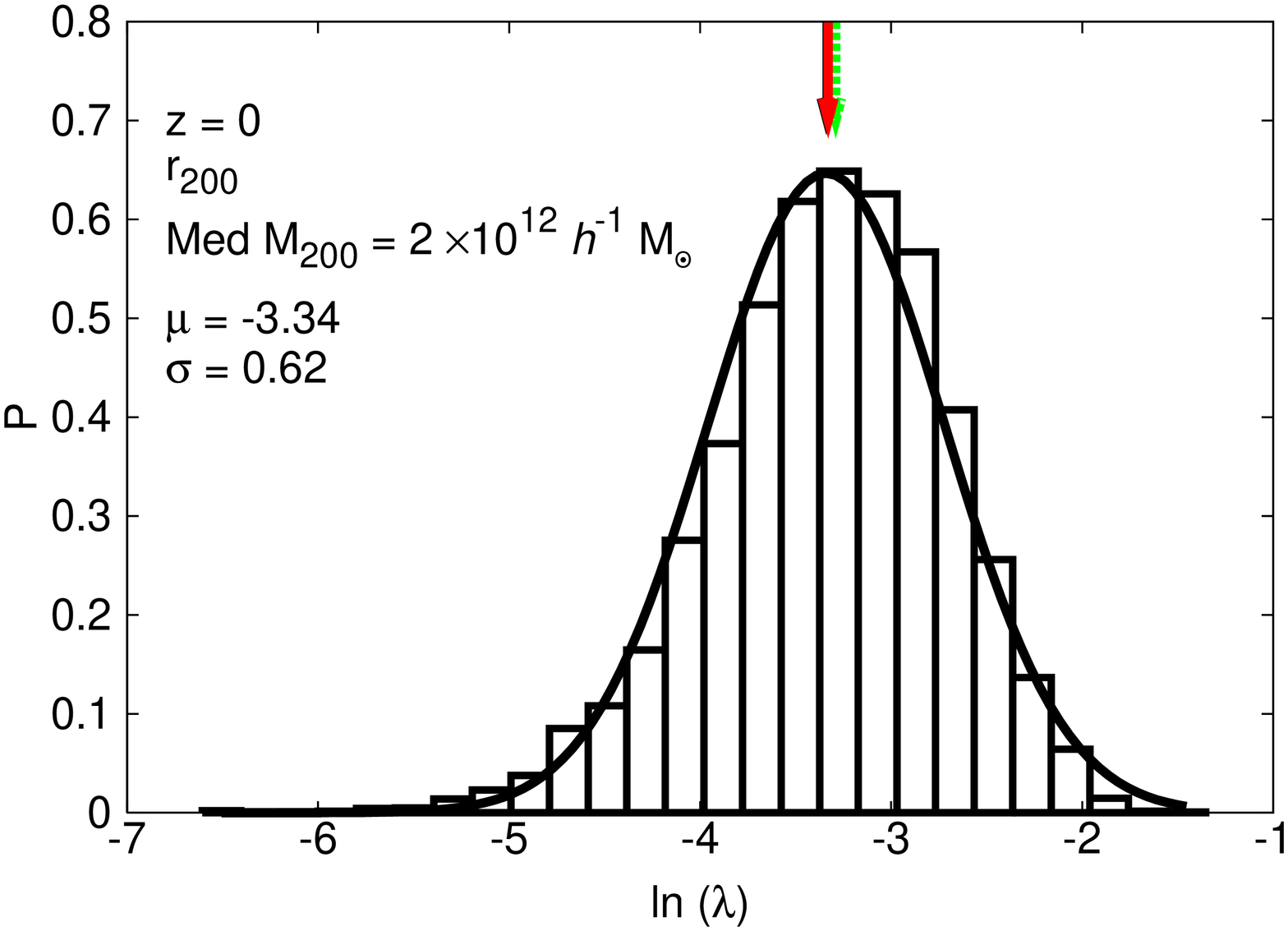} &
\includegraphics[width=8.4cm,height=8.1cm,angle=-90,keepaspectratio]{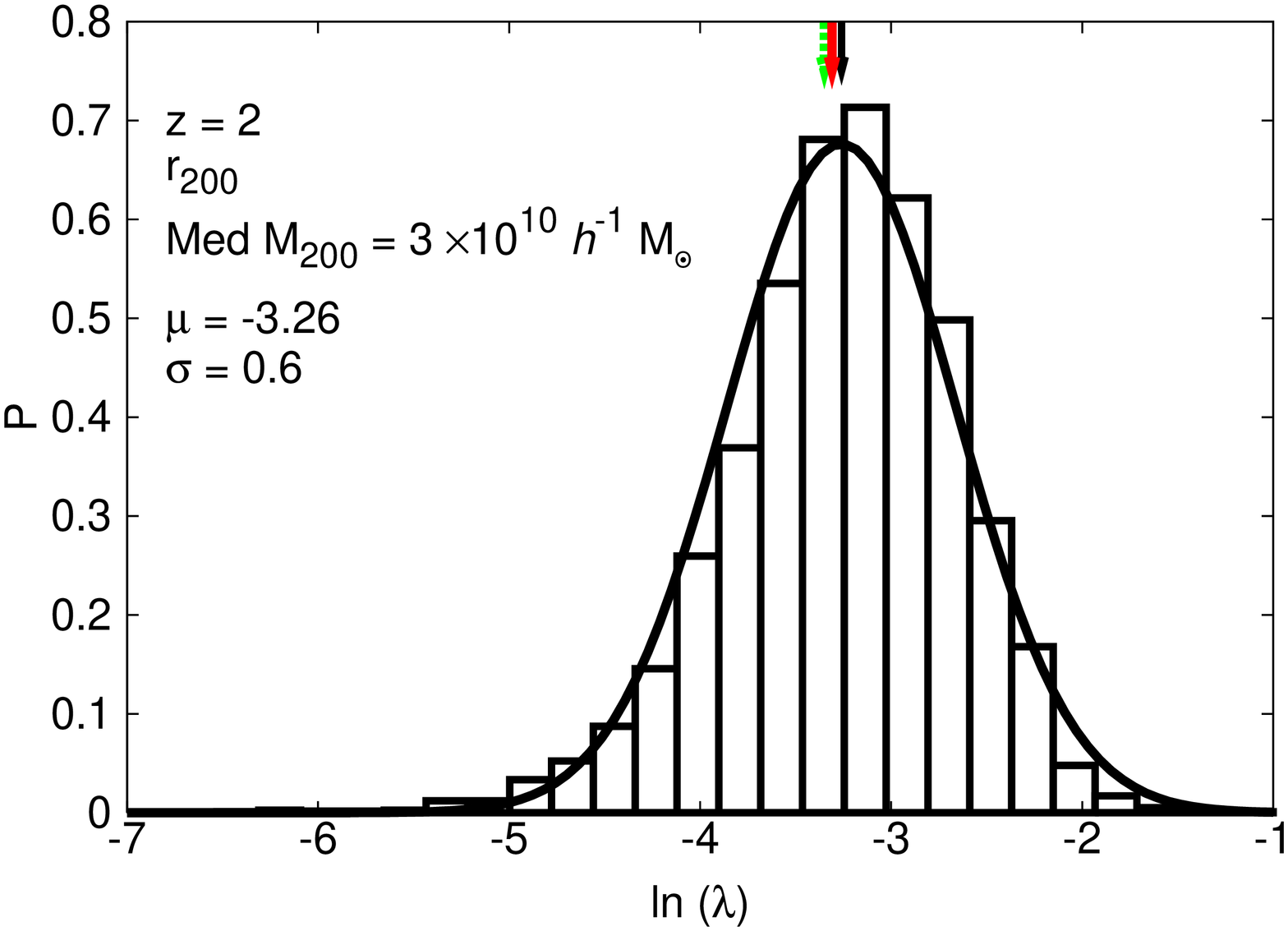} \\
\includegraphics[width=8.4cm,height=8.1cm,angle=-90,keepaspectratio]{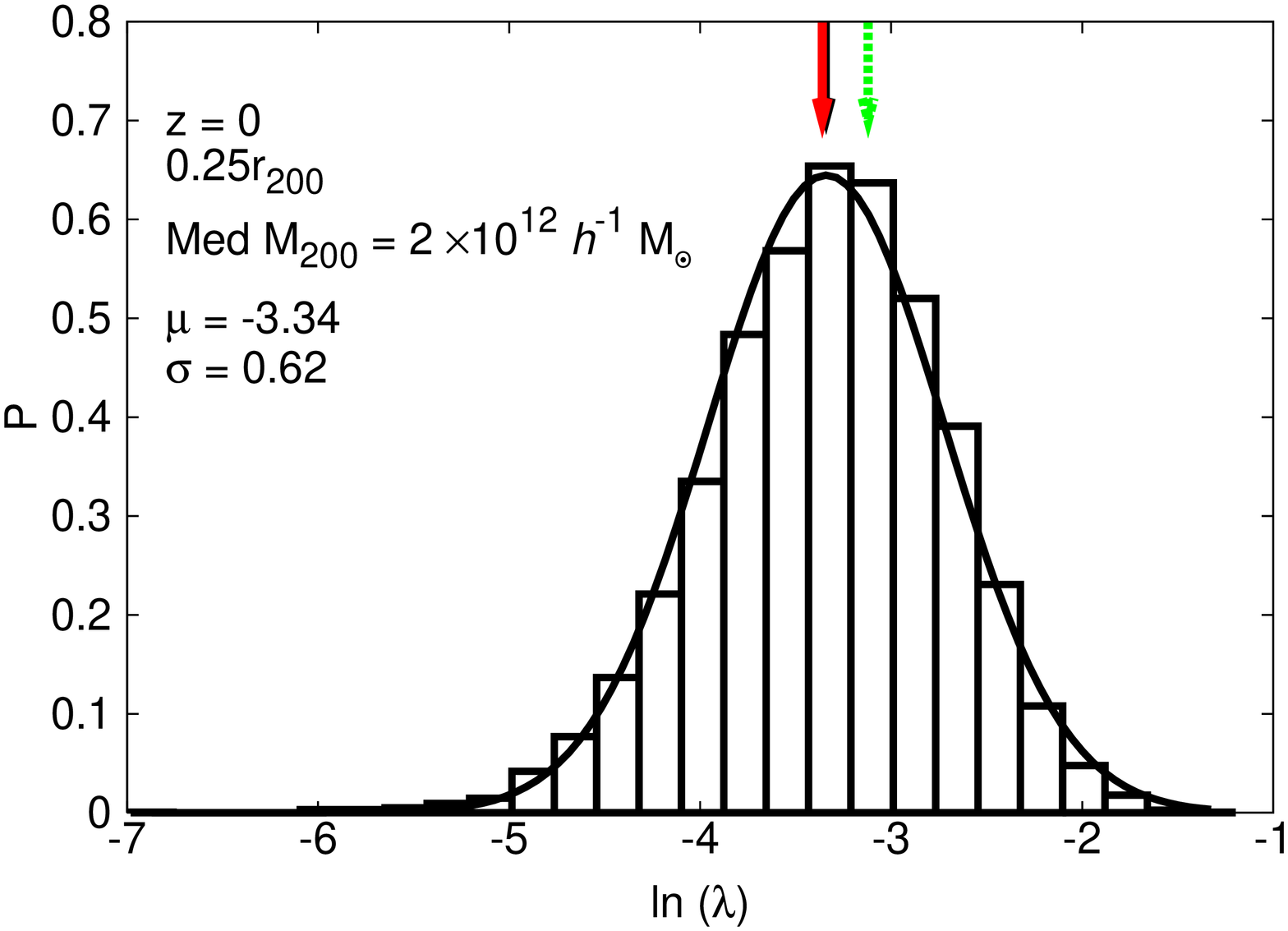} &
\includegraphics[width=8.4cm,height=8.1cm,angle=-90,keepaspectratio]{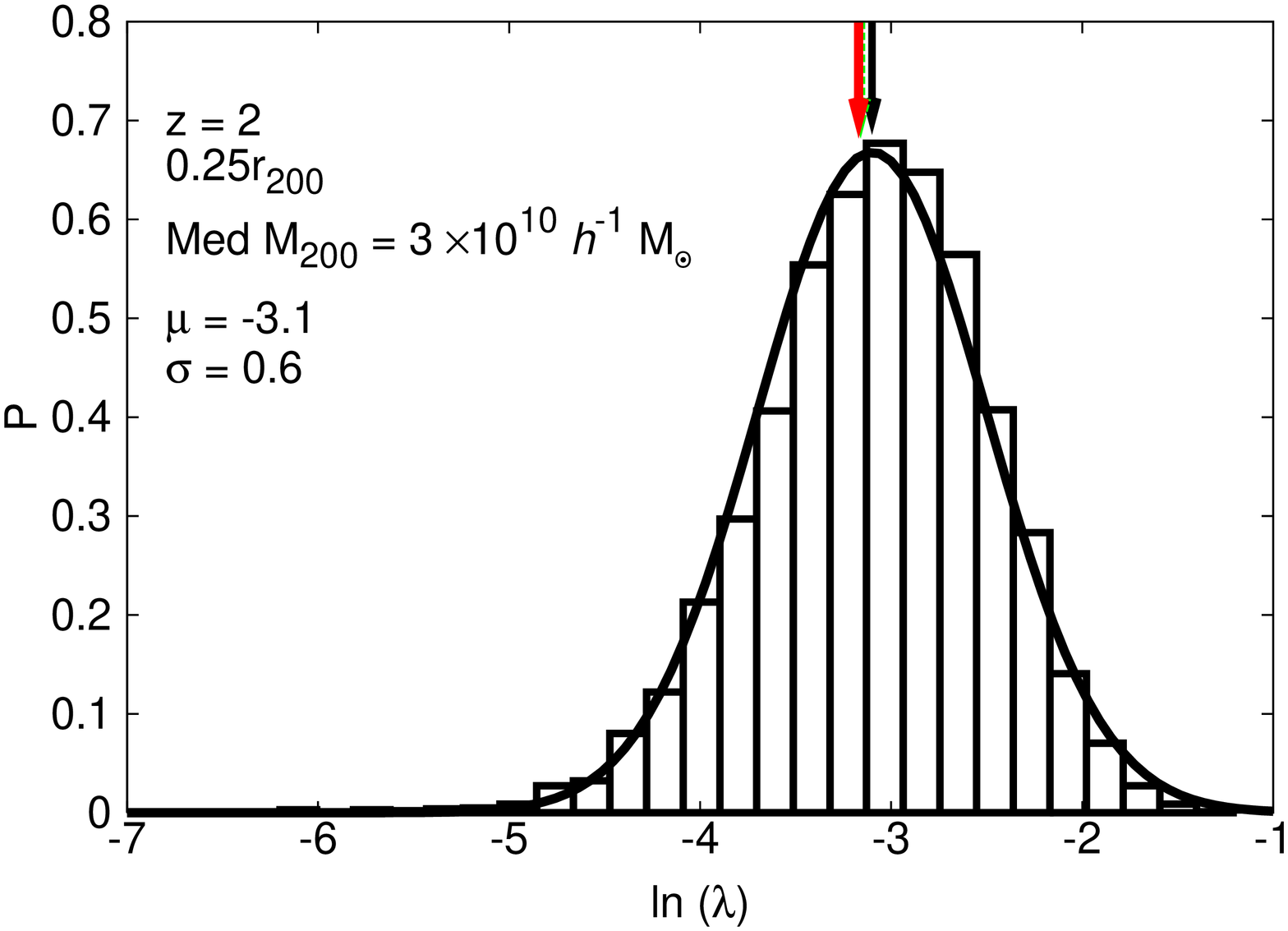} \\
\end{tabular} 
\caption[Effect of feedback on spin parameter distribution.]{\label{spindistall} 
The distribution of log halo spin parameter for dark matter particles in haloes at $z = 0$ (left) and $z = 2$ (right), taken
from the DMONLY {\it WMAP}\,3 runs. The top (bottom) panels show the spin distribution computed using all particles 
within $r_{200}$ $(0.25r_{200})$. The mean and standard deviation of the distribution are computed for each sample 
and are listed in Table \ref{spintable}. The best-fitting Gaussian curve, assuming these parameters, is overlaid.
For comparison, we over-plot arrows representing the mean of the distributions for the two baryon runs with 
the most extreme feedback prescriptions -  the run with no feedback (NOSN\_NOZCOOL; green arrow) and the 
AGN feedback run (AGN; red arrow).  
It is evident from this figure that efficient cooling results in an increased DM spin parameter, especially
within the central region of the halo, while strong (AGN) feedback results in values that are indistinguishable from the 
dark matter only case.}

\end{figure*}

\begin{table*}
\caption[Log-normal fits to the dark matter only spin distributions.]{\label{spintable}  {
The mean, median, standard deviation and skewness of the distribution of $\ln(\lambda)$ parameters for all haloes in each simulation run at $z = 0$, $1$ and $2$. We compare the distribution of spin parameters computed using dark matter particles within $r_{200}$ and that computed using only the 
dark matter particles within the central $0.25r_{200}$ region.  Haloes are required to contain at least 1000 particles within $r_{200}$.  Here $\ln(\lambda_0)$ denotes the mean,
$\ln(\lambda_{\rm med})$ the median and $\sigma$ the standard deviation of the distribution of $\ln\lambda$, see equation (\ref{lognormal}).  Errors represent the 1$\sigma$ confidence intervals and have been determined using bootstrap resampling (1000 bootstrap realisations have been used). \\
 }}
 \centering

 \begin{tabular} {@{}l cccc cccc@{}}% centered columns (4 columns)
 \hline%\hline
&  \multicolumn{4}{c}{Within $r_{200}$} &    \multicolumn{4}{c}{Within $0.25r_{200}$} \\
Simulation & ${\ln({\lambda_0})}$ & $\ln({\lambda_{\rm med}})$ & ${\sigma}$ & skew  &  ${\ln(\lambda_0)}$  & $\ln(\lambda_{\rm med})$ & ${\sigma}$ & skew\\
\hline

$z = 0$   \\
DMONLY {\it WMAP}\,1       &   -3.41      $\,\,\pm{0.03}$   &   -3.38      $\,\,\pm{0.03}$   &    0.60      $\,\,^ {+   0.02           }_{-   0.03             }$             &   -0.47      $\,\,^ {+   0.15           }_{-   0.12             }$             &   -3.50      $\,\,\pm{0.03}$   &   -3.51      $\,\,^ {+   0.05           }_{-   0.02             }$             &    0.63      $\,\,\pm{0.02}$   &   -0.19      $\,\,^ {+   0.16           }_{-   0.14             }$\\

DMONLY {\it WMAP}\,3   &   -3.34      $\,\,\pm{0.01}$   &   -3.29      $\,\,\pm{0.01}$   &    0.62      $\,\,\pm{0.01}$   &   -0.49      $\,\,\pm{0.04}$   &   -3.34      $\,\,\pm{0.01}$   &   -3.31      $\,\,\pm{0.01}$   &    0.62      $\,\,^ {+   0.00           }_{-   0.01             }$             &   -0.36      $\,\,^ {+   0.05           }_{-   0.04             }$
 \\

DMONLY {\it WMAP}\,5 &   -3.33      $\,\,\pm{0.01}$   &   -3.30      $\,\,\pm{0.01}$   &    0.62      $\,\,\pm{0.01}$   &   -0.46      $\,\,\pm{0.03}$   &   -3.37      $\,\,\pm{0.01}$   &   -3.33      $\,\,\pm{0.01}$   &    0.63      $\,\,\pm{0.01}$   &   -0.40      $\,\,^{+   0.03           }_{-   0.04             }$     \\

AGN        &   -3.33      $\,\,\pm{0.01}$             &   -3.28      $\,\,\pm{0.01}$             &    0.60      $\,\,\pm{0.01}$             &   -0.50      $\,\,^ {+   0.06           }_{-   0.07             }$             &   -3.36      $\,\,\pm{0.01}$             &   -3.32      $\,\,\pm{0.01}$             &    0.63      $\,\,\pm{0.01}$             &   -0.49      $\,\,\pm{0.07}$
 
   \\

WDENS        &   -3.32      $\,\,\pm{0.01}$             &   -3.27      $\,\,\pm{0.01}$             &    0.60      $\,\,\pm{0.01}$             &   -0.58      $\,\,\pm{0.06}$             &   -3.34      $\,\,\pm{0.01}$             &   -3.29      $\,\,\pm{0.01}$             &    0.62      $\,\,\pm{0.01}$             &   -0.46      $\,\,\pm{0.04}$  \\

REF        &   -3.33      $\,\,\pm{0.01}$             &   -3.28      $\,\,\pm{0.01}$             &    0.61      $\,\,\pm{0.01}$             &   -0.62      $\,\,^ {+   0.07           }_{-   0.06             }$             &   -3.29      $\,\,\pm{0.01}$             &   -3.23      $\,\,\pm{0.01}$             &    0.61      $\,\,\pm{0.01}$             &   -0.54      $\,\,^ {+   0.04           }_{-   0.05             }$ \\

NOSN\_NOZCOOL     &   -3.29      $\,\,\pm{0.01}$             &   -3.24      $\,\,\pm{0.01}$             &    0.61      $\,\,\pm{0.01}$             &   -0.65      $\,\,\pm{0.05}$             &   -3.12      $\,\,\pm{0.01}$             &   -3.05      $\,\,\pm{0.01}$             &    0.60      $\,\,\pm{0.01}$             &   -0.69      $\,\,\pm{0.04}$\\

\ \\
$z = 1$ \\
DMONLY {\it WMAP}\,3    &   -3.26      $\,\,\pm{0.01}$   &   -3.22      $\,\,\pm{0.01}$   &    0.62      $\,\,\pm{0.01}$   &   -0.62      $\,\,^ {+   0.06           }_{-   0.07             }$             &   -3.18      $\,\,\pm{0.01}$   &   -3.14      $\,\,\pm{0.01}$   &    0.60      $\,\,\pm{0.01}$   &   -0.34 $\,\,\pm{0.05}$ \\

AGN  &   -3.30      $\,\,\pm{0.01}$             &   -3.24      $\,\,\pm{0.01}$             &    0.59      $\,\,\pm{0.01}$             &   -0.51      $\,\,\pm{0.06}$             &   -3.22      $\,\,\pm{0.01}$             &   -3.16      $\,\,\pm{0.01}$             &    0.60      $\,\,\pm{0.01}$             &   -0.50      $\,\,\pm{0.06}$   \\
WDENS           &   -3.30      $\,\,\pm{0.01}$             &   -3.26      $\,\,^ \pm{0.01}$             &    0.60      $\,\,\pm{0.01}$             &   -0.58      $\,\,^ {+   0.07           }_{-   0.06             }$             &   -3.24      $\,\,\pm{0.01}$             &   -3.20      $\,\,\pm{0.01}$             &    0.61      $\,\,\pm{0.01}$             &   -0.52      $\,\,^ {+   0.08           }_{-   0.07             }$\\
REF    &   -3.32      $\,\,\pm{0.01}$             &   -3.27      $\,\,\pm{0.01}$             &    0.61      $\,\,\pm{0.01}$             &   -0.63      $\,\,^ {+   0.08           }_{-   0.07             }$             &   -3.21      $\,\,\pm{0.01}$             &   -3.17      $\,\,\pm{0.01}$             &    0.58      $\,\,\pm{0.01}$             &   -0.41      $\,\,\pm{0.01}$   \\
NOSN\_NOZCOOL      &   -3.31      $\,\,\pm{0.01}$             &   -3.25      $\,\,\pm{0.01}$             &    0.59      $\,\,\pm{0.01}$             &   -0.56      $\,\,\pm{0.05}$             &   -3.12      $\,\,\pm{0.01}$             &   -3.08      $\,\,\pm{0.01}$             &    0.57      $\,\,\pm{0.01}$             &   -0.57      $\,\,\pm{0.05}$ \\
\ \\

$z = 2$\\
DMONLY {\it WMAP}\,3  &   -3.26      $\,\,\pm{0.01}$   &   -3.22      $\,\,\pm{0.01}$   &    0.60      $\,\,\pm{0.01}$   &   -0.54      $\,\,\pm{0.05}$   &   -3.10      $\,\,\pm{0.01}$   &   -3.07      $\,\,\pm{0.01}$   &    0.60      $\,\,\pm{0.01}$   &   -0.40 $\,\,\pm{0.05} $  \\

AGN  &   -3.31      $\,\,\pm{0.01}$             &   -3.27      $\,\,\pm{0.01}$             &    0.58      $\,\,\pm{0.01}$             &   -0.47      $\,\,^ {+   0.04           }_{-   0.05             }$             &   -3.17      $\,\,\pm{0.01}$             &   -3.12      $\,\,\pm{0.01}$             &    0.59      $\,\,\pm{0.01}$             &   -0.55      $\,\,^ {+   0.06           }_{-   0.05             }$ \\
WDENS   &   -3.31      $\,\,\pm{0.01}$             &   -3.26      $\,\,\pm{0.01}$             &    0.59      $\,\,\pm{0.01}$             &   -0.54      $\,\,\pm{0.05}$             &   -3.19      $\,\,\pm{0.01}$             &   -3.16      $\,\,\pm{0.01}$             &    0.58      $\,\,\pm{0.01}$             &   -0.49      $\,\,\pm{0.06}$ \\  
REF  &   -3.32      $\,\,\pm{0.01}$             &   -3.28      $\,\,\pm{0.01}$             &    0.59      $\,\,\pm{0.01}$             &   -0.53      $\,\,\pm{0.05}$             &   -3.20      $\,\,\pm{0.01}$             &   -3.15      $\,\,\pm{0.01}$             &    0.59      $\,\,\pm{0.01}$             &   -0.54      $\,\,\pm{0.04}$    \\
NOSN\_NOZCOOL &   -3.35      $\,\,\pm{0.01}$             &   -3.32      $\,\,\pm{0.01}$             &    0.59      $\,\,\pm{0.01}$             &   -0.62      $\,\,^ {+   0.10           }_{-   0.09             }$             &   -3.16      $\,\,\pm{0.01}$             &   -3.11      $\,\,\pm{0.01}$             &    0.59      $\,\,\pm{0.01}$             &   -0.71      $\,\,\pm{0.06}$ \\ 
\hline
 \end{tabular}
 \end{table*}

\subsubsection{Dark Matter Simulations}

The distribution of dark matter halo spin is typically found to be well characterised by a log-normal distribution
\footnote{Although we note that a more accurate fitting formula for the distribution of halo spins is presented 
in \cite{bib:Bett07}. }
\begin{equation}
\label{lognormal}
P(\lambda) = \frac{1}{\lambda \sqrt{2\pi} \sigma} \exp\left(\frac{-\ln^2(\lambda/\lambda_0)}{2\sigma^2} \right),
\end{equation}
where $\lambda_0$ and $\sigma$ are free parameters, determining the mean and standard deviation respectively.
We confirm this result in Fig.~\ref{spindistall}, where the distribution from the DMONLY{\it WMAP}\,3 
haloes is shown as a histogram. Values for the mean and standard deviation of the sample are also given, 
and are used together with Equation~\ref{lognormal}, to predict the equivalent log-normal spin distribution 
(shown as the solid curve). The left panels show the results for haloes at $z = 0$ while results for the 
$z = 2$ haloes are presented in the right panels. Top panels correspond to spin parameters computed using 
all dark matter particles within $r_{200}$, while the bottom panels are for the inner region ($0.25 \, r_{200}$). 
Table~\ref{spintable} lists values for the mean, standard deviation, median and skewness of $\lambda$ for all runs, and 
at $z=0,1$ and 2.

At $z = 0$ we find $\ln(\lambda_0$) = $-3.34 \pm 0.03$ (or $\lambda_0 = 0.036$) and $\sigma = 0.62\pm 0.01$, for 
our DMONLY{\it WMAP}\,3 run, when all dark matter particles within $r_{200}$ are considered. These values are in 
good agreement with those found in previous analyses such as \cite{bib:Bullock01}, who obtained best-fit values of 
$\lambda_0 = 0.035 $ and $\sigma = 0.5$; \cite{bib:Bailin05}, who measured ${\lambda_0} = 0.035$  and $\sigma = 0.58$; 
and \cite{bib:Maccio08}, who found a mean value of ${\lambda_0} = 0.034$ and $\sigma = 0.59$. Our median spin parameter 
($\lambda_{\rm med}=0.037$) is also in good agreement with the analysis of \cite{bib:Bett07}, who found 
$\lambda_{\rm med}=0.0367 - 0.0429$ (depending on the definition of the halo).  
The spin parameters of the higher redshift ($z = 1$ and 2) samples are found to be slightly higher 
than those at $z=0$. For example, at $z=2$, the mean spin parameter of the DMONLY{\it WMAP}\,3 haloes
is $\lambda_0=0.038$, around 8 per cent higher. At least part of this (statistically significant) shift can be 
explained by a very weak dependence of spin with halo mass (see below and \citealt{bib:Munoz11}), given that the mean mass of our haloes
at $z=2$ is lower than at $z=0$. 

We also checked whether excluding unrelaxed haloes (centroid shift $\geq$ $0.07\, r_{200}$) from our sample made any difference to our results, as 
\cite{bib:Maccio06} found that this reduced the mean spin parameter by $\sim 15$ per cent and \cite{bib:JeesonDaniel11} found that spin correlates with the relaxedness of the halo\footnote{A low value of relaxedness corresponds to a relaxed halo.} (they demonstrated that this is due to the strong anti-correlation of both spin and relaxedness with concentration). Our results are
broadly in agreement: at $z=0$ we find a reduction in $\lambda_0$ 
of around 12 per cent while at $z=2$, this increases to 17 per cent.

At $z = 0$, we find that the inner region ($r<0.25 \, r_{200}$, shown in the bottom left panel of 
Fig.~\ref{spindistall}) is characterised by nearly the same distribution as the overall
 halo. However, this does not hold at $z = 2$ (bottom right panel) where we find that the mean spin 
parameter is significantly higher  ($\ln (\lambda_0)$ = $-3.10 \pm 0.01$) than that of the whole halo 
($\ln(\lambda_0)$ = $-3.26 \pm 0.01$).  Interestingly, this difference is even  stronger for the relaxed 
sample at $z = 2$, where the inner region of the relaxed dark matter haloes exhibits a mean spin, 
$\ln(\lambda_0)=-3.21\pm0.01$ while the mean spin computed over the whole halo is found  to be $\ln(\lambda_0)=-3.42\pm0.01$.  The difference is likely due to the increased merger rate at $z = 2$.  The lower mass haloes at $z = 2$ are strongly affected by frequent mergers.  Subhaloes transfer their angular momentum to the inner region as they fall into the centre of the halo.  After a major merger the inner part of the halo loses angular momentum to the outer part of the halo due to dynamical friction (inside-out transfer of angular momentum) as discussed in \cite{bib:Sharma12}.  

\begin{figure*}
\begin{center}
\begin{tabular}{c c}
\includegraphics[width=8cm,height=8cm,angle=-90,keepaspectratio]{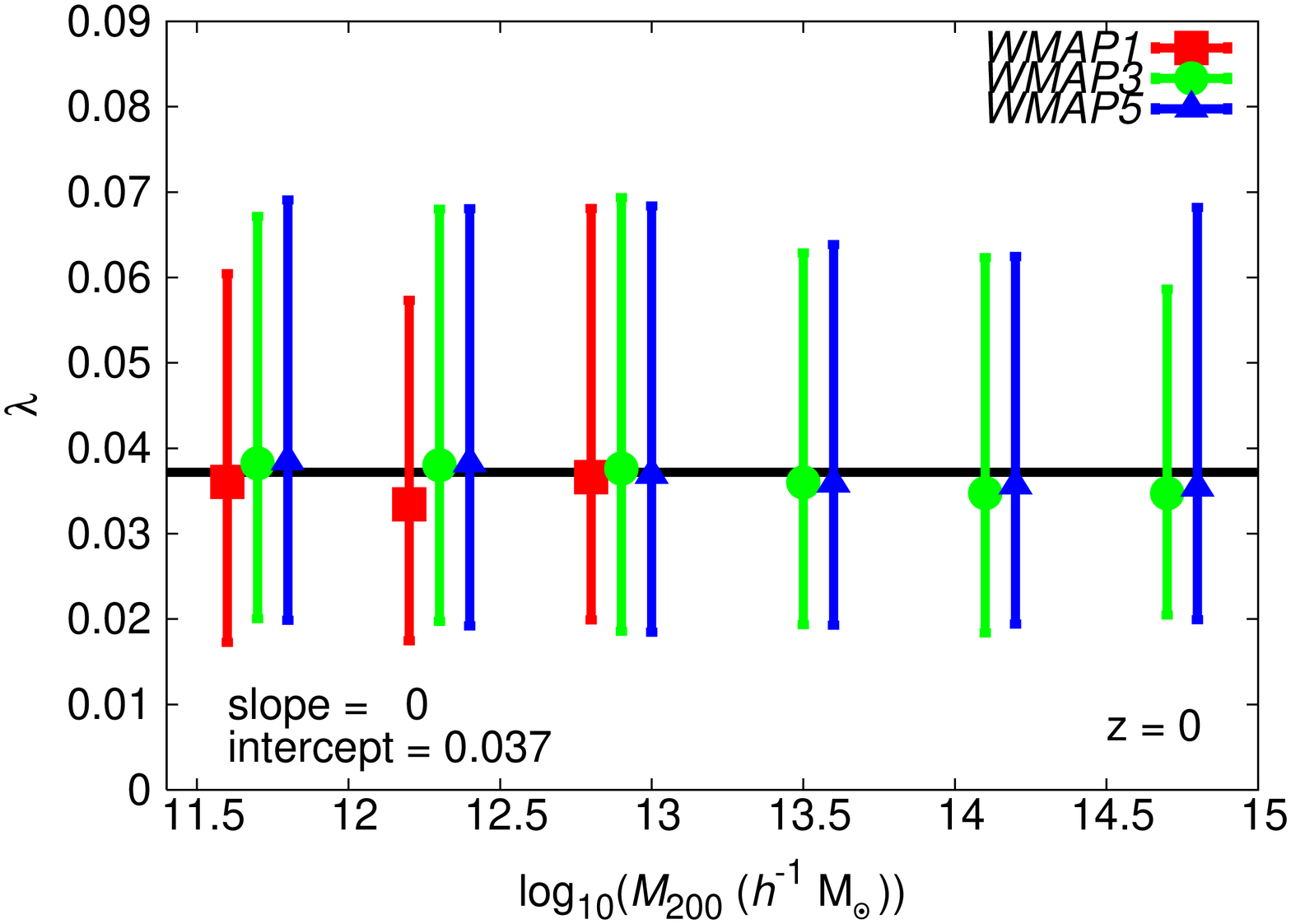} &
\includegraphics[width=8cm,height=8cm,angle=-90,keepaspectratio]{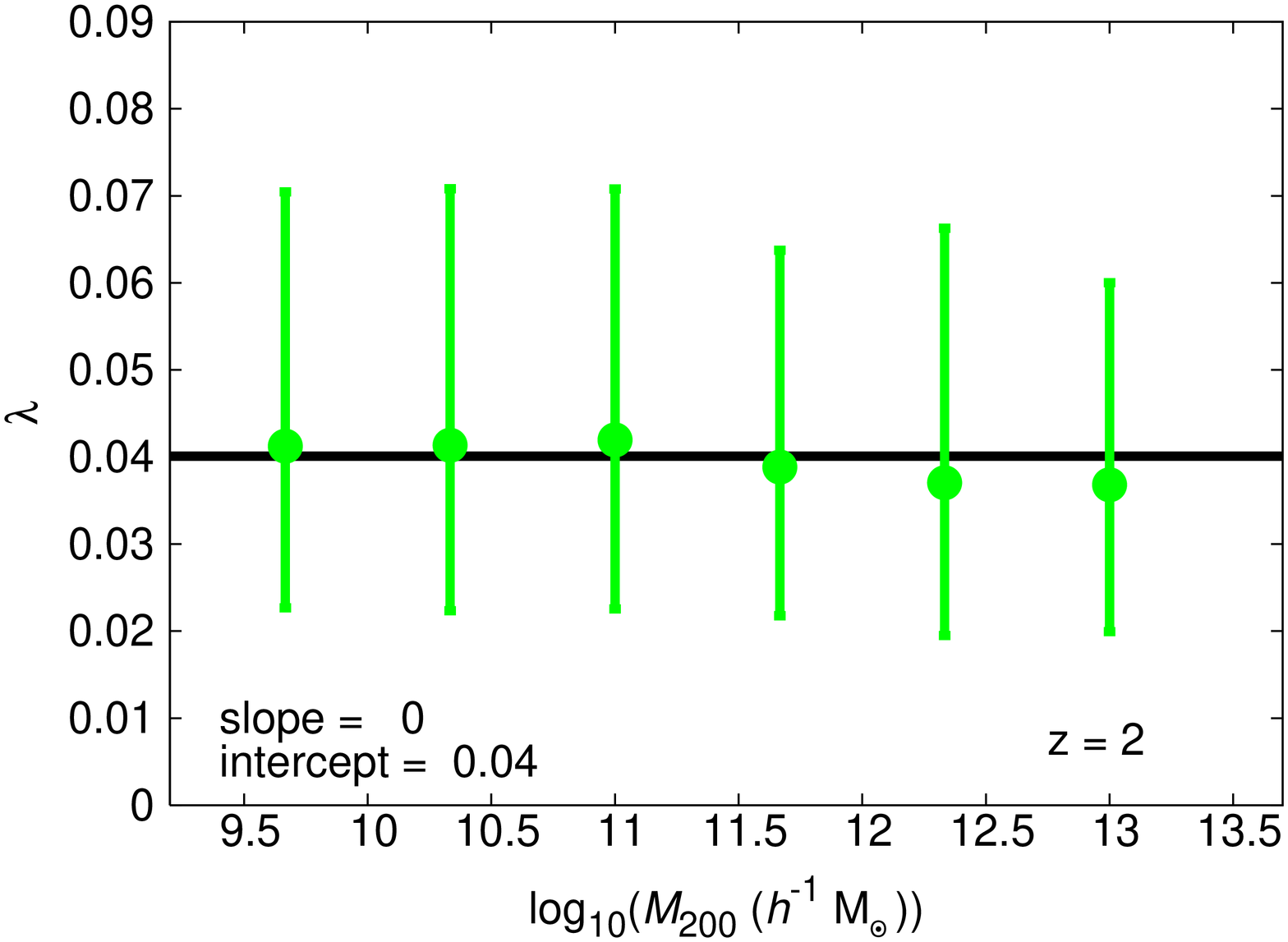} \\
\end{tabular}
\end{center}
\caption[Mass dependence of DM haloes]{\label{dmvm_all_spin} The left (right) panel shows the spin parameter versus $M_{200}$ in the dark matter
only simulations at $z = 0$ (2). In these panels we show the median value within each mass bin and errors represent the 1$\sigma$ scatter in the mass bin.  Haloes are required to contain at least 1000 particles and mass bins at least 20 haloes.  We plot the results from the three cosmologies we have considered: {\it WMAP}\,1 is depicted using red squares, {\it WMAP}\,3 using green circles and {\it WMAP}\,5 using blue triangles.  For clarity we show the median of the whole {\it WMAP}\,3 sample as a solid line and present this value in each panel.  We note that the spin parameter remains unchanged as we vary the cosmological parameters and that there is also no clear dependence on mass.
}
\end{figure*}

\begin{figure*}
\begin{tabular}{l l}
\includegraphics[width=8cm,height=8cm,angle=-90,keepaspectratio]{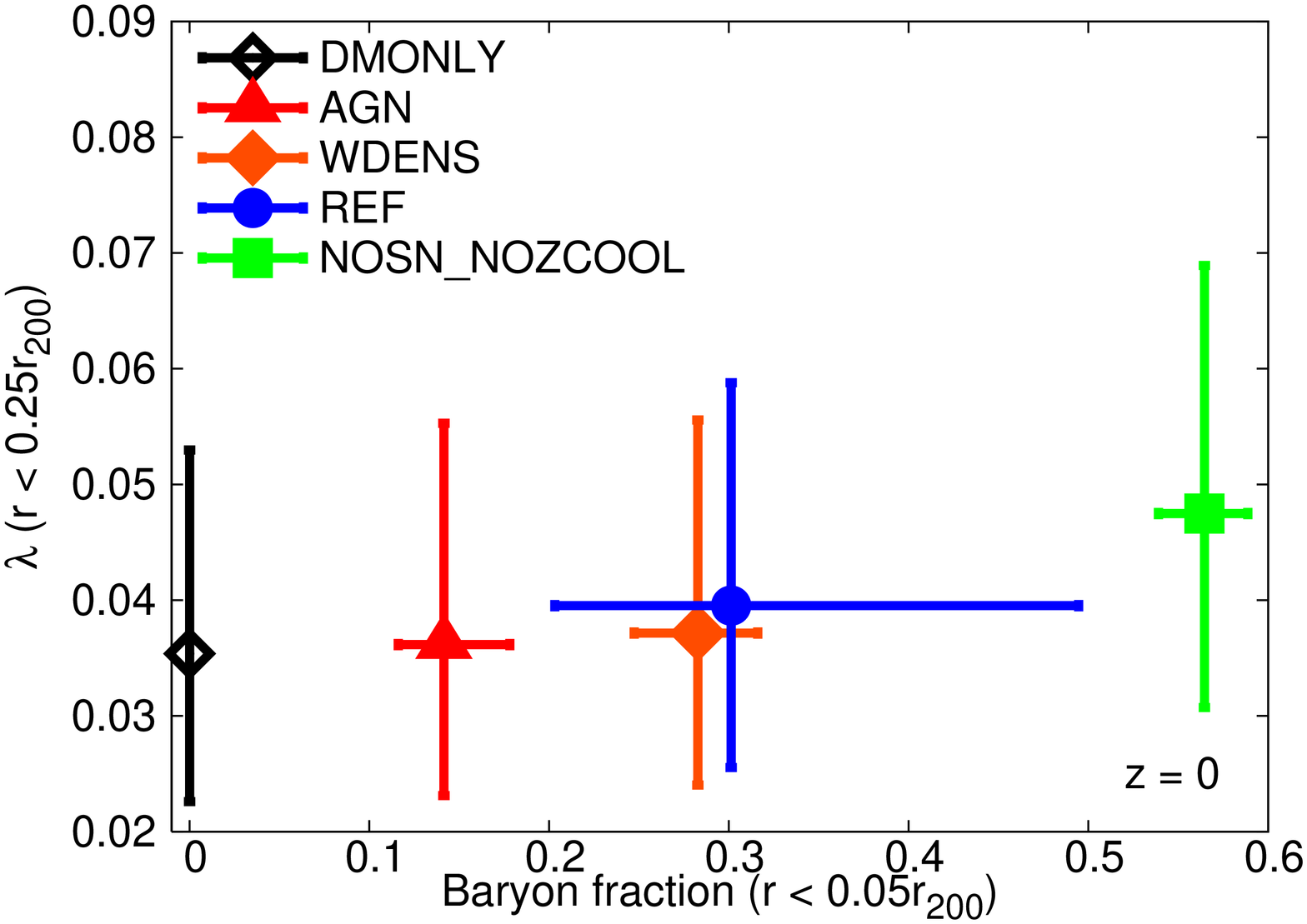} &

\includegraphics[width=8cm,height=8cm,angle=-90,keepaspectratio]{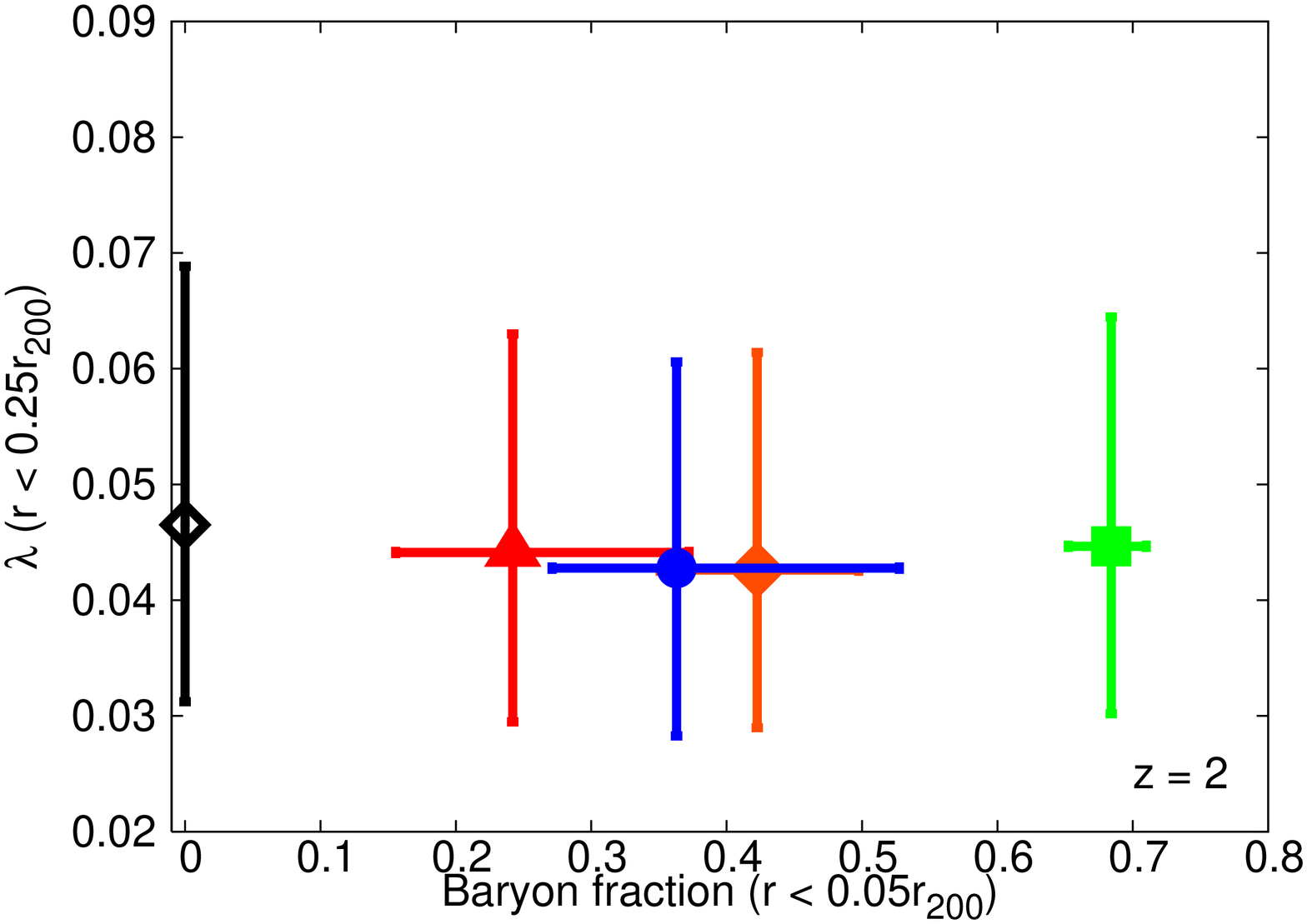} \\

\end{tabular}
\caption[Effect of baryons on the shape distribution of
  haloes.]{\label{spinvBF_all} This figure illustrates how the DM halo spin depends on the central baryon fraction of the halo (used as a proxy for galaxy formation efficiency).  The $z = 0$ (2)  haloes are shown in the left (right) panel; the median halo mass for the sample is $\sim 8 \times 10^{11}$ ($3 \times 10^{10}$) $ h^{-1} {\rm M}_{\odot}$.  We plot the median values and error bars show the $1\sigma$ halo-to-halo scatter.  The DM spin parameters are computed within $0.25\, r_{200}$ while the baryon fraction is calculated within $0.05\, r_{200}$. The two quantities are positively correlated 
at $z=0$. }
\end{figure*}

We have also considered the effects of cosmology on the spin parameter distribution of dark matter haloes, comparing the spin distributions from dark matter only simulations run with the {\it WMAP}\,1, {\it WMAP}\,3 and {\it WMAP}\,5 cosmological parameters at $z = 0$. The mean, median, standard deviation and skewness for the distributions of the log of the spin parameters are also listed in Table \ref{spintable}.  It is evident from this table (as in \citealt{bib:Maccio08}) that the spin distribution is not sensitive to the exact choice of cosmological parameters.  We note that the largest change in the models we have considered is the value of $\sigma_8$, which varies from 0.7 in {\it WMAP}\,3 to 0.9 in {\it WMAP}\,1.  Haloes in a lower $\sigma_8$ cosmology are expected to form later and be less concentrated than those in a higher $\sigma_8$ cosmology (\citealt{bib:Navarro97}).  The spin distribution itself remains almost unchanged for the range of parameters explored here, although we note a slight decrease in the spin parameter for the {\it WMAP}\,1 haloes, consistent with the anti-correlation of spin with concentration (as discussed in \citealt{bib:Maccio08} and \citealt{bib:JeesonDaniel11}).  Haloes formed in this cosmology are expected to have the highest concentrations and hence may be expected to have the lowest spin values.

We explore the relationship between spin and mass in Fig. \ref{dmvm_all_spin}.  In this figure we show the
median spin value and the $1\sigma$ halo-to-halo scatter for several mass bins, for each of the three cosmologies 
that we have considered at $z = 0$ (left panel) and for the {\it WMAP}\,3 cosmology only at $z = 2$ (right panel). 
We use least-squares fitting of the median in each mass bin to determine the slope and intercept of the $\lambda-\log_{10}(M_{200})$ relation in each simulation 
set, assuming a pivot mass of $M_{\rm pivot} = 10^{12}\, h^{-1}\, \mbox{M}_\odot$. The spin parameter
is largely insensitive to halo mass; a weak trend is seen (which may explain the slight redshift dependence seen earlier, due to our two samples covering a different mass range), but is not statistically significant (we find the best-fitting slope to be consistent with zero at the $1\sigma$ level in all cases - where the $1\sigma$ confidence is determined using bootstrap resampling). 
Once again, we can clearly see that the effect of cosmology on the spin parameter is negligible.

\subsubsection{Effect of baryons on DM halo spin}
\label{baryonspin}

Following on from the dark matter only case, we now consider the distribution of DM halo spin parameters computed from the baryon runs. We only
include dark matter particles in our determination of specific angular momentum to directly assess the effects of baryons on the dark matter component. 
The mean, median, standard deviation and skewness for the distribution of spin parameters in each run is presented in Table \ref{spintable}.  For ease of comparison with the dark matter only case, we show in Fig. \ref{spindistall} the mean value for two of the most extreme feedback schemes we have considered: the no feedback case (NOSN\_NOZCOOL; green arrow) and the stellar and AGN feedback run (AGN; red arrow).   

From Fig.~\ref{spindistall} and Table~\ref{spintable} we see that, at $z = 0$, the baryons induce a small increase in the mean (and median) value
of $\lambda$ within $r_{200}$. For NOSN\_NOZCOOL (where feedback is absent), the increase is around 5 per cent, but for the more realistic\footnote{The AGN run reproduces the $z=0$ observed relations between black holes and the mass and velocity dispersion of their host galaxies 
 (\citealt{bib:Booth09}), as well as the observed optical and X-ray properties of the groups in which they reside (\citealt{bib:McCarthy10}). Whether the AGN model is also the most realistic simulation at high redshift is unclear.}
AGN run (where feedback is strongest), 
it is only $\sim$1 per cent. At higher redshift ($z=1$ and 2), the effect of baryons on the mean spin parameter is weak but tends to decrease the spin in all cases. It is also interesting to note that, within the central region (0.25$r_{200}$), the baryon runs have distributions that are skewed to lower values than the DMONLY run (also true for the whole halo at $z = 0$).

The effects of baryons are stronger in runs with weak or no feedback, when we consider dark matter particles within the central region
(consistent with the findings of \citealt{bib:Bett10}, who studied the effects of baryons using a more restricted set of simulations).
At $z = 0$, the mean spin parameter computed within the central region increases by 25 per cent in NOSN\_NOZCOOL (but again, is almost unchanged in AGN). This effect is shown
more clearly in Fig. \ref{spinvBF_all}, where we plot\footnote{Note: To ensure that mass dependent trends do not influence our comparison we do not include the 400$h^{-1}\, \rm{Mpc}$ box dark matter simulation when directly comparing runs at $z = 0$.  This leaves us with a sample of 4329 haloes in the DMONLY simulation at $z = 0$ with a median mass of $8.3\times 10^{11} \,  h^{-1}\, \mbox{M}_\odot$ comparable to the median mass in the baryon runs.} the median spin parameter for the central region ($0.25 \, r_{200}$) against the 
baryon fraction within the very centre ($0.05 \, r_{200}$). The latter is used as a proxy for galaxy formation efficiency, as weaker feedback
leads to stronger gas cooling and a larger central baryon fraction, associated with the galaxy at the centre of the halo. 
It is clear that there is a positive correlation between the two quantities
at $z=0$ but this correlation has all but disappeared at $z=2$.  

We find that baryons do not influence the mass dependence of the spin parameter (in almost all cases, the slope is consistent with zero at $1\sigma$
and in all cases it is consistent with zero at $3\sigma$). The trends found in this analysis are also preserved when only the sub-set of relaxed haloes is considered.

\subsubsection{Origin of spin variations in the DM}

\begin{figure*}
\begin{tabular}{l l}
\includegraphics[width=8cm,height=8cm,angle=-90,keepaspectratio]{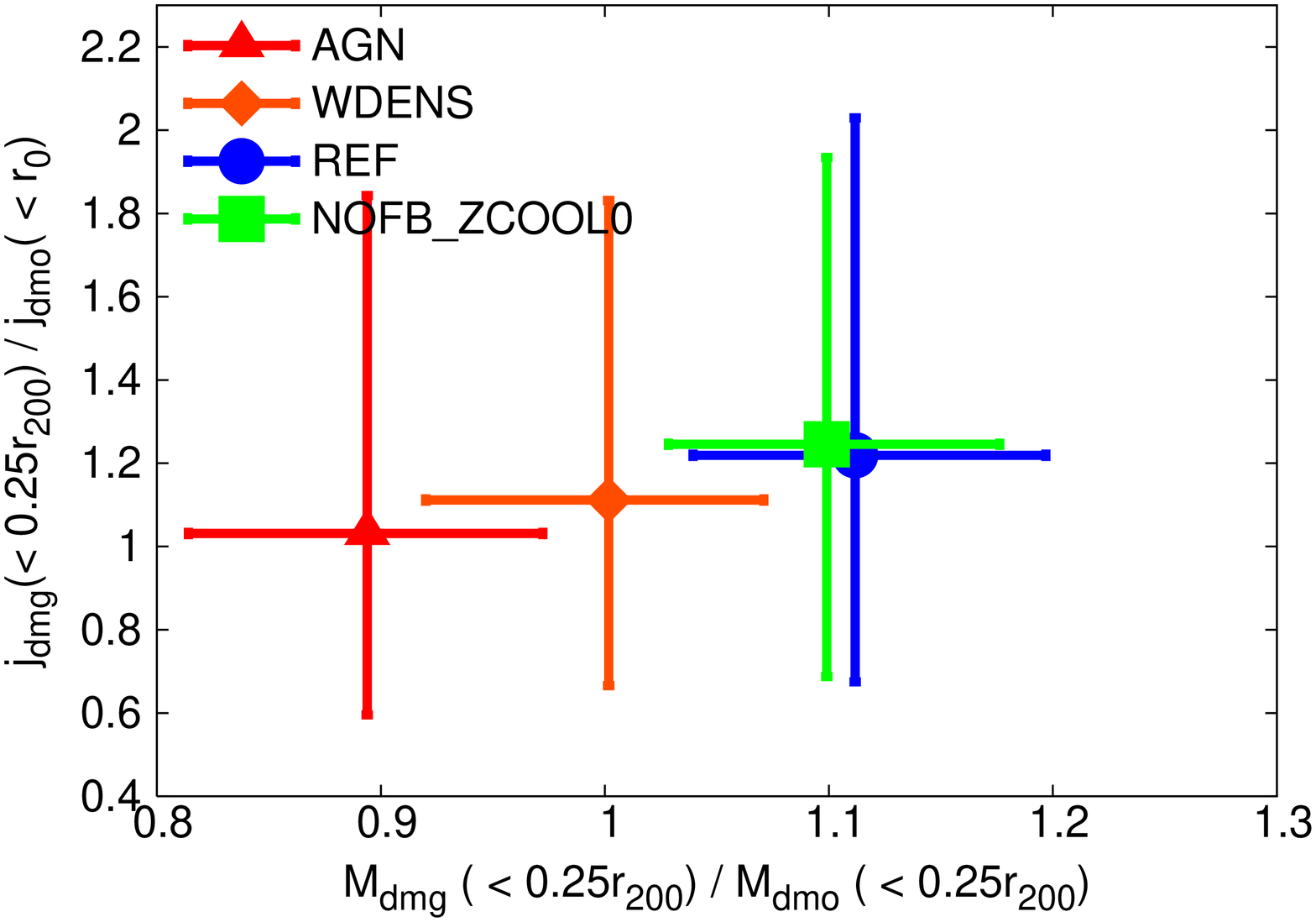} &
\includegraphics[width=8cm,height=8cm,angle=-90,keepaspectratio]{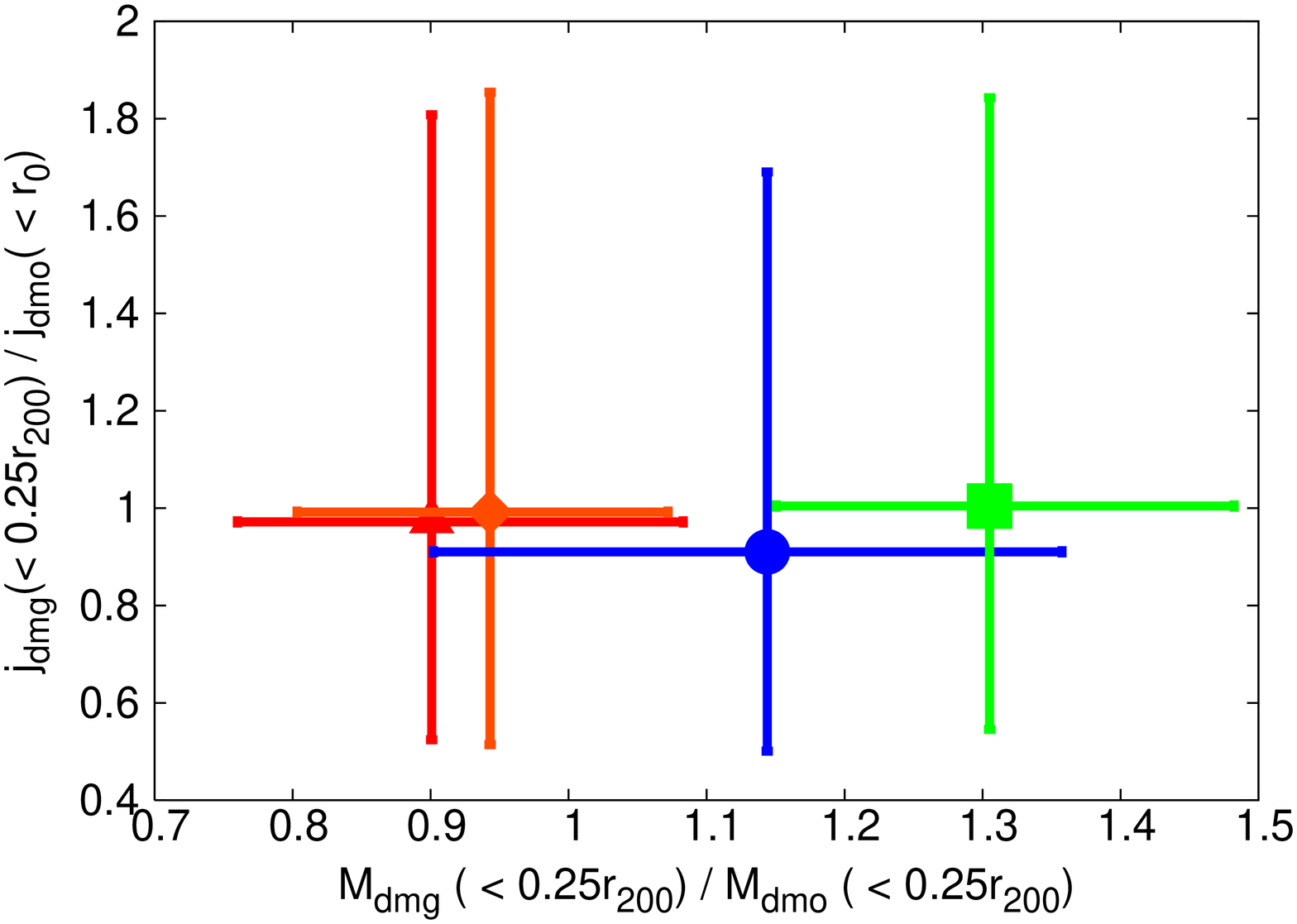} \\

\end{tabular}
\caption[]{The ratio of specific angular momentum of the dark matter within $0.25r_{200}$ from the baryon runs $j_{\rm dmg}$ $(< 0.25r_{200})$ to that of the 
corresponding dark matter only run within a radius containing an equivalent mass ($j_{\rm dmo}( < r0)$).  In the left (right) panel this ratio is shown against the DM mass ratio within $0.25r_{200}$ for the 500 (1000) most massive haloes that have been matched between runs at $z = 0$ ($z = 2$).  The median halo mass for the $z = 0$ sample is $\sim 6 - 8 \times 10^{12}$$ h^{-1} {\rm M}_{\odot}$ while at $z = 2$ the median mass is $\sim 7 - 9 \times 10^{11}$$ h^{-1} {\rm M}_{\odot}$.   Symbols depict the median values while the error bars represent the $1\sigma$ scatter.  \label{jratiovBF_all}}
\end{figure*}

From equation \ref{lambdadef}, it is clear that the increase in $\lambda$ must be due to an increase of the specific angular momentum of the dark matter,
or a decrease in the enclosed mass. As discussed in \cite{bib:Bett10}, an increase in specific angular momentum can result from two factors. The first is the transfer of angular momentum from gas to dark matter via tidal torques and dynamical friction (see for example \citealt{bib:Tonini06b}).  The second is the contraction of the dark matter 
in response to the deepened potential well of the system (assuming angular momentum is conserved). 

To explore this further, we apply the approach of \cite{bib:Bett10} to our haloes at $z = 0$ and $z = 2$.
We begin by matching haloes between the DMONLY and baryon runs; a match is identified by selecting the halo from the baryon run that contains the most of the $1000$ most-bound particles from the dark matter simulation. We then 
compute the DM mass ratio (within $0.25 \, r_{200}$) for each matched pair of haloes. If the DM halo has contracted due to the presence of baryons, the mass should
increase with respect to the mass in the DMONLY halo. We then compare the specific angular momentum of the dark matter within $0.25 \, r_{200}$ from the 
baryon run $j_{\rm dmg} (< 0.25 \, r_{200})$, to the corresponding halo in DMONLY, within a radius $r_0$ containing the equivalent mass, $j_{\rm dmo} (< r_0)$.  If the halo contracts due to the presence of baryons and conserves its angular momentum (i.e. if there is no transfer from the baryons) then this ratio will be unity.  

The results for each baryon run are shown in Fig. \ref{jratiovBF_all}, where we show the ratio of specific angular momentum versus the mass ratio within 
$0.25 \, r_{200}$ for the 500 (1000; 500 from each box) most massive haloes that we are able to match\footnote{A match is found in 92 per cent of the cases (and at least 72 per cent of these matches have $M_{200}$ ratios within 30 per cent and minimum potential positions within 30 per cent of $r_{200}$ of each other).} between runs at $z = 0$ ($z = 2$). 
From this figure, it can clearly be seen that the AGN feedback acts to reduce the mass in the central regions, but as galaxy formation efficiency increases, 
we see the effect of halo contraction (see \citealt{bib:Eggen62, bib:Zeldovich80}). This effect was studied in detail for the same set of runs by \cite{bib:Duffy10}.

If we now consider the ratio of specific angular momentum at $z = 0$, we find that the AGN result is consistent with no transfer, but as galaxy formation 
(and thus halo contraction) is more efficient, a net transfer from the baryons to the dark matter does occur (this is also seen in \citealt{bib:Kaufmann07, bib:Abadi10, bib:Bett10}).  At $z = 2$ there is no evidence for a trend in the transfer of angular momentum with galaxy formation efficiency.

In summary, we find that in all baryon runs with efficient galaxy formation at $z = 0$ there is evidence for the transfer of angular momentum from baryons to dark matter.  This is likely due to the fact that satellites are able to survive infall into the host halo.  The effect of feedback on the halo central density and on satellites is discussed in, for example, \cite{bib:Pedrosa10}, \cite{bib:Duffy10} and \cite{bib:Libeskind10}.  The haloes in the weak/no feedback runs also experience a net increase in the dark matter within the central region (contraction).  These two effects combine, resulting in an increased spin parameter in the baryon runs when compared to dark matter only simulations.
In the lower mass haloes at $z = 2$ there is no evidence for the transfer angular momentum within the central regions.  The slight decrease we see in the spin parameter at $z = 2$ can be understood as arising from the decreased specific angular momentum (shallower potential well) in the case of expansion in the strong feedback runs (AGN and WDENS) and as a result of the increased circular velocity in the case of contraction in the weak/no feedback runs (NOSN\_NOZCOOL and REF).

\subsection{Halo Shapes}
\label{shaperesults}

\begin{figure*}
\begin{center}
\begin{tabular}{c c}
\includegraphics[width=8cm,height=8cm,angle=-90,keepaspectratio]{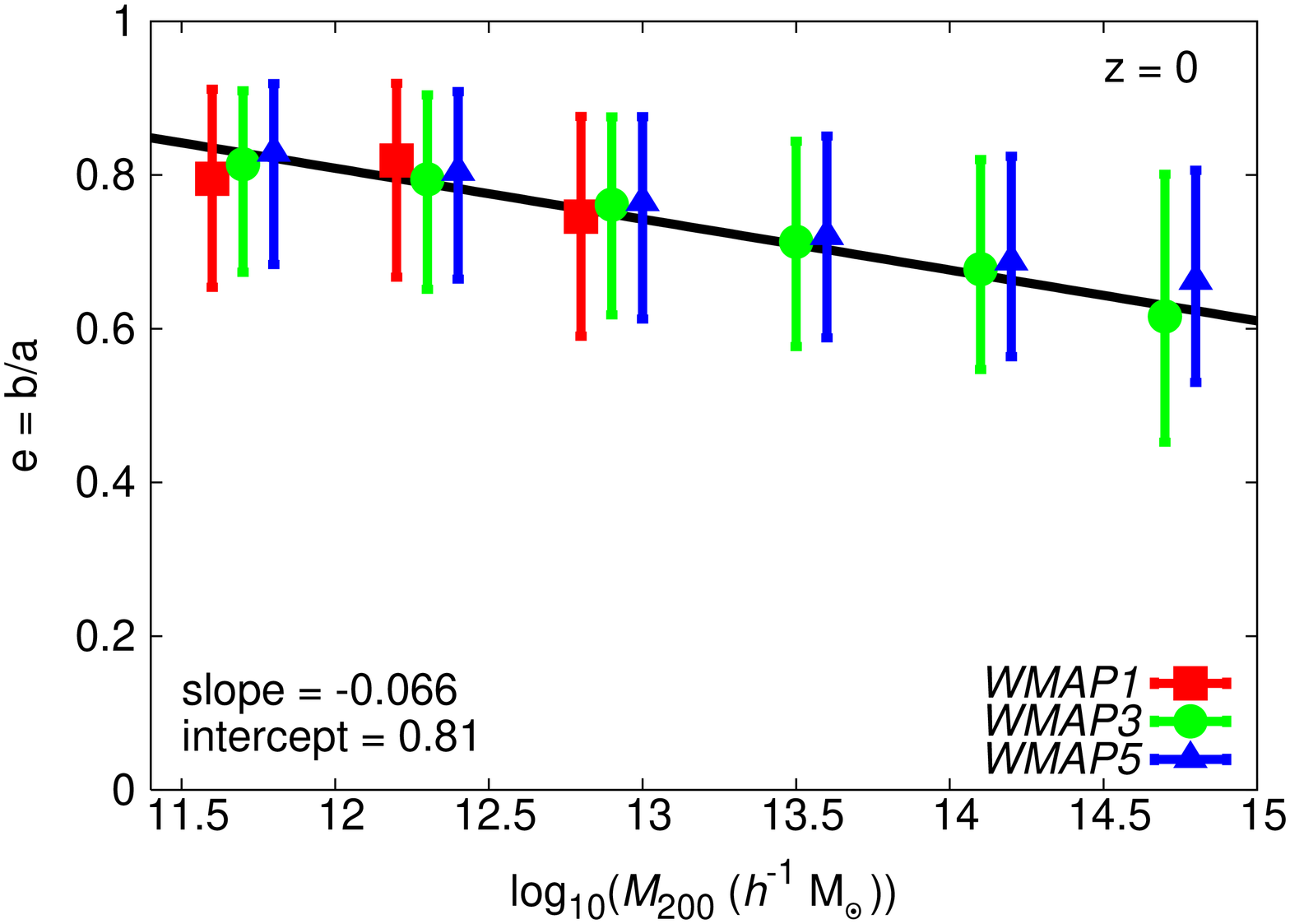} &
\includegraphics[width=8cm,height=8cm,angle=-90,keepaspectratio]{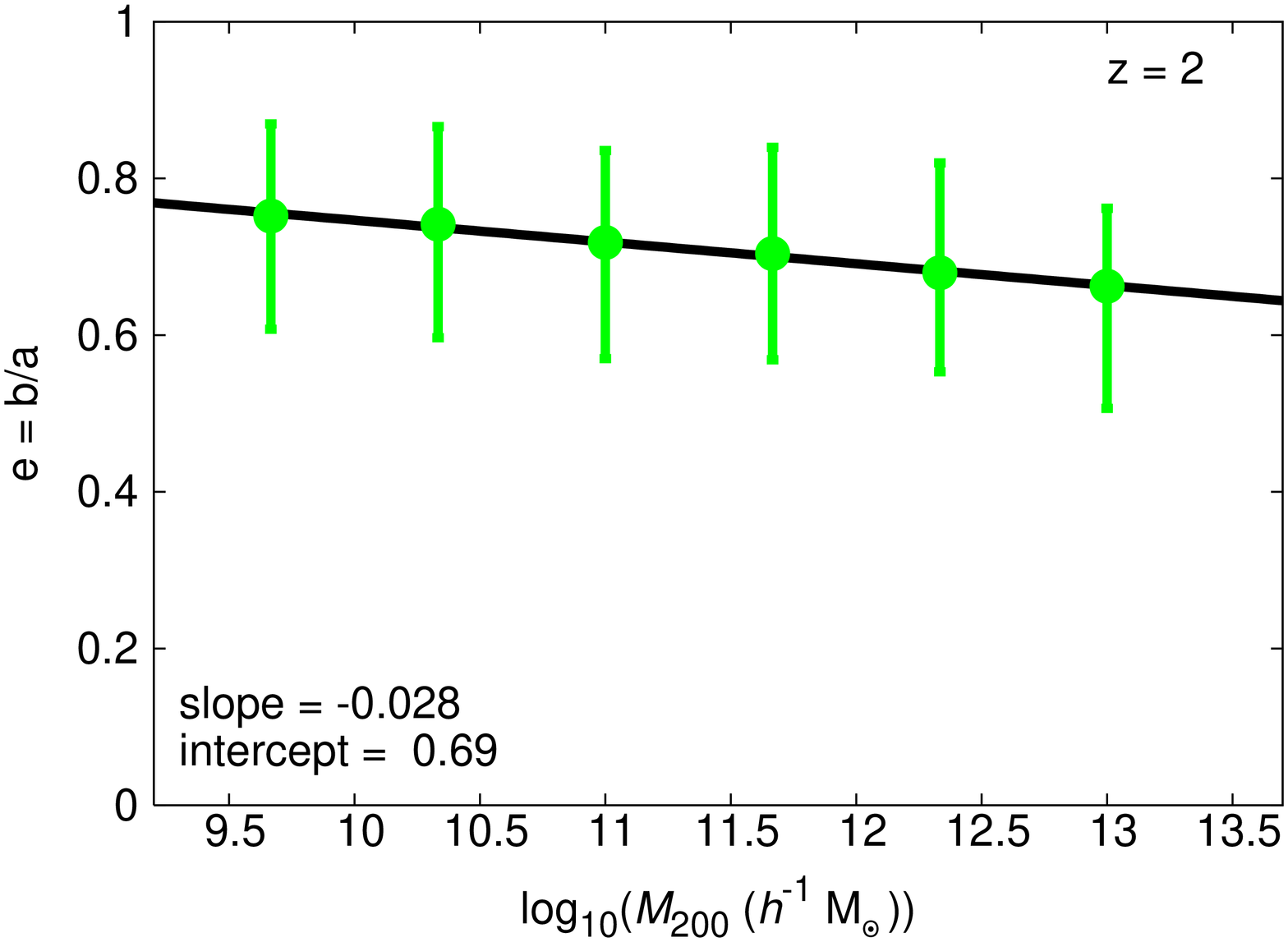} \\

\includegraphics[width=8cm,height=8cm,angle=-90,keepaspectratio]{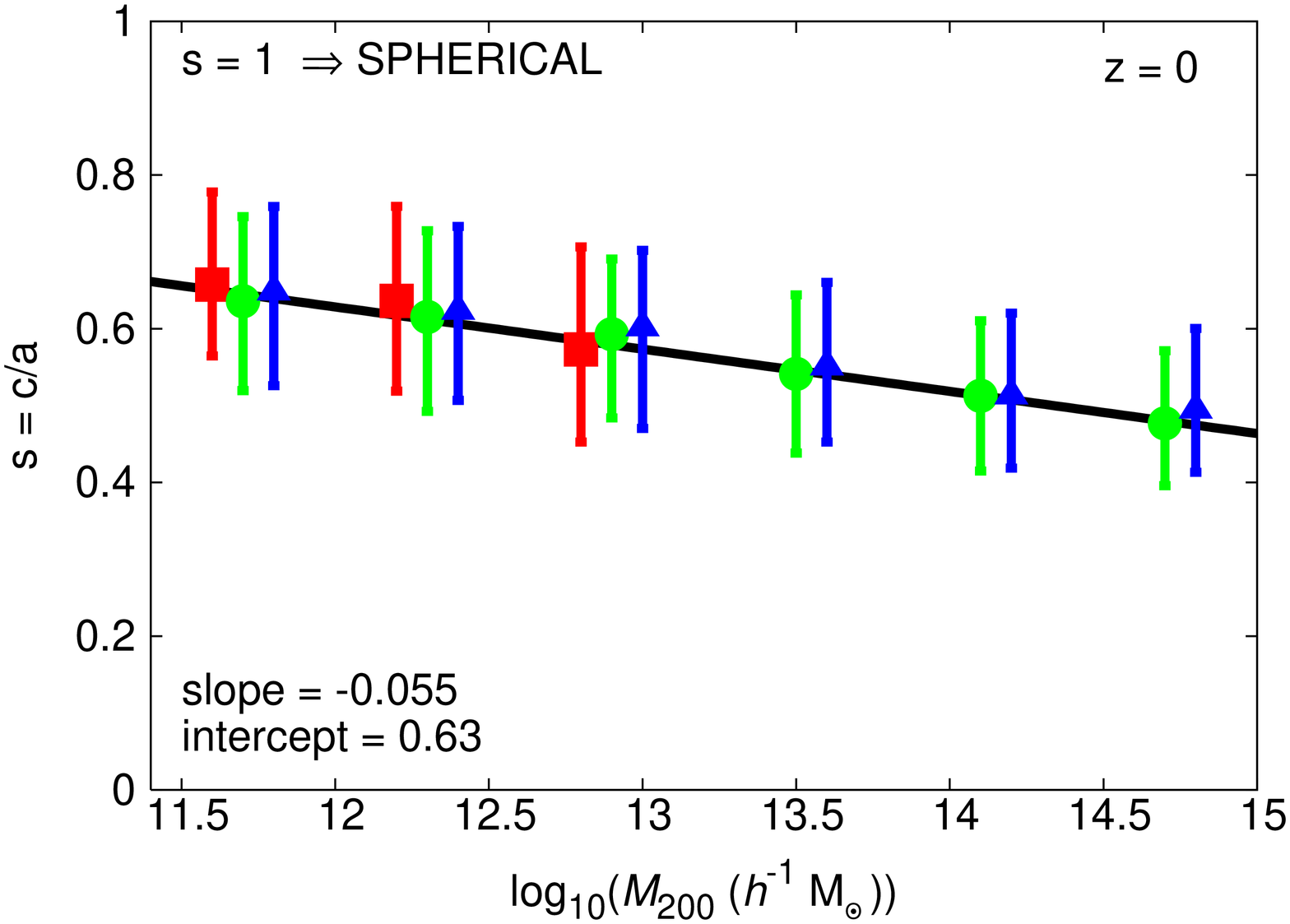} &
\includegraphics[width=8cm,height=8cm,angle=-90,keepaspectratio]{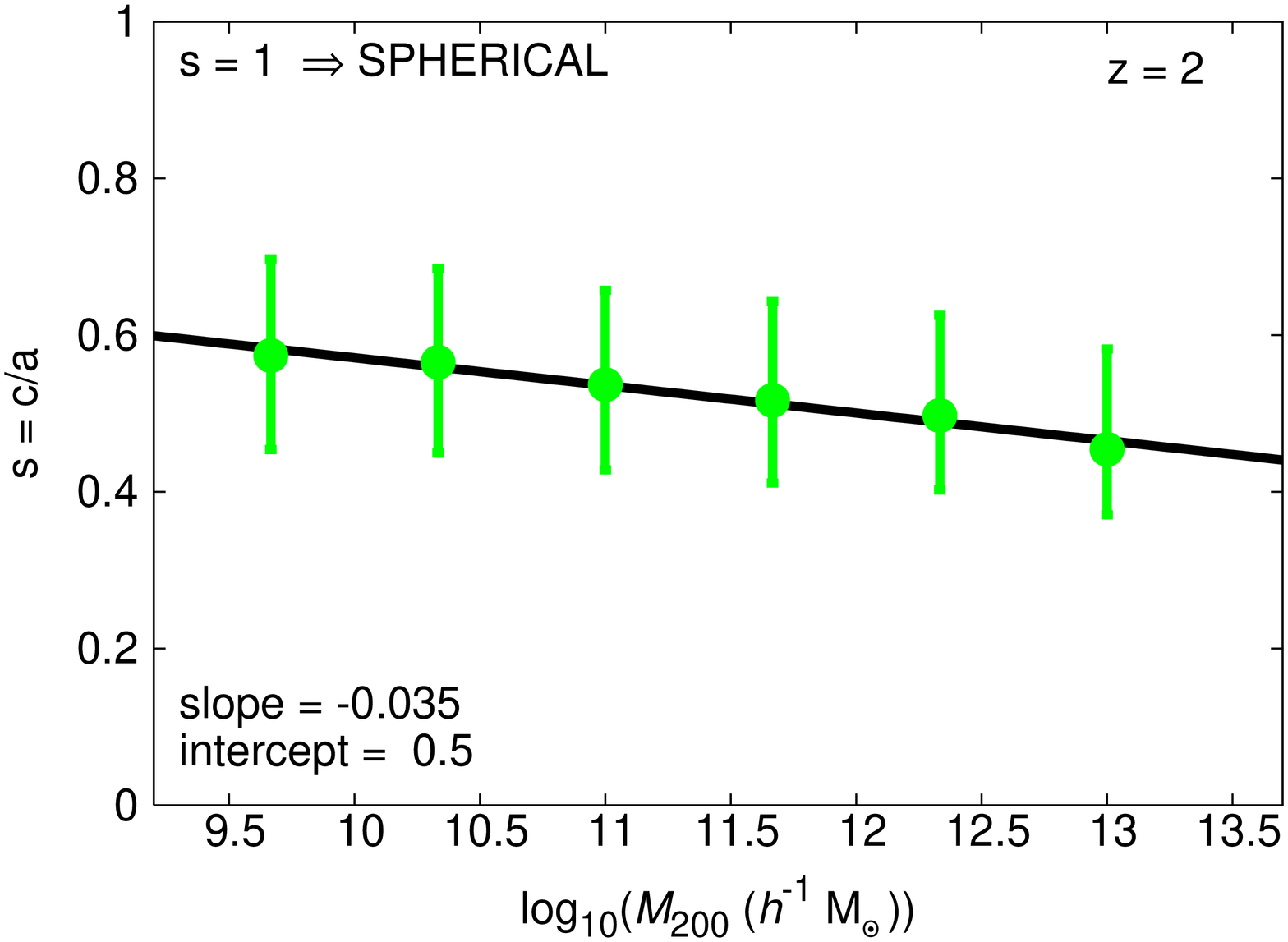} \\

\includegraphics[width=8cm,height=8cm,angle=-90,keepaspectratio]{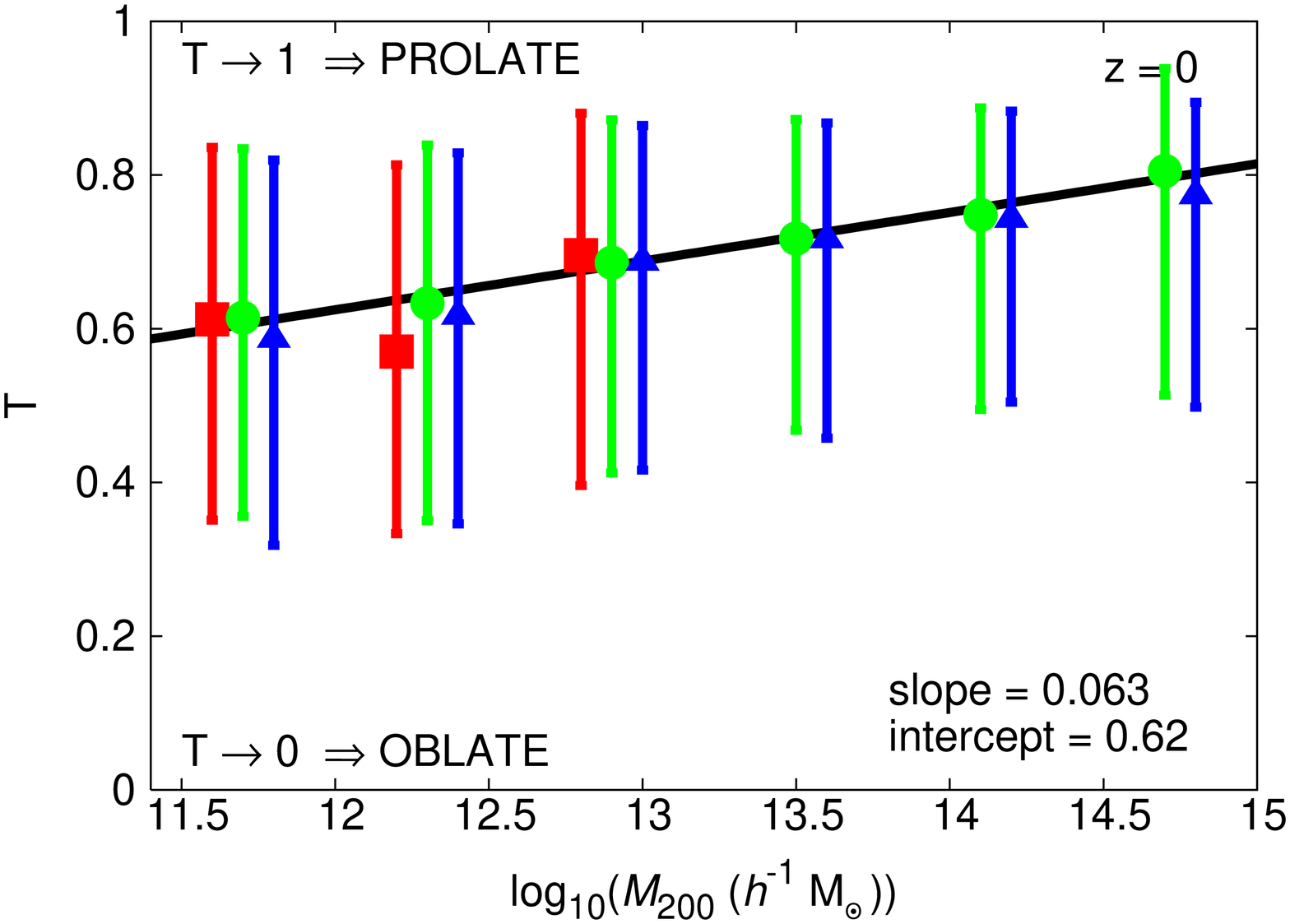} &
\includegraphics[width=8cm,height=8cm,angle=-90,keepaspectratio]{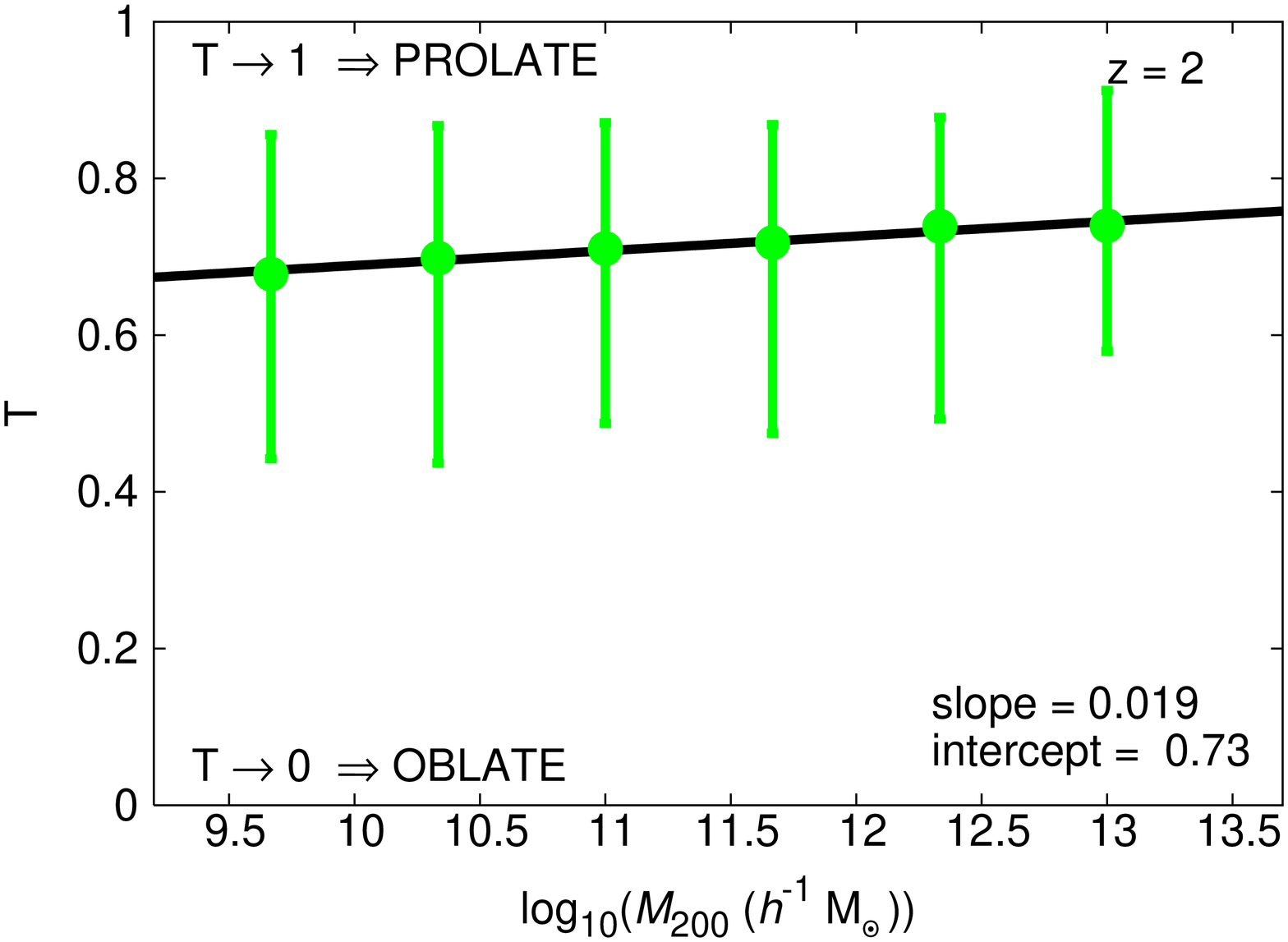} \\
\end{tabular}
\end{center}
\caption[Mass dependence of DM haloes]{\label{dmvm_all} The panels show, from top to bottom, how the axis ratios ($e=b/a$, $s=c/a$) and triaxiality ($T$) 
of dark matter haloes in the DMONLY runs scale with halo mass ($M_{200}$) 
at $z = 0$ (left) and at $z = 2$ (right). 
We show the median value within each mass bin and the error bars represent the 1$
\sigma$ intrinsic scatter. The results from the {\it WMAP}1 haloes are depicted using red squares, 
{\it WMAP}3 using green circles and {\it WMAP}5 using blue triangles. 
 Haloes are required to contain at least 1000 particles and mass bins at least 20 haloes.  Least-squares lines of best-fit (assuming a pivot mass of $10^{12}\,
  h^{-1}\, \mbox{M}_\odot$) are computed for each simulation and are presented in Table \ref{dmmasstable1_all}.  The best-fit line for the {\it WMAP}\,3 haloes 
  is shown in each panel.}
\end{figure*}

\begin{table*}
  \caption[mass relation parameters.]{ {\label{dmmasstable1_all}}
Best-fit slope and intercept values for the mass dependence of halo shape parameters ($e, s$ and $T$), assuming 
a linear relationship between each shape parameter and $\log_{10}(M_{200})$, about a pivot mass of  $10^{12}\, h^{-1}\mbox{M}_\odot$.
Fits are performed for dark matter particles within $r_{200}$, for all runs considered at $z=0,1$ and $2$.  
The errors correspond to 1$\sigma$ confidence intervals, determined by bootstrap resampling 1000 realisations of the data.}
 \vspace{0.15cm}
 \begin{tabular} {@{}l cccccc |@{}}% centered columns (4 columns)
   \hline
  % & \multicolumn{4}{c} {$z = 0$}  \\
   & \multicolumn{2}{c}{$e=b/a$} & \multicolumn{2}{c}{$s=c/a$} & \multicolumn{2}{c}{$T=(a^2-b^2)/(a^2-c^2)$}  \\
    & Slope & Intercept & Slope & Intercept & Slope & Intercept  \\
\hline
$z = 0$ \\
   DMONLY {\it WMAP}\,1   & $  -0.041   \,\,^{+  0.016      }_{-  0.041  } $ & $   0.799   \,\,^{+  0.027      }_{-  0.007      } $                 & $  -0.070   \,\,^{+  0.012      }_{-  0.019  } $ & $   0.643   \,\,^{+  0.013      }_{-  0.008   } $ & $  0.070   \,\,^{+  0.049      }_{-  0.071  } $ & $   0.605   \,\,^{+  0.045      }_{-  0.028      } $ \\
   DMONLY {\it WMAP}\,3   & $  -0.066   \,\,^{+  0.008      }_{-  0.012  } $ & $   0.809   \,\,^{+  0.006      }_{-  0.005      } $                 & $  -0.055   \,\,^{+  0.002      }_{-  0.002  } $ & $   0.628   \,\,^{+  0.003      }_{-  0.003   } $ & $  0.063   \,\,^{+  0.014      }_{-  0.008  } $ & $   0.625   \,\,^{+  0.007      }_{-  0.009      } $                 \\
   DMONLY {\it WMAP}\,5   & $  -0.058   \,\,^{+  0.008      }_{-  0.003  } $ & $   0.814   \,\,^{+  0.003      }_{-  0.004      } $                 & $  -0.055   \,\,^{+  0.006      }_{-  0.004  } $ & $   0.637   \,\,^{+  0.003      }_{-  0.004   } $ & $  0.064   \,\,^{+  0.006      }_{-  0.010  } $ & $   0.610   \,\,^{+  0.007      }_{-  0.005      } $           \\
   AGN  & $  -0.069   \,\,^{+  0.006      }_{-  0.022  } $ & $   0.836   \,\,^{+  0.004      }_{-  0.002      } $                 & $  -0.078   \,\,^{+  0.008      }_{-  0.005  } $ & $   0.661   \,\,^{+  0.002      }_{-  0.003   } $ & $  0.114   \,\,^{+  0.015      }_{-  0.023  } $ & $   0.579   \,\,^{+  0.007      }_{-  0.005      } $          \\
   WDENS  & $  -0.058   \,\,^{+  0.016      }_{-  0.018  } $ & $   0.847   \,\,^{+  0.003      }_{-  0.004      } $                 & $  -0.063   \,\,^{+  0.008      }_{-  0.006  } $ & $   0.671   \,\,^{+  0.003      }_{-  0.004   } $ & $  0.088   \,\,^{+  0.012      }_{-  0.033  } $ & $   0.569   \,\,^{+  0.008      }_{-  0.005      } $          \\
   REF   & $  -0.027   \,\,^{+  0.010      }_{-  0.036  } $ & $   0.855   \,\,^{+  0.011      }_{-  0.002      } $                 & $  -0.034   \,\,^{+  0.008      }_{-  0.026  } $ & $   0.682   \,\,^{+  0.008      }_{-  0.002   } $ & $  0.046   \,\,^{+  0.025      }_{-  0.027  } $ & $   0.557   \,\,^{+  0.005      }_{-  0.009      } $      \\
   NOSN\_NOZCOOL   & $  -0.037   \,\,^{+  0.007      }_{-  0.029  } $ & $   0.862   \,\,^{+  0.007      }_{-  0.002      } $                 & $  -0.059   \,\,^{+  0.007      }_{-  0.015  } $ & $   0.707   \,\,^{+  0.005      }_{-  0.002   } $ & $  0.043   \,\,^{+  0.033      }_{-  0.012  } $ & $   0.561   \,\,^{+  0.004      }_{-  0.009      } $       \\  
 \ \\
$z = 1$\\
DMONLY {\it WMAP}\,3  & $  -0.067   \,\,^{+  0.023      }_{-  0.019  } $ & $   0.726   \,\,^{+  0.004      }_{-  0.004      } $                 & $  -0.056   \,\,^{+  0.024      }_{-  0.014  } $ & $   0.549   \,\,^{+  0.002      }_{-  0.003   } $ & $  0.058   \,\,^{+  0.007      }_{-  0.031  } $ & $   0.710   \,\,^{+  0.004      }_{-  0.005      } $ \\
AGN  & $  -0.070   \,\,^{+  0.024      }_{-  0.008  } $ & $   0.756   \,\,^{+  0.003      }_{-  0.004      } $                 & $  -0.045   \,\,^{+  0.004      }_{-  0.017  } $ & $   0.577   \,\,^{+  0.003      }_{-  0.002   } $ & $  0.046   \,\,^{+  0.023      }_{-  0.024  } $ & $   0.677   \,\,^{+  0.007      }_{-  0.004      } $   \\
WDENS & $  -0.066   \,\,^{+  0.029      }_{-  0.021  } $ & $   0.766   \,\,^{+  0.004      }_{-  0.004      } $                 & $  -0.045   \,\,^{+  0.005      }_{-  0.014  } $ & $   0.586   \,\,^{+  0.004      }_{-  0.002   } $ & $  0.050   \,\,^{+  0.028      }_{-  0.014  } $ & $   0.672   \,\,^{+  0.007      }_{-  0.007      } $ \\
REF    & $  -0.067   \,\,^{+  0.017      }_{-  0.007  } $ & $   0.781   \,\,^{+  0.004      }_{-  0.004      } $                 & $  -0.057   \,\,^{+  0.024      }_{-  0.014  } $ & $   0.606   \,\,^{+  0.003      }_{-  0.003   } $ & $  0.071   \,\,^{+  0.027      }_{-  0.040  } $ & $   0.660   \,\,^{+  0.006      }_{-  0.006      } $ \\
NOSN\_NOZCOOL & $  -0.077   \,\,^{+  0.022      }_{-  0.019  } $ & $   0.791   \,\,^{+  0.003      }_{-  0.003      } $                 & $  -0.057   \,\,^{+  0.005      }_{-  0.034  } $ & $   0.623   \,\,^{+  0.004      }_{-  0.002   } $ & $  0.065   \,\,^{+  0.038      }_{-  0.037  } $ & $   0.654   \,\,^{+  0.007      }_{-  0.005      } $ \\
\ \\
$z = 2$\\
DMONLY {\it WMAP}\,3  & $  -0.028   \,\,^{+  0.004      }_{-  0.006  } $ & $   0.691   \,\,^{+  0.006      }_{-  0.009      } $                 & $  -0.035   \,\,^{+  0.007      }_{-  0.007  } $ & $   0.501   \,\,^{+  0.010      }_{-  0.010   } $ & $  0.019   \,\,^{+  0.007      }_{-  0.014  } $ & $   0.726   \,\,^{+  0.010      }_{-  0.021      } $    \\
AGN  & $  -0.032   \,\,^{+  0.003      }_{-  0.005  } $ & $   0.703   \,\,^{+  0.004      }_{-  0.007      } $                 & $  -0.022   \,\,^{+  0.003      }_{-  0.012  } $ & $   0.539   \,\,^{+  0.004      }_{-  0.016   } $ & $  0.036   \,\,^{+  0.005      }_{-  0.007  } $ & $   0.732   \,\,^{+  0.009      }_{-  0.009      } $  \\
WDENS  & $  -0.035   \,\,^{+  0.006      }_{-  0.005  } $ & $   0.707   \,\,^{+  0.009      }_{-  0.007      } $                 & $  -0.027   \,\,^{+  0.001      }_{-  0.010  } $ & $   0.541   \,\,^{+  0.003      }_{-  0.014   } $ & $  0.034   \,\,^{+  0.006      }_{-  0.010  } $ & $   0.727   \,\,^{+  0.008      }_{-  0.016      } $\\
REF   & $  -0.025   \,\,^{+  0.002      }_{-  0.007  } $ & $   0.718   \,\,^{+  0.004      }_{-  0.009      } $                 & $  -0.018   \,\,^{+  0.002      }_{-  0.004  } $ & $   0.550   \,\,^{+  0.004      }_{-  0.006   } $ & $  0.028   \,\,^{+  0.009      }_{-  0.005  } $ & $   0.719   \,\,^{+  0.014      }_{-  0.008      } $ \\
NOSN\_NOZCOOL  & $  -0.051   \,\,^{+  0.003      }_{-  0.006  } $ & $   0.723   \,\,^{+  0.005      }_{-  0.009      } $                 & $  -0.047   \,\,^{+  0.002      }_{-  0.006  } $ & $   0.571   \,\,^{+  0.002      }_{-  0.010   } $ & $  0.047   \,\,^{+  0.004      }_{-  0.008  } $ & $   0.724   \,\,^{+  0.006      }_{-  0.011      } $  \\

\hline
 \end{tabular}
 \end{table*}

\subsubsection{Dark matter only simulations}

Before turning our attention to the effect of baryons on halo shape, we will analyse the DMONLY runs. Of particular interest is the relationship 
between the shape parameters, $e=b/a$, $s=c/a$ and $T=(a^2-b^2)/(a^2-c^2)$, and mass, $M_{200}$, of the dark matter halo. Previous
work has shown that $e$ and $s$ are negatively correlated with mass, and $T$ positively correlated (e.g. \citealt{bib:Bett07, bib:Maccio08, bib:JeesonDaniel11}). 
Our dark matter only results extend this work by studying the dependence of these correlations on both cosmology and redshift, as well as 
establishing the baseline for the runs with baryons.

The mass dependence of the shape parameters is shown in Fig. \ref{dmvm_all}.  In each panel we 
show the median value for $e$, $s$ and $T$, within several mass bins, while the error bars represent 
the $1\sigma$ halo-to-halo scatter.  Again, haloes are required to contain at least 1000 particles 
and the mass bins at least 20 haloes.  We use least-squares fitting to determine the slope and 
intercept of the mass relation for each parameter, using a pivot mass of 
$M_{\rm pivot} = 10^{12}\, h^{-1}\, \mbox{M}_\odot$. In the left panels we show the 
median values for all three cosmological models at $z=0$, but for clarity, only plot the fit to the 
DMONLY {\it WMAP}\,3 haloes. The corresponding results for the DMONLY {\it WMAP}\,3 run at $z=2$
are shown in the right panels. The best-fit parameters for the mass dependence of all simulation 
sets are presented in Table \ref{dmmasstable1_all}.  
 
As with the spin parameter, it is clear from the left panels of Fig. \ref{dmvm_all}, that there is 
no significant difference between the different cosmological models studied here. At $z = 0$ there 
is a strong trend for more massive haloes to have smaller axis ratios. Quantitatively, the sphericity 
varies from $\sim 0.6$ for haloes of mass $10^{12}$ $h^{-1} \,M_\odot$, to $\sim 0.5$ for haloes of mass $>10^{14}$ $h^{-1} \,M_\odot$. Since $e=b/a$ decreases faster with mass 
than $s=c/a$ does, the triaxiality parameter increases with mass; over the same range 
in mass, the triaxiality parameter increases from $\sim 0.6$ to $\sim0.8$. In short, high-mass 
haloes are less spherical and more prolate than their lower-mass counterparts. This is likely a 
consequence of their more recent formation time (as discussed in \citealt{bib:Springel04, bib:JeesonDaniel11}). 
Our results are in agreement with previous studies (\citealt{bib:Bett07, bib:Maccio08, bib:JeesonDaniel11}).  

At $z = 2$, we find that the correlations between the shape parameters and mass are much weaker, 
but still present. For example, $s \propto M_{200}^{-0.055}$ at $z=0$ but $s \propto M_{200}^{-0.035}$ 
at $z=2$. A similar trend  is seen in the evolution of the concentration-mass relation 
\citep{bib:Zhao03,bib:Duffy08} and again, is likely to reflect that haloes have had less time 
to collapse, so exhibit a narrower range in formation times (and therefore internal properties).  This explanation is consistent with the findings of \cite{bib:JeesonDaniel11}, who used a principal components analysis to show that concentration, which they found to be essentially equivalent to formation time, is the most fundamental property of dark matter haloes in that scatter in this quantity accounts for much of the scatter in the other dimensionless properties of dark matter haloes, including the shape parameters investigated here.

\subsubsection{Effect of baryons on DM halo shape}
\label{baryonshape}

\begin{figure*}
\begin{tabular}{c c c}

\includegraphics[width=8cm,height=8cm,angle=-90,keepaspectratio]{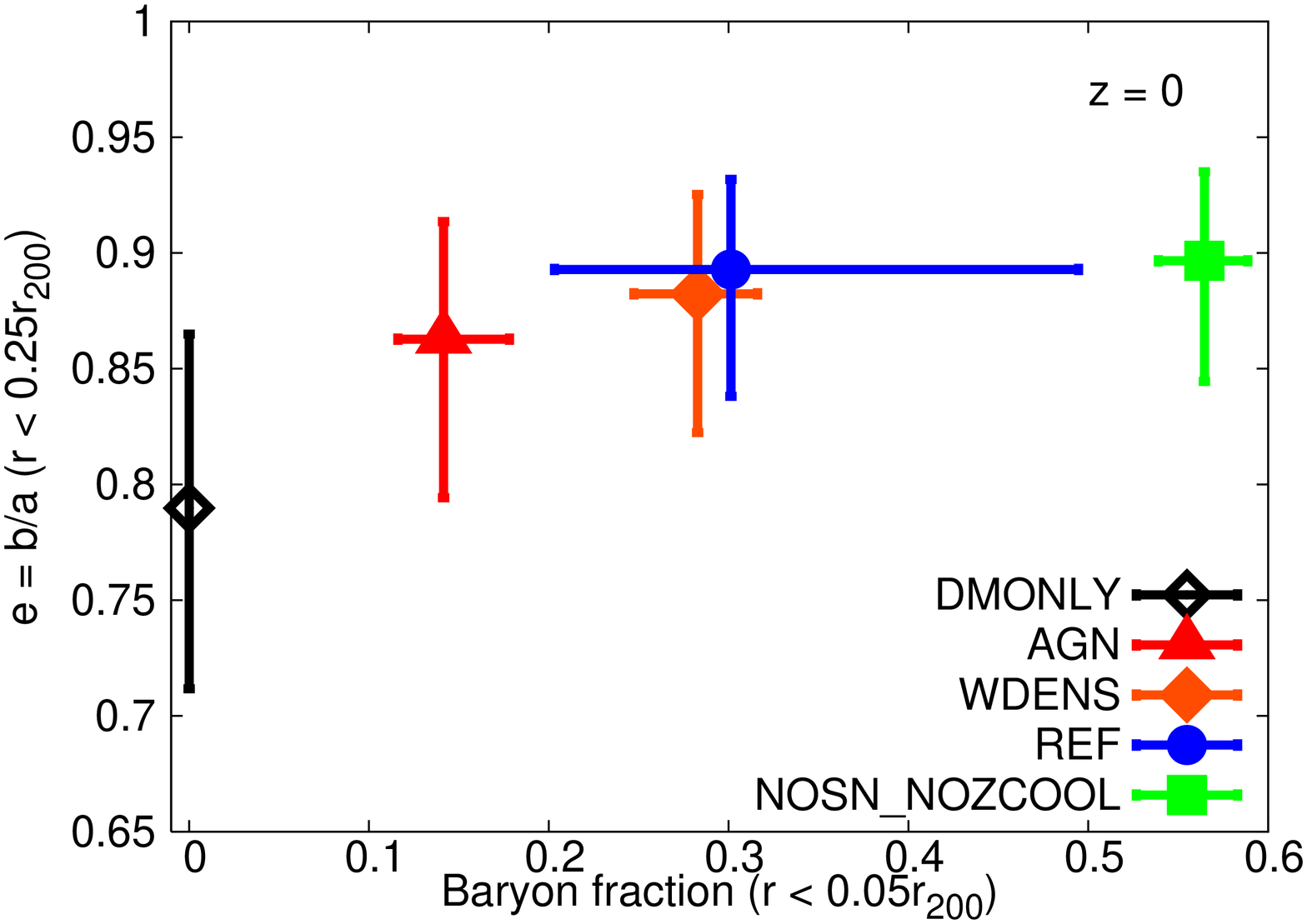} &
\includegraphics[width=8cm,height=8cm,angle=-90,keepaspectratio]{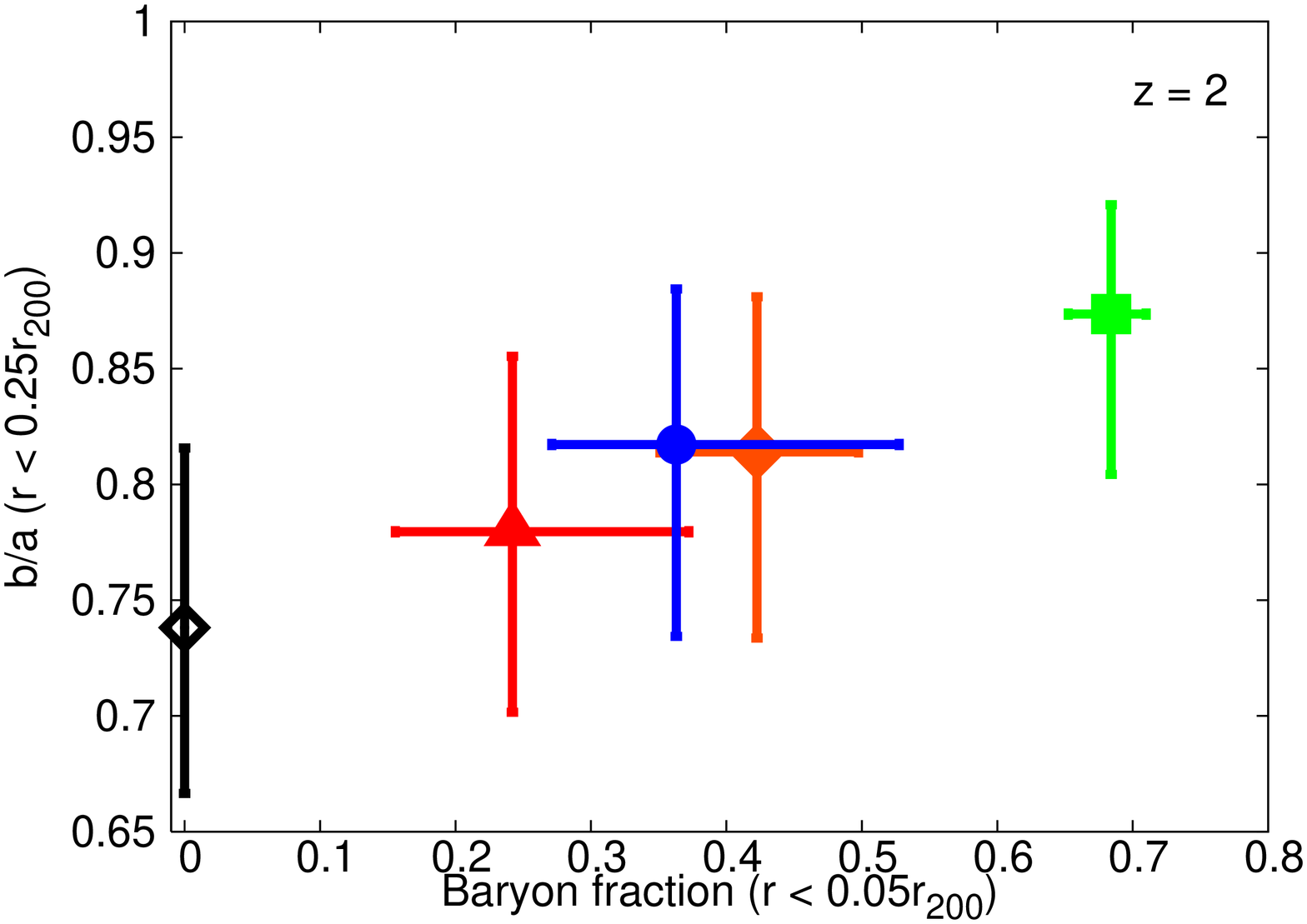} \\

\includegraphics[width=8cm,height=8cm,angle=-90,keepaspectratio]{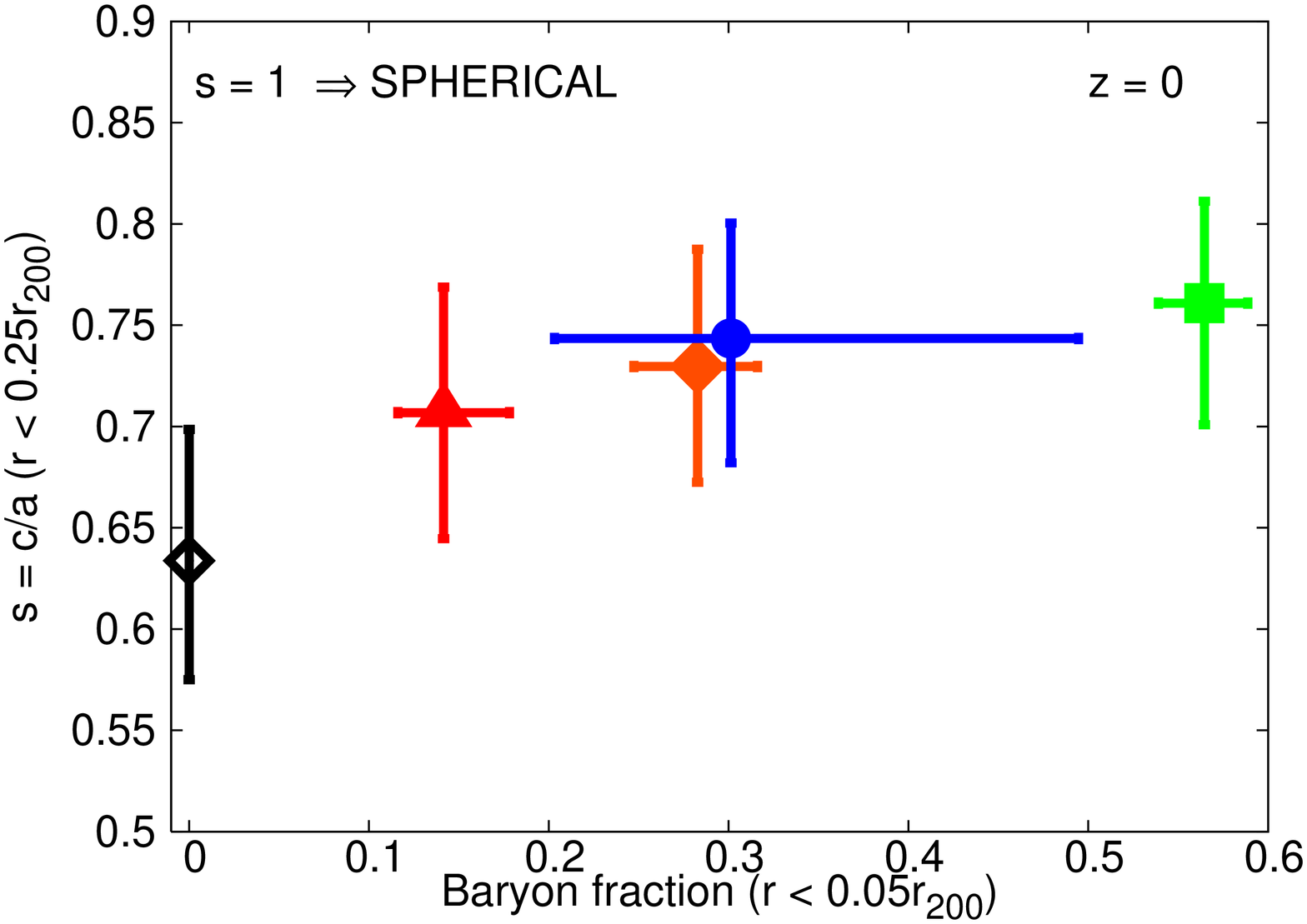} &
\includegraphics[width=8cm,height=8cm,angle=-90,keepaspectratio]{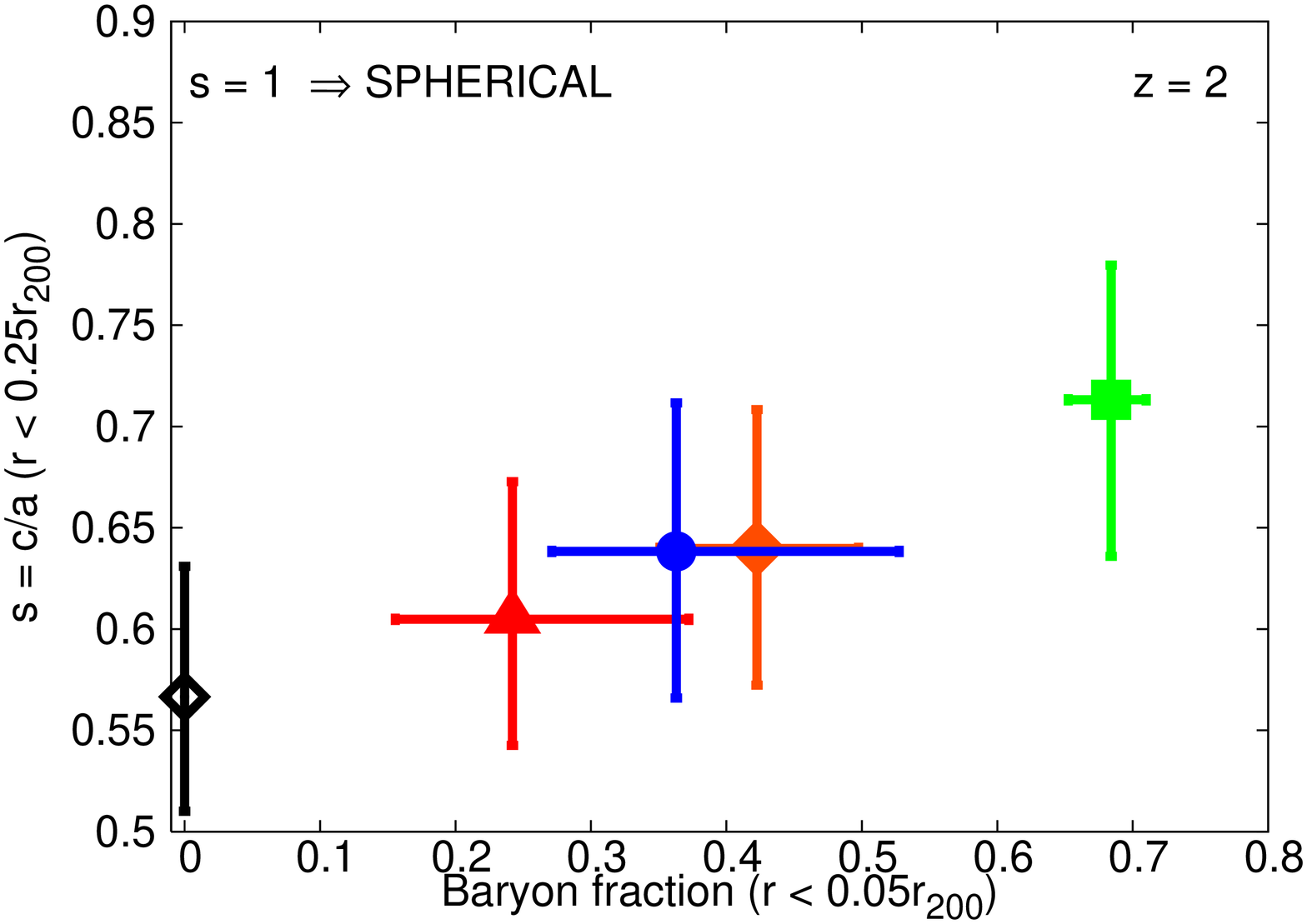} \\

\includegraphics[width=8cm,height=8cm,angle=-90,keepaspectratio]{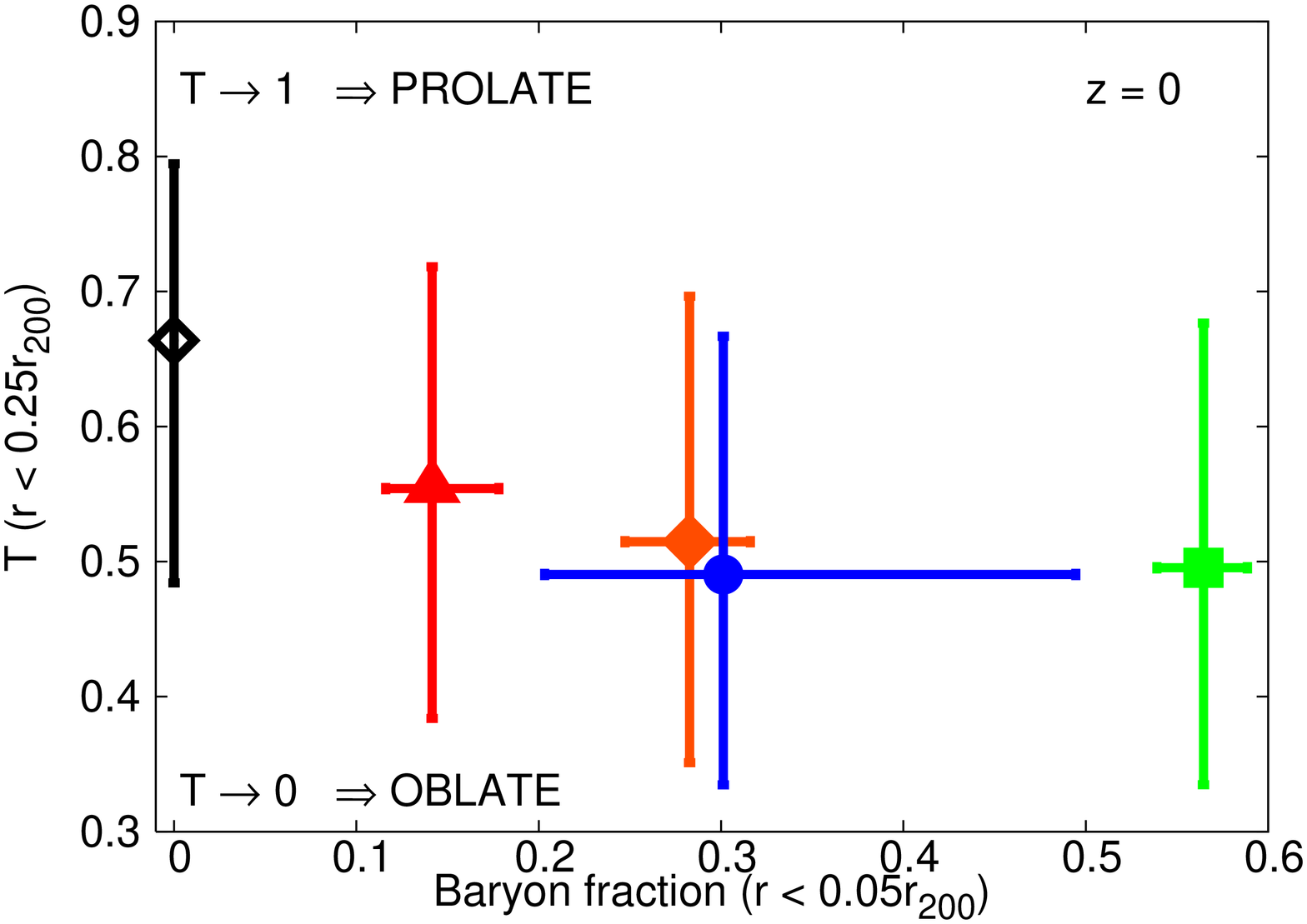} &
\includegraphics[width=8cm,height=8cm,angle=-90,keepaspectratio]{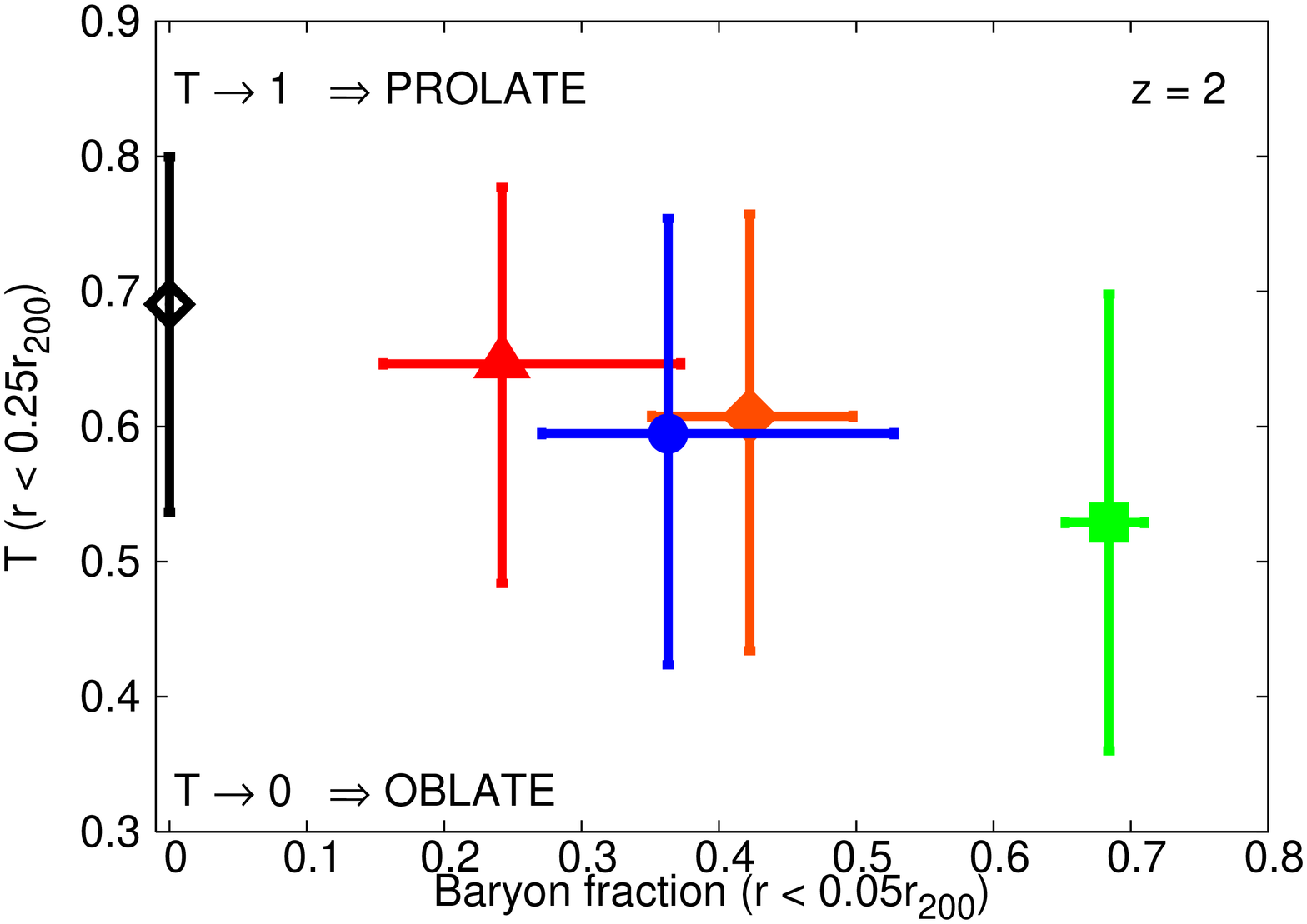} \\

\end{tabular}
\caption[Effect of baryons on the shape distribution of
  haloes.]{\label{SvBF_all} The dependence of the shape of the central halo on galaxy formation efficiency (as measured by the baryon fraction in the inner region of the halo).  Here we show how the halo shape (computed within $0.25r_{200}$) depends on the baryon fraction contained within $0.05r_{200}$.  We show the median and $1\sigma$ halo-to-halo scatter for the each of the feedback implementations.  All  haloes contain at least 1000 particles.  Results at $z = 0$ (2) are shown in the left (right) panels; the median halo mass is $\sim 8 \times 10^{11}$ ($3 \times 10^{10}$) $ h^{-1} {\rm M}_{\odot}$.}  Runs with more efficient galaxy
formation (weaker feedback) contain haloes that are more spherical and less prolate than runs with
stronger feedback.
\end{figure*}

Baryons are known to have a significant effect on the shapes of haloes.  As gas cools and condenses
within the centre of a halo, it reduces the fraction of box orbits present in the dark matter and
stars, resulting in a more spherical structure.  Feedback prevents cooling and acts to expel gas 
from the central region, resulting in a lower central mass concentration, more box orbits and a 
more triaxial shape (see, for example, \citealt{bib:Debattista08}; \citealt{bib:Bryan11}). Here, we 
quantify the impact of these effects for our range of feedback prescriptions, over a wide range in 
halo mass, and at low and high redshift.

The mass dependence of the halo shape parameters in the baryon runs is summarised in 
Table \ref{dmmasstable1_all}. Again, this is for the shape determined within $r_{200}$ using only the dark matter particles. From this we see that the trends seen in the DMONLY runs are still present in the
runs with baryons. However, it is interesting to note that the baryons are able to affect the shape of the halo out to $r_{200}$, albeit in a minor way. 
The axis ratio intercepts (both $b/a$ and $c/a$) systematically increase with increasing galaxy formation efficiency, while the triaxiality parameter 
decreases accordingly. 

The baryons have a larger impact on the shape of the central region of the dark matter halo. To quantify this, we compare the median values for 
$e, s$ and $T$ within $0.25r_{200}$ to the central baryon fraction (within $0.05r_{200}$) for each of the simulation runs.  Note: To ensure that mass dependent trends do not influence our comparison we do not include the 400$h^{-1}\, \rm{Mpc}$ box dark matter simulation at $z = 0$ when directly comparing runs.  These results are 
presented in Fig. \ref{SvBF_all}.  From this figure we can see that there is a clear trend for $e$ and $s$ to increase with increasing baryon 
fraction, while the triaxiality decreases with increasing baryon fraction. This trend is apparent both at $z = 0$ and $z = 2$. At $z=0$, the median sphericity 
is increased by $\sim 25$ per cent when comparing the extreme (DMONLY and NOSN\_NOZCOOL) cases, while the triaxiality is reduced by 
approximately the same factor. The change is still significant in the more realistic AGN model, where the sphericity increases by around 10-15 per cent
and the triaxiality decreases by a similar factor. 
As expected, these results are consistent with the idea that gas cooling to the centre of haloes results in more spherical haloes. It also drives down the value
of the triaxiality parameter, tending to make the haloes considerably less prolate than seen in dark matter only runs.  Feedback reduces the amount of 
gas that is able to cool and condense to the centre of the halo (resulting in a lower central baryon fraction) and hence reduces the impact on the shape of the halo.

\section{Summary}
\label{summaryshapes}
In this paper, we have exploited a subset of runs from the OverWhelmingly Large Simulations (OWLS; 
\citealt{bib:Schaye10}) to investigate the impact of baryons (through gas cooling, star formation
and feedback from stars and black holes) on the spin and shape of dark matter haloes. Our results
allow statistically meaningful conclusions to be drawn regarding the impact of baryons on these
properties, due to the large number of haloes spanning a wide dynamic range in mass. We have also
checked whether our results depend on cosmology and redshift. Our main results are summarised below.

\begin{enumerate}
\item
The spin distribution of dark matter haloes in simulations without baryons is characterised by a 
log-normal curve, with best-fit values of $\lambda_0= 0.036$ $(0.038)$ and $\sigma = 0.62$ $(0.60)$ at $z = 0$ (2), 
in agreement with previous work \citep{bib:Bullock01,bib:Bailin05,bib:Bett07,bib:Maccio08}.  The distribution is 
very similar for the {\it WMAP}\,1, {\it WMAP}\,3  and {\it WMAP}\,5 cosmologies, suggesting that there is no 
strong dependence on $\sigma_8$ (the parameter that varies the most between the three models). No significant
dependence of spin with mass is seen, both at $z=0$ and $z=2$.
At $z = 0$ the spin parameter remains essentially unchanged if computed using only mass within the central region 
($0.25r_{200}$), as found by \cite{bib:Bailin05}.  However, at $z = 2$ the inner region of haloes has a higher mean 
spin than that computed over the whole halo. Restricting our sample to relaxed haloes causes a small (10-15 per 
cent) decrease in the mean value (in agreement with \citealt{bib:Maccio06} and \citealt{bib:JeesonDaniel11}).
\item
At $z = 0$ the spin distribution of dark matter haloes extracted from the baryon runs is not significantly different 
to that of dark matter only haloes when computed using all dark matter particles within $r_{200}$.  However, 
in the central regions ($0.25r_{200}$), where baryons are expected to play an important role, haloes in runs
with absent or weak stellar feedback tend to have higher median spin values than those from stronger feedback runs
(which are very similar to the dark matter only case). We showed that this is, at least in part, due to the 
transfer of angular momentum from the baryons to the dark matter in the former runs.  At $z = 2$ the baryon runs exhibit slightly lower median spin values than the dark matter only case, an effect that is likely due to the increased circular velocity in weak feedback runs and decreased specific angular momentum within the central regions in the strong feedback runs.

\item
Dark matter only haloes extracted from OWLS typically have sphericities of $\sim 0.5 $ to $0.6$ and 
triaxialities of between 0.6 and 0.8 (indicating triaxial to prolate shapes) over the mass and 
redshift ranges we have explored. More massive haloes have less spherical and more prolate shapes.  
Again, we find that halo shape is insensitive to the choice of cosmological model.  
Galaxies and groups at $z = 2$ show the same trends with mass as the groups and clusters at $z = 0$, but weaker.

\item 
When baryons are included, we find that the mass dependent trends remain, and that the intercepts of the relation between 
sphericity (triaxiality) and mass slowly increase (decrease) with increasing galaxy formation efficiency. At
$M_{200}=10^{12}h^{-1}{\rm M}_{\odot}$, baryons increase the dark matter shape parameters by around 10 per cent in the most extreme
case (no feedback). A similar result is seen at higher redshift. Larger differences are again seen when we
consider only the central regions of the halo.
\end{enumerate}

In conclusion, we find that the baryons have a very minor effect on the spin and overall shape of the entire dark matter halo
when the feedback is strong enough to match observed stellar mass fractions.  In particular, the model with AGN feedback can reproduce several observational properties on galactic and groups scale at $z = 0$ by removing gas, suppressing the baryonic impact on the dark matter halo shape.  It should therefore be safe to assume results from 
dark matter only simulations when considering the overall halo properties, at least on the scales resolved by our simulations.  However, even when feedback is strong we find that baryons have a significant effect on the shape of the {\it inner} halo.

\section*{Acknowledgements}
We thank Volker Springel for the use of {\sc gadget} and {\sc subfind} and Mareike Haberichter for useful discussions.  The simulations presented here were run on Stella, the LOFAR Blue Gene/L system in Groningen, on the Cosmology Machine at the
Institute for Computational Cosmology in Durham (which is part of the DiRAC Facility jointly funded by STFC, the Large Facilities Capital Fund of BIS, and Durham University) as part of the Virgo Consortium research programme, and on Darwin in Cambridge.  This work was sponsored by the National Computing Facilities Foundation (NCF) for the use of supercomputer facilities, with financial support from the Netherlands Organization for Scientific Research (NWO).  This work was supported by an NWO VIDI grant and by the Marie Curie Initial Training Network CosmoComp (PITN-GA-2009-238356).  SEB acknowledges support provided by the EU Framework 6 Marie Curie Early Stage Training Programme under contract number MEST-CT-2005-19669 ``ESTRELA''.  STK was supported by the Science and Technology Facilities Council (STFC) through grant ST/G002592/1.

\bibliographystyle{mn2e}
\bibliography{ms}

\begin{thebibliography}{87}
\expandafter\ifx\csname natexlab\endcsname\relax\def\natexlab#1{#1}\fi

\bibitem[{{Abadi} {et~al}\mbox{.}(2010){Abadi}, {Navarro}, {Fardal}, {Babul},
  \& {Steinmetz}}]{bib:Abadi10}
{Abadi} M.~G., {Navarro} J.~F., {Fardal} M., {Babul} A., {Steinmetz} M., 2010,
  \mnras, 407, 435

\bibitem[{{Allgood} {et~al}\mbox{.}(2006){Allgood}, {Flores}, {Primack},
  {Kravtsov}, {Wechsler}, {Faltenbacher}, \& {Bullock}}]{bib:Allgood06}
{Allgood} B., {Flores} R.~A., {Primack} J.~R., {Kravtsov} A.~V., {Wechsler}
  R.~H., {Faltenbacher} A., {Bullock} J.~S., 2006, \mnras, 367, 1781

\bibitem[{{Bailin} \& {Steinmetz}(2004)}]{bib:Bailin04}
{Bailin} J., {Steinmetz} M., 2004, \apj, 616, 27

\bibitem[{{Bailin} \& {Steinmetz}(2005)}]{bib:Bailin05}
{Bailin} J., {Steinmetz} M., 2005, \apj, 627, 647

\bibitem[{{Bardeen} {et~al}\mbox{.}(1986){Bardeen}, {Bond}, {Kaiser}, \&
  {Szalay}}]{bib:Bardeen86}
{Bardeen} J.~M., {Bond} J.~R., {Kaiser} N., {Szalay} A.~S., 1986, \apj, 304, 15

\bibitem[{{Barnes} \& {Hernquist}(1996)}]{bib:Barnes96}
{Barnes} J.~E., {Hernquist} L., 1996, \apj, 471, 115

\bibitem[{{Becker} \& {Kravtsov}(2011)}]{bib:Becker11}
{Becker} M.~R., {Kravtsov} A.~V., 2011, \apj, 740, 25

\bibitem[{{Bett}(2012)}]{bib:Bett12}
{Bett} P., 2012, \mnras, 420, 3303

\bibitem[{{Bett} {et~al}\mbox{.}(2007){Bett}, {Eke}, {Frenk}, {Jenkins},
  {Helly}, \& {Navarro}}]{bib:Bett07}
{Bett} P., {Eke} V., {Frenk} C.~S., {Jenkins} A., {Helly} J., {Navarro} J.,
  2007, \mnras, 376, 215

\bibitem[{{Bett} {et~al}\mbox{.}(2010){Bett}, {Eke}, {Frenk}, {Jenkins}, \&
  {Okamoto}}]{bib:Bett10}
{Bett} P., {Eke} V., {Frenk} C.~S., {Jenkins} A., {Okamoto} T., 2010, \mnras,
  404, 1137

\bibitem[{{Booth} \& {Schaye}(2009)}]{bib:Booth09}
{Booth} C.~M., {Schaye} J., 2009, \mnras, 398, 53

\bibitem[{{Bryan} {et~al}\mbox{.}(2012){Bryan}, {Mao}, {Kay}, {Schaye}, {Dalla
  Vecchia}, \& {Booth}}]{bib:Bryan11}
{Bryan} S.~E., {Mao} S., {Kay} S.~T., {Schaye} J., {Dalla Vecchia} C., {Booth}
  C.~M., 2012, \mnras, 422, 183

\bibitem[{{Bullock}(2002)}]{bib:Bullock02}
{Bullock} J.~S., 2002, in The Shapes of Galaxies and their Dark Matter Haloes,
  {P.~Natarajan}, ed., pp. 109--113

\bibitem[{{Bullock} {et~al}\mbox{.}(2001){Bullock}, {Dekel}, {Kolatt},
  {Kravtsov}, {Klypin}, {Porciani}, \& {Primack}}]{bib:Bullock01}
{Bullock} J.~S., {Dekel} A., {Kolatt} T.~S., {Kravtsov} A.~V., {Klypin} A.~A.,
  {Porciani} C., {Primack} J.~R., 2001, \apj, 555, 240

\bibitem[{{Cole} \& {Lacey}(1996)}]{bib:Cole96}
{Cole} S., {Lacey} C., 1996, \mnras, 281, 716

\bibitem[{{Dalla Vecchia} \& {Schaye}(2008)}]{bib:DallaVecchia08}
{Dalla Vecchia} C., {Schaye} J., 2008, \mnras, 387, 1431

\bibitem[{{Davis} {et~al}\mbox{.}(1985){Davis}, {Efstathiou}, {Frenk}, \&
  {White}}]{bib:Davis85}
{Davis} M., {Efstathiou} G., {Frenk} C.~S., {White} S.~D.~M., 1985, \apj, 292,
  371

\bibitem[{{de Zeeuw} \& {Franx}(1989)}]{bib:deZeeuw89}
{de Zeeuw} T., {Franx} M., 1989, \apj, 343, 617

\bibitem[{{Debattista} {et~al}\mbox{.}(2008){Debattista}, {Moore}, {Quinn},
  {Kazantzidis}, {Maas}, {Mayer}, {Read}, \& {Stadel}}]{bib:Debattista08}
{Debattista} V.~P., {Moore} B., {Quinn} T., {Kazantzidis} S., {Maas} R.,
  {Mayer} L., {Read} J., {Stadel} J., 2008, \apj, 681, 1076

\bibitem[{{Debattista} \& {Sellwood}(1999)}]{bib:Debattista99}
{Debattista} V.~P., {Sellwood} J.~A., 1999, \apjl, 513, L107

\bibitem[{{Dolag} {et~al}\mbox{.}(2008){Dolag}, {Borgani}, {Schindler},
  {Diaferio}, \& {Bykov}}]{bib:Dolag08}
{Dolag} K., {Borgani} S., {Schindler} S., {Diaferio} A., {Bykov} A.~M., 2008,
  \ssr, 134, 229

\bibitem[{{Dubinski}(1994)}]{bib:Dubinski94}
{Dubinski} J., 1994, \apj, 431, 617

\bibitem[{{Dubinski} \& {Carlberg}(1991)}]{bib:Dubinski91}
{Dubinski} J., {Carlberg} R.~G., 1991, \apj, 378, 496

\bibitem[{{Duffy} {et~al}\mbox{.}(2008){Duffy}, {Schaye}, {Kay}, \& {Dalla
  Vecchia}}]{bib:Duffy08}
{Duffy} A.~R., {Schaye} J., {Kay} S.~T., {Dalla Vecchia} C., 2008, \mnras, 390,
  L64

\bibitem[{{Duffy} {et~al}\mbox{.}(2010){Duffy}, {Schaye}, {Kay}, {Dalla
  Vecchia}, {Battye}, \& {Booth}}]{bib:Duffy10}
{Duffy} A.~R., {Schaye} J., {Kay} S.~T., {Dalla Vecchia} C., {Battye} R.~A.,
  {Booth} C.~M., 2010, \mnras, 405, 2161

\bibitem[{{Eggen} {et~al}\mbox{.}(1962){Eggen}, {Lynden-Bell}, \&
  {Sandage}}]{bib:Eggen62}
{Eggen} O.~J., {Lynden-Bell} D., {Sandage} A.~R., 1962, \apj, 136, 748

\bibitem[{{Evrard} {et~al}\mbox{.}(1994){Evrard}, {Summers}, \&
  {Davis}}]{bib:Evrard94}
{Evrard} A.~E., {Summers} F.~J., {Davis} M., 1994, \apj, 422, 11

\bibitem[{{Franx} {et~al}\mbox{.}(1991){Franx}, {Illingworth}, \& {de
  Zeeuw}}]{bib:Franx91}
{Franx} M., {Illingworth} G., {de Zeeuw} T., 1991, \apj, 383, 112

\bibitem[{{Frenk} {et~al}\mbox{.}(1988){Frenk}, {White}, {Davis}, \&
  {Efstathiou}}]{bib:Frenk88}
{Frenk} C.~S., {White} S.~D.~M., {Davis} M., {Efstathiou} G., 1988, \apj, 327,
  507

\bibitem[{{Holmberg}(1969)}]{bib:Holmberg74}
{Holmberg} E., 1969, Arkiv for Astronomi, 5, 305

\bibitem[{{Hopkins} {et~al}\mbox{.}(2005){Hopkins}, {Bahcall}, \&
  {Bode}}]{bib:Hopkins05}
{Hopkins} P.~F., {Bahcall} N.~A., {Bode} P., 2005, \apj, 618, 1

\bibitem[{{Jeeson-Daniel} {et~al}\mbox{.}(2011){Jeeson-Daniel}, {Dalla
  Vecchia}, {Haas}, \& {Schaye}}]{bib:JeesonDaniel11}
{Jeeson-Daniel} A., {Dalla Vecchia} C., {Haas} M.~R., {Schaye} J., 2011,
  \mnras, 415, L69

\bibitem[{{Jing} \& {Suto}(2002)}]{bib:Jing02}
{Jing} Y.~P., {Suto} Y., 2002, \apj, 574, 538

\bibitem[{{Kang} {et~al}\mbox{.}(2005){Kang}, {Mao}, {Gao}, \&
  {Jing}}]{bib:Kang05}
{Kang} X., {Mao} S., {Gao} L., {Jing} Y.~P., 2005, \aap, 437, 383

\bibitem[{{Katz} \& {Gunn}(1991)}]{bib:Katz91}
{Katz} N., {Gunn} J.~E., 1991, \apj, 377, 365

\bibitem[{{Kaufmann} {et~al}\mbox{.}(2007){Kaufmann}, {Mayer}, {Wadsley},
  {Stadel}, \& {Moore}}]{bib:Kaufmann07}
{Kaufmann} T., {Mayer} L., {Wadsley} J., {Stadel} J., {Moore} B., 2007, \mnras,
  375, 53

\bibitem[{{Kazantzidis} {et~al}\mbox{.}(2004){Kazantzidis}, {Kravtsov},
  {Zentner}, {Allgood}, {Nagai}, \& {Moore}}]{bib:Kazantzidis04}
{Kazantzidis} S., {Kravtsov} A.~V., {Zentner} A.~R., {Allgood} B., {Nagai} D.,
  {Moore} B., 2004, \apjl, 611, L73

\bibitem[{{Knebe} {et~al}\mbox{.}(2004){Knebe}, {Gill}, {Gibson}, {Lewis},
  {Ibata}, \& {Dopita}}]{bib:Knebe04}
{Knebe} A., {Gill} S.~P.~D., {Gibson} B.~K., {Lewis} G.~F., {Ibata} R.~A.,
  {Dopita} M.~A., 2004, \apj, 603, 7

\bibitem[{{Knebe} {et~al}\mbox{.}(2010){Knebe}, {Libeskind}, {Knollmann},
  {Yepes}, {Gottl{\"o}ber}, \& {Hoffman}}]{bib:Knebe10}
{Knebe} A., {Libeskind} N.~I., {Knollmann} S.~R., {Yepes} G., {Gottl{\"o}ber}
  S., {Hoffman} Y., 2010, \mnras, 405, 1119

\bibitem[{{Komatsu} {et~al}\mbox{.}(2009){Komatsu}, {Dunkley}, {Nolta},
  {Bennett}, {Gold}, {Hinshaw}, {Jarosik}, {Larson}, {Limon}, {Page},
  {Spergel}, {Halpern}, {Hill}, {Kogut}, {Meyer}, {Tucker}, {Weiland},
  {Wollack}, \& {Wright}}]{bib:Komatsu09}
{Komatsu} E. {et~al.}, 2009, \apjs, 180, 330

\bibitem[{{Libeskind} {et~al}\mbox{.}(2007){Libeskind}, {Cole}, {Frenk},
  {Okamoto}, \& {Jenkins}}]{bib:Libeskind07}
{Libeskind} N.~I., {Cole} S., {Frenk} C.~S., {Okamoto} T., {Jenkins} A., 2007,
  \mnras, 374, 16

\bibitem[{{Libeskind} {et~al}\mbox{.}(2005){Libeskind}, {Frenk}, {Cole},
  {Helly}, {Jenkins}, {Navarro}, \& {Power}}]{bib:Libeskind05}
{Libeskind} N.~I., {Frenk} C.~S., {Cole} S., {Helly} J.~C., {Jenkins} A.,
  {Navarro} J.~F., {Power} C., 2005, \mnras, 363, 146

\bibitem[{{Libeskind} {et~al}\mbox{.}(2010){Libeskind}, {Yepes}, {Knebe},
  {Gottl{\"o}ber}, {Hoffman}, \& {Knollmann}}]{bib:Libeskind10}
{Libeskind} N.~I., {Yepes} G., {Knebe} A., {Gottl{\"o}ber} S., {Hoffman} Y.,
  {Knollmann} S.~R., 2010, \mnras, 401, 1889

\bibitem[{{Macci{\`o}} {et~al}\mbox{.}(2008){Macci{\`o}}, {Dutton}, \& {van den
  Bosch}}]{bib:Maccio08}
{Macci{\`o}} A.~V., {Dutton} A.~A., {van den Bosch} F.~C., 2008, \mnras, 391,
  1940

\bibitem[{{Macci{\`o}} {et~al}\mbox{.}(2007){Macci{\`o}}, {Dutton}, {van den
  Bosch}, {Moore}, {Potter}, \& {Stadel}}]{bib:Maccio07}
{Macci{\`o}} A.~V., {Dutton} A.~A., {van den Bosch} F.~C., {Moore} B., {Potter}
  D., {Stadel} J., 2007, \mnras, 378, 55

\bibitem[{{Macci{\`o}} {et~al}\mbox{.}(2006){Macci{\`o}}, {Moore}, {Stadel}, \&
  {Diemand}}]{bib:Maccio06}
{Macci{\`o}} A.~V., {Moore} B., {Stadel} J., {Diemand} J., 2006, \mnras, 366,
  1529

\bibitem[{{McCarthy} {et~al}\mbox{.}(2010){McCarthy}, {Schaye}, {Ponman},
  {Bower}, {Booth}, {Dalla Vecchia}, {Crain}, {Springel}, {Theuns}, \&
  {Wiersma}}]{bib:McCarthy10}
{McCarthy} I.~G. {et~al.}, 2010, \mnras, 406, 822

\bibitem[{{Merrifield}(2004)}]{bib:Merrifield04}
{Merrifield} M.~R., 2004, in IAU Symposium, Vol. 220, Dark Matter in Galaxies,
  {S.~Ryder, D.~Pisano, M.~Walker, \& K.~Freeman}, ed., p. 431

\bibitem[{{Mo} {et~al}\mbox{.}(1998){Mo}, {Mao}, \& {White}}]{bib:Mo98}
{Mo} H.~J., {Mao} S., {White} S.~D.~M., 1998, \mnras, 295, 319

\bibitem[{{Mu{\~n}oz-Cuartas} {et~al}\mbox{.}(2011){Mu{\~n}oz-Cuartas},
  {Macci{\`o}}, {Gottl{\"o}ber}, \& {Dutton}}]{bib:Munoz11}
{Mu{\~n}oz-Cuartas} J.~C., {Macci{\`o}} A.~V., {Gottl{\"o}ber} S., {Dutton}
  A.~A., 2011, \mnras, 411, 584

\bibitem[{{Navarro} {et~al}\mbox{.}(1997){Navarro}, {Frenk}, \&
  {White}}]{bib:Navarro97}
{Navarro} J.~F., {Frenk} C.~S., {White} S.~D.~M., 1997, \apj, 490, 493

\bibitem[{{Neto} {et~al}\mbox{.}(2007){Neto}, {Gao}, {Bett}, {Cole}, {Navarro},
  {Frenk}, {White}, {Springel}, \& {Jenkins}}]{bib:Neto07}
{Neto} A.~F. {et~al.}, 2007, \mnras, 381, 1450

\bibitem[{{O'Brien} {et~al}\mbox{.}(2010){O'Brien}, {Freeman}, \& {van der
  Kruit}}]{bib:Obrien10}
{O'Brien} J.~C., {Freeman} K.~C., {van der Kruit} P.~C., 2010, \aap, 515, A63

\bibitem[{{Oguri} {et~al}\mbox{.}(2003){Oguri}, {Lee}, \& {Suto}}]{bib:Oguri03}
{Oguri} M., {Lee} J., {Suto} Y., 2003, \apj, 599, 7

\bibitem[{{Ostriker} \& {Binney}(1989)}]{bib:Ostriker89}
{Ostriker} E.~C., {Binney} J.~J., 1989, \mnras, 237, 785

\bibitem[{{Pedrosa} {et~al}\mbox{.}(2010){Pedrosa}, {Tissera}, \&
  {Scannapieco}}]{bib:Pedrosa10}
{Pedrosa} S., {Tissera} P.~B., {Scannapieco} C., 2010, \mnras, 402, 776

\bibitem[{{Peebles}(1969)}]{bib:Peebles69}
{Peebles} P.~J.~E., 1969, \apj, 155, 393

\bibitem[{{Power} {et~al}\mbox{.}(2003){Power}, {Navarro}, {Jenkins}, {Frenk},
  {White}, {Springel}, {Stadel}, \& {Quinn}}]{bib:Power03}
{Power} C., {Navarro} J.~F., {Jenkins} A., {Frenk} C.~S., {White} S.~D.~M.,
  {Springel} V., {Stadel} J., {Quinn} T., 2003, \mnras, 338, 14

\bibitem[{{Sackett}(1999)}]{bib:Sackett99}
{Sackett} P.~D., 1999, in Astronomical Society of the Pacific Conference
  Series, Vol. 182, Galaxy Dynamics - A Rutgers Symposium, {D.~R.~Merritt,
  M.~Valluri, \& J.~A.~Sellwood}, ed., p. 393

\bibitem[{{Sales} {et~al}\mbox{.}(2012){Sales}, {Navarro}, {Theuns}, {Schaye},
  {White}, {Frenk}, {Crain}, \& {Dalla Vecchia}}]{bib:Sales12}
{Sales} L.~V., {Navarro} J.~F., {Theuns} T., {Schaye} J., {White} S.~D.~M.,
  {Frenk} C.~S., {Crain} R.~A., {Dalla Vecchia} C., 2012, \mnras, 423, 1544

\bibitem[{{Schaye} \& {Dalla Vecchia}(2008)}]{bib:Schaye08}
{Schaye} J., {Dalla Vecchia} C., 2008, \mnras, 383, 1210

\bibitem[{{Schaye} {et~al}\mbox{.}(2010){Schaye}, {Dalla Vecchia}, {Booth},
  {Wiersma}, {Theuns}, {Haas}, {Bertone}, {Duffy}, {McCarthy}, \& {van de
  Voort}}]{bib:Schaye10}
{Schaye} J. {et~al.}, 2010, \mnras, 402, 1536

\bibitem[{{Seljak} \& {Zaldarriaga}(1996)}]{bib:Seljak96}
{Seljak} U., {Zaldarriaga} M., 1996, \apj, 469, 437

\bibitem[{{Sharma} \& {Steinmetz}(2005)}]{bib:Sharma05}
{Sharma} S., {Steinmetz} M., 2005, \apj, 628, 21

\bibitem[{{Sharma} {et~al}\mbox{.}(2012){Sharma}, {Steinmetz}, \&
  {Bland-Hawthorn}}]{bib:Sharma12}
{Sharma} S., {Steinmetz} M., {Bland-Hawthorn} J., 2012, \apj, 750, 107

\bibitem[{{Spergel} {et~al}\mbox{.}(2007){Spergel}, {Bean}, {Dor{\'e}},
  {Nolta}, {Bennett}, {Dunkley}, {Hinshaw}, {Jarosik}, {Komatsu}, {Page},
  {Peiris}, {Verde}, {Halpern}, {Hill}, {Kogut}, {Limon}, {Meyer}, {Odegard},
  {Tucker}, {Weiland}, {Wollack}, \& {Wright}}]{bib:WMAP3}
{Spergel} D.~N. {et~al.}, 2007, \apjs, 170, 377

\bibitem[{{Spergel} {et~al}\mbox{.}(2003){Spergel}, {Verde}, {Peiris},
  {Komatsu}, {Nolta}, {Bennett}, {Halpern}, {Hinshaw}, {et~al.}}]{bib:wmap1_03}
{Spergel} D.~N. {et~al.}, 2003, ApJS, 148, 175

\bibitem[{{Springel}(2005)}]{bib:Gadget2}
{Springel} V., 2005, \mnras, 364, 1105

\bibitem[{{Springel} {et~al}\mbox{.}(2004){Springel}, {White}, \&
  {Hernquist}}]{bib:Springel04}
{Springel} V., {White} S.~D.~M., {Hernquist} L., 2004, in IAU Symposium, Vol.
  220, Dark Matter in Galaxies, {S.~Ryder, D.~Pisano, M.~Walker, \&
  K.~Freeman}, ed., p. 421

\bibitem[{{Springel} {et~al}\mbox{.}(2005){Springel}, {White}, {Jenkins},
  {Frenk}, {Yoshida}, {Gao}, {Navarro}, {Thacker}, {et~al.}}]{bib:Springel05}
{Springel} V. {et~al.}, 2005, Nature, 435, 629

\bibitem[{{Springel} {et~al}\mbox{.}(2001){Springel}, {White}, {Tormen}, \&
  {Kauffmann}}]{bib:Springel01}
{Springel} V., {White} S.~D.~M., {Tormen} G., {Kauffmann} G., 2001, \mnras,
  328, 726

\bibitem[{{Steiman-Cameron} {et~al}\mbox{.}(1992){Steiman-Cameron}, {Kormendy},
  \& {Durisen}}]{bib:Steiman92}
{Steiman-Cameron} T.~Y., {Kormendy} J., {Durisen} R.~H., 1992, \aj, 104, 1339

\bibitem[{{Tissera} \& {Dominguez-Tenreiro}(1998)}]{bib:Tissera98}
{Tissera} P.~B., {Dominguez-Tenreiro} R., 1998, \mnras, 297, 177

\bibitem[{{Tissera} {et~al}\mbox{.}(2010){Tissera}, {White}, {Pedrosa}, \&
  {Scannapieco}}]{bib:Tissera10}
{Tissera} P.~B., {White} S.~D.~M., {Pedrosa} S., {Scannapieco} C., 2010,
  \mnras, 406, 922

\bibitem[{{Tonini} {et~al}\mbox{.}(2006{\natexlab{a}}){Tonini}, {Lapi}, \&
  {Salucci}}]{bib:Tonini06a}
{Tonini} C., {Lapi} A., {Salucci} P., 2006{\natexlab{a}}, \apj, 649, 591

\bibitem[{{Tonini} {et~al}\mbox{.}(2006{\natexlab{b}}){Tonini}, {Lapi},
  {Shankar}, \& {Salucci}}]{bib:Tonini06b}
{Tonini} C., {Lapi} A., {Shankar} F., {Salucci} P., 2006{\natexlab{b}}, \apjl,
  638, L13

\bibitem[{{Vera-Ciro} {et~al}\mbox{.}(2011){Vera-Ciro}, {Sales}, {Helmi},
  {Frenk}, {Navarro}, {Springel}, {Vogelsberger}, \& {White}}]{bib:Vera-Ciro11}
{Vera-Ciro} C.~A., {Sales} L.~V., {Helmi} A., {Frenk} C.~S., {Navarro} J.~F.,
  {Springel} V., {Vogelsberger} M., {White} S.~D.~M., 2011, \mnras, 416, 1377

\bibitem[{{Warren} {et~al}\mbox{.}(1992){Warren}, {Quinn}, {Salmon}, \&
  {Zurek}}]{bib:Warren92}
{Warren} M.~S., {Quinn} P.~J., {Salmon} J.~K., {Zurek} W.~H., 1992, \apj, 399,
  405

\bibitem[{{White}(1996)}]{bib:White96}
{White} S.~D.~M., 1996, in Cosmology and large scale Structure, {R.~Schaeffer,
  J.~Silk, M.~Spiro, \& J.~Zinn-Justin}, ed., p. 349

\bibitem[{{Wiersma} {et~al}\mbox{.}(2009{\natexlab{a}}){Wiersma}, {Schaye}, \&
  {Smith}}]{bib:Wiersma09}
{Wiersma} R.~P.~C., {Schaye} J., {Smith} B.~D., 2009{\natexlab{a}}, \mnras,
  393, 99

\bibitem[{{Wiersma} {et~al}\mbox{.}(2009{\natexlab{b}}){Wiersma}, {Schaye},
  {Theuns}, {Dalla Vecchia}, \& {Tornatore}}]{bib:Wiersma09b}
{Wiersma} R.~P.~C., {Schaye} J., {Theuns} T., {Dalla Vecchia} C., {Tornatore}
  L., 2009{\natexlab{b}}, \mnras, 399, 574

\bibitem[{{Zel'dovich}(1970)}]{bib:Zeldovich70}
{Zel'dovich} Y.~B., 1970, \aap, 5, 84

\bibitem[{{Zeldovich} {et~al}\mbox{.}(1980){Zeldovich}, {Klypin}, {Khlopov}, \&
  {Chechetkin}}]{bib:Zeldovich80}
{Zeldovich} Y.~B., {Klypin} A.~A., {Khlopov} M.~Y., {Chechetkin} V.~M., 1980,
  Sov. J. Nucl. Phys., 31, 664

\bibitem[{{Zemp} {et~al}\mbox{.}(2011){Zemp}, {Gnedin}, {Gnedin}, \&
  {Kravtsov}}]{bib:Zemp11}
{Zemp} M., {Gnedin} O.~Y., {Gnedin} N.~Y., {Kravtsov} A.~V., 2011, \apjs, 197,
  30

\bibitem[{{Zemp} {et~al}\mbox{.}(2012){Zemp}, {Gnedin}, {Gnedin}, \&
  {Kravtsov}}]{bib:Zemp12}
{Zemp} M., {Gnedin} O.~Y., {Gnedin} N.~Y., {Kravtsov} A.~V., 2012, \apj, 748,
  54

\bibitem[{{Zentner} {et~al}\mbox{.}(2005){Zentner}, {Kravtsov}, {Gnedin}, \&
  {Klypin}}]{bib:Zentner05}
{Zentner} A.~R., {Kravtsov} A.~V., {Gnedin} O.~Y., {Klypin} A.~A., 2005, \apj,
  629, 219

\bibitem[{{Zhao} {et~al}\mbox{.}(2003){Zhao}, {Jing}, {Mo}, \&
  {B{\"o}rner}}]{bib:Zhao03}
{Zhao} D.~H., {Jing} Y.~P., {Mo} H.~J., {B{\"o}rner} G., 2003, \apjl, 597, L9

\end{thebibliography}

\begin{appendix}

\section{Resolution Tests}
\label{resolutionshapes}

\begin{figure*}
\begin{center}
\begin{tabular}{cc}

\includegraphics[width=7cm,height=7cm,angle=-90,keepaspectratio]{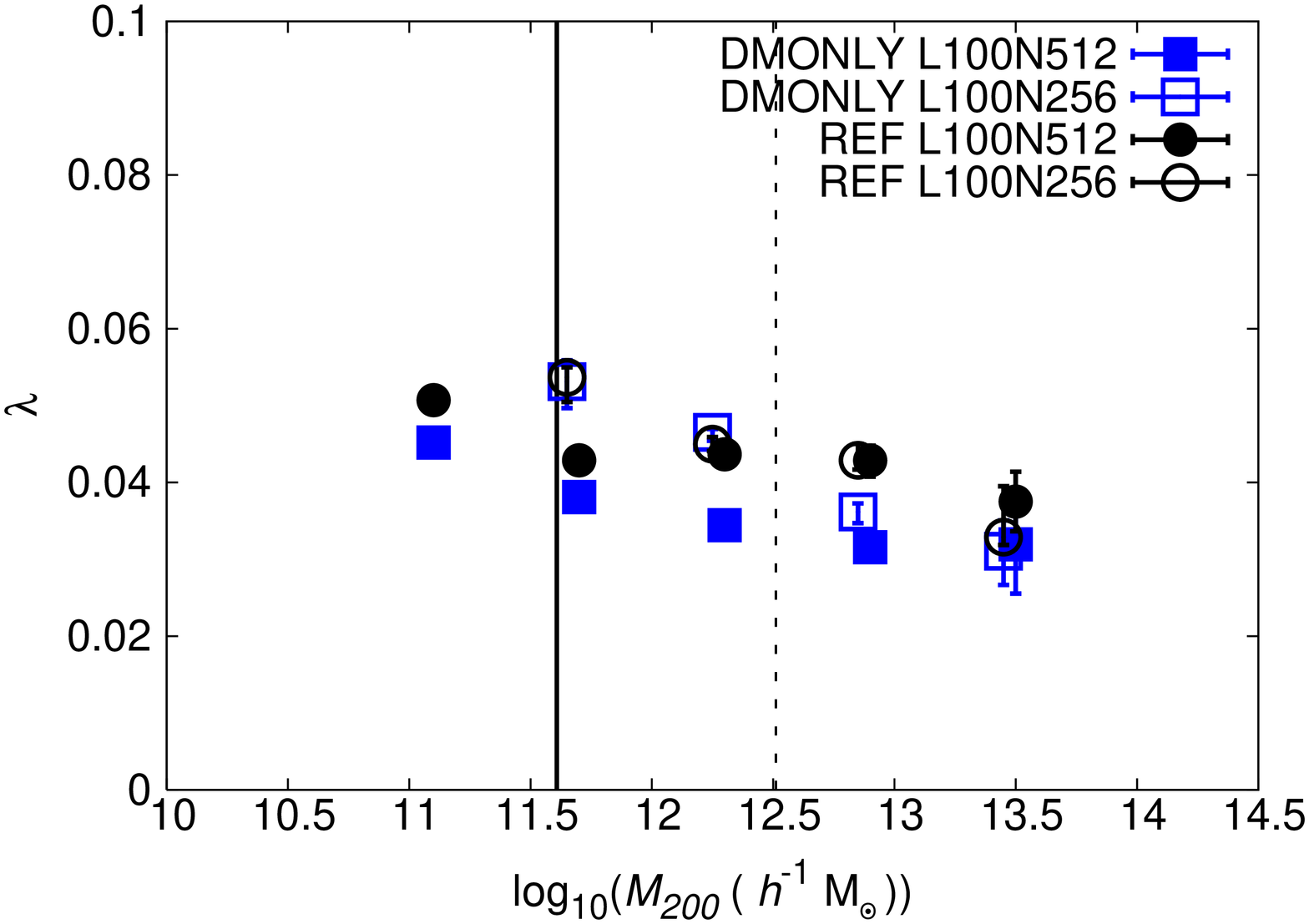}&
\includegraphics[width=7cm,height=7cm,angle=-90,keepaspectratio]{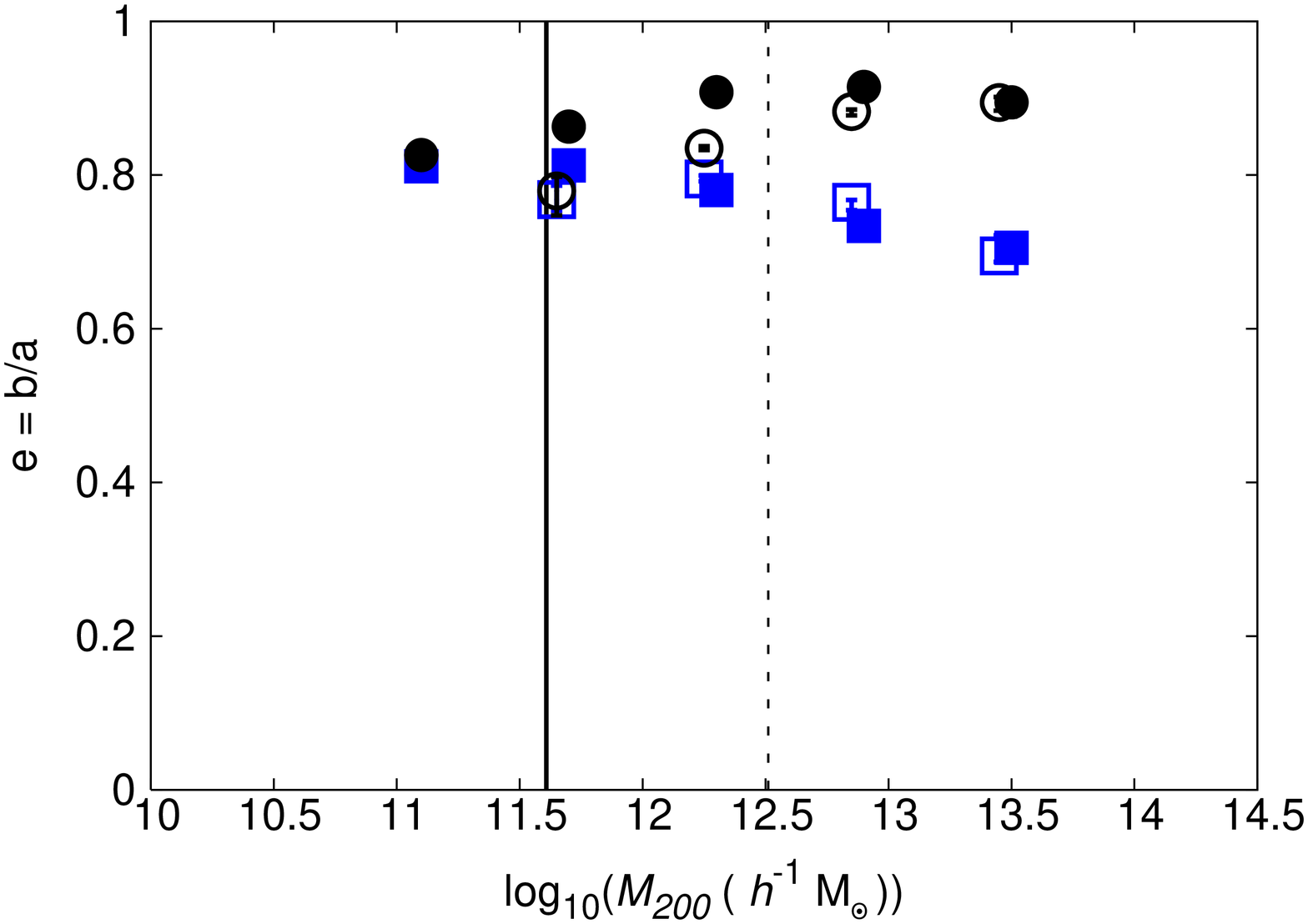}\\
\includegraphics[width=7cm,height=7cm,angle=-90,keepaspectratio]{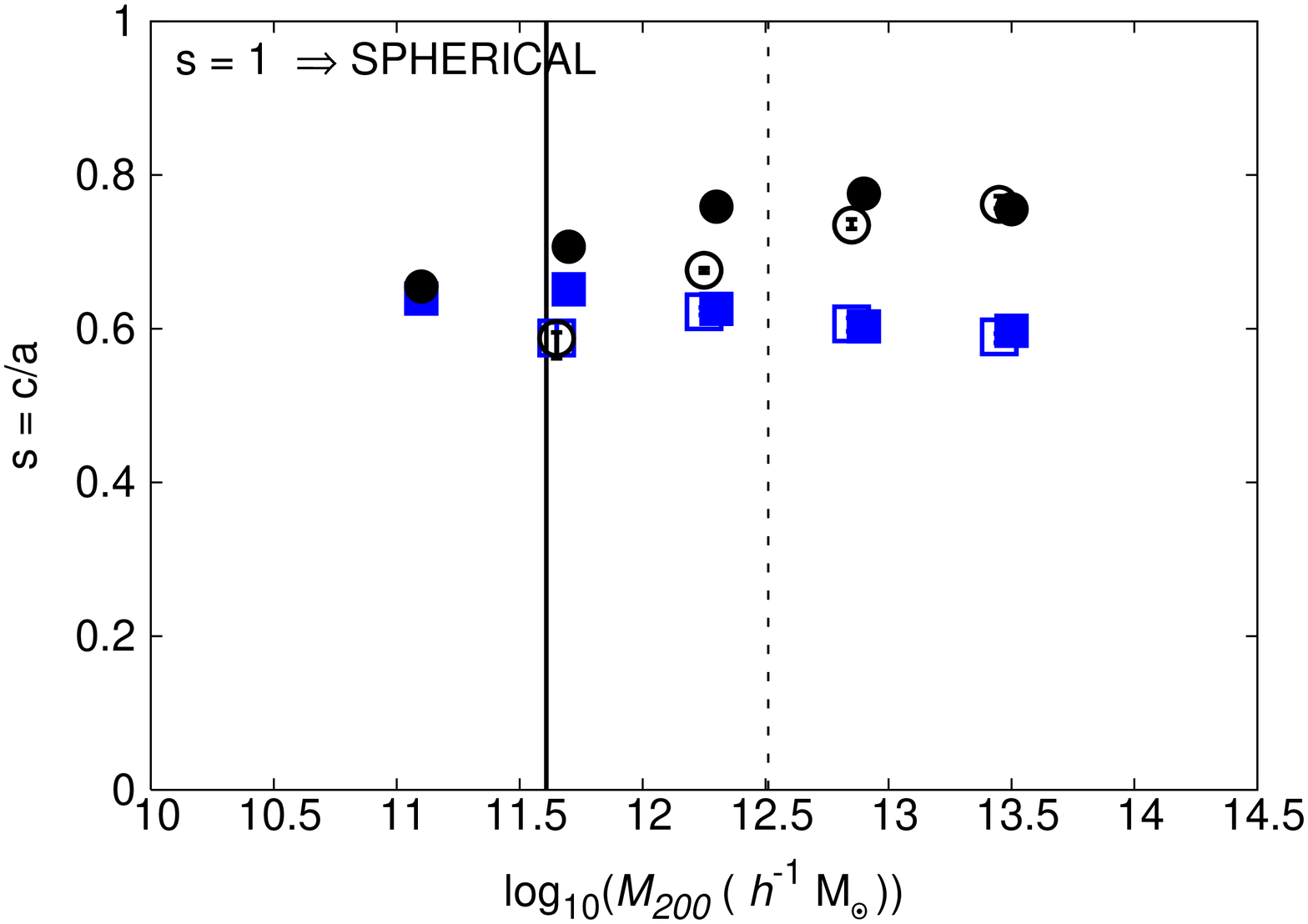} &
\includegraphics[width=7cm,height=7cm,angle=-90,keepaspectratio]{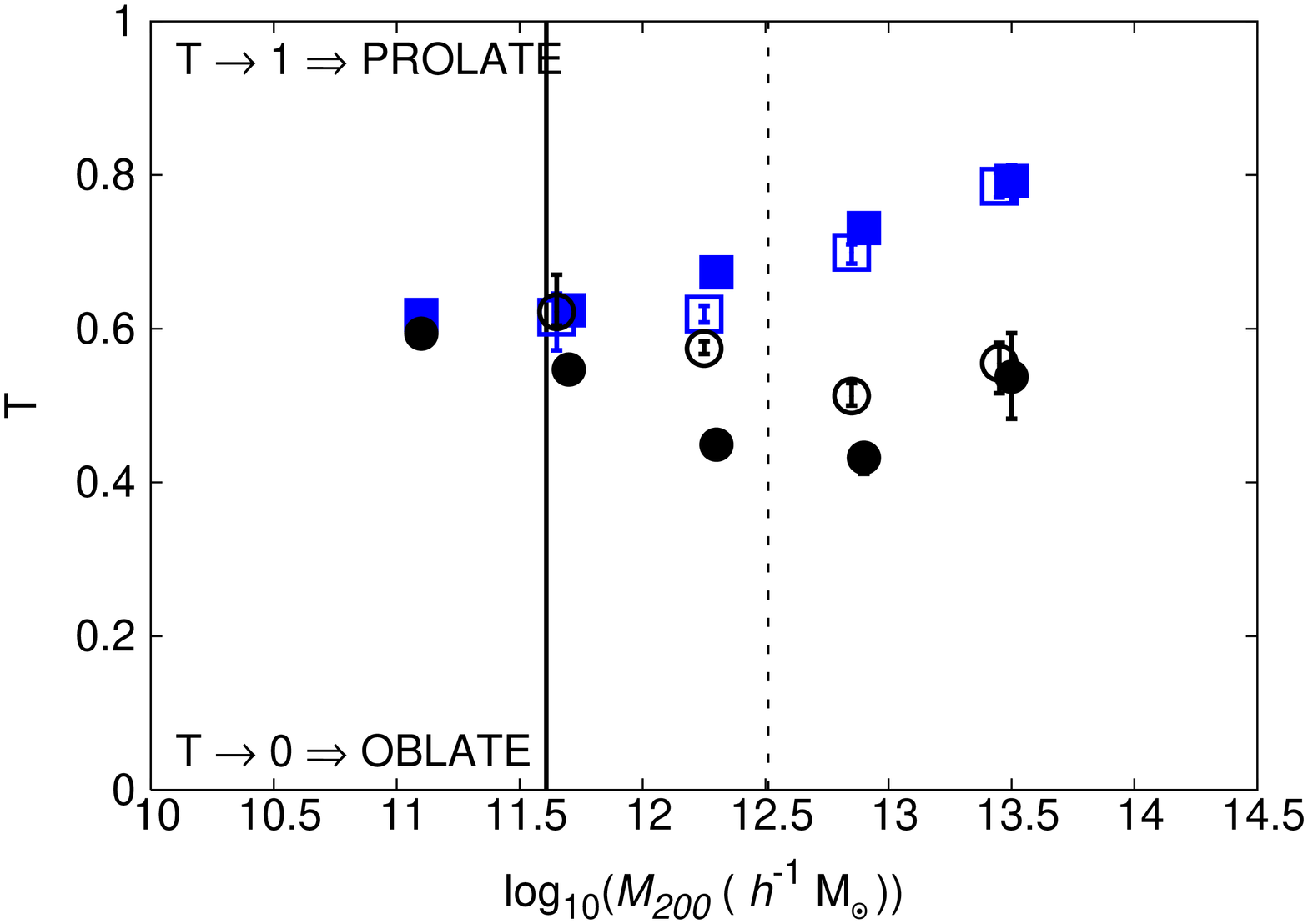}

\end{tabular} 
\end{center}
\caption[Resolution tests for DM halo triaxiality.]{\label{dmres_c} Resolution
  tests for halo properties determined within the central 0.25$r_{200}$. Properties of haloes extracted from the $512^3$ ( $256^3$) particle simulations are shown as filled (open) symbols. Dark matter only (weak stellar feedback) runs are shown as blue squares (black circles).  The error bars show the $1\sigma$ bootstrap resampled median distributions (1000 samples).  We note that above 1000 particles (vertical lines), all quantities are well converged with respect to the numerical resolution, and we therefore adopt this resolution limit.}
\end{figure*}

\begin{figure*}
\begin{center}
\begin{tabular}{cc}

\includegraphics[width=7cm,height=7cm,angle=-90,keepaspectratio]{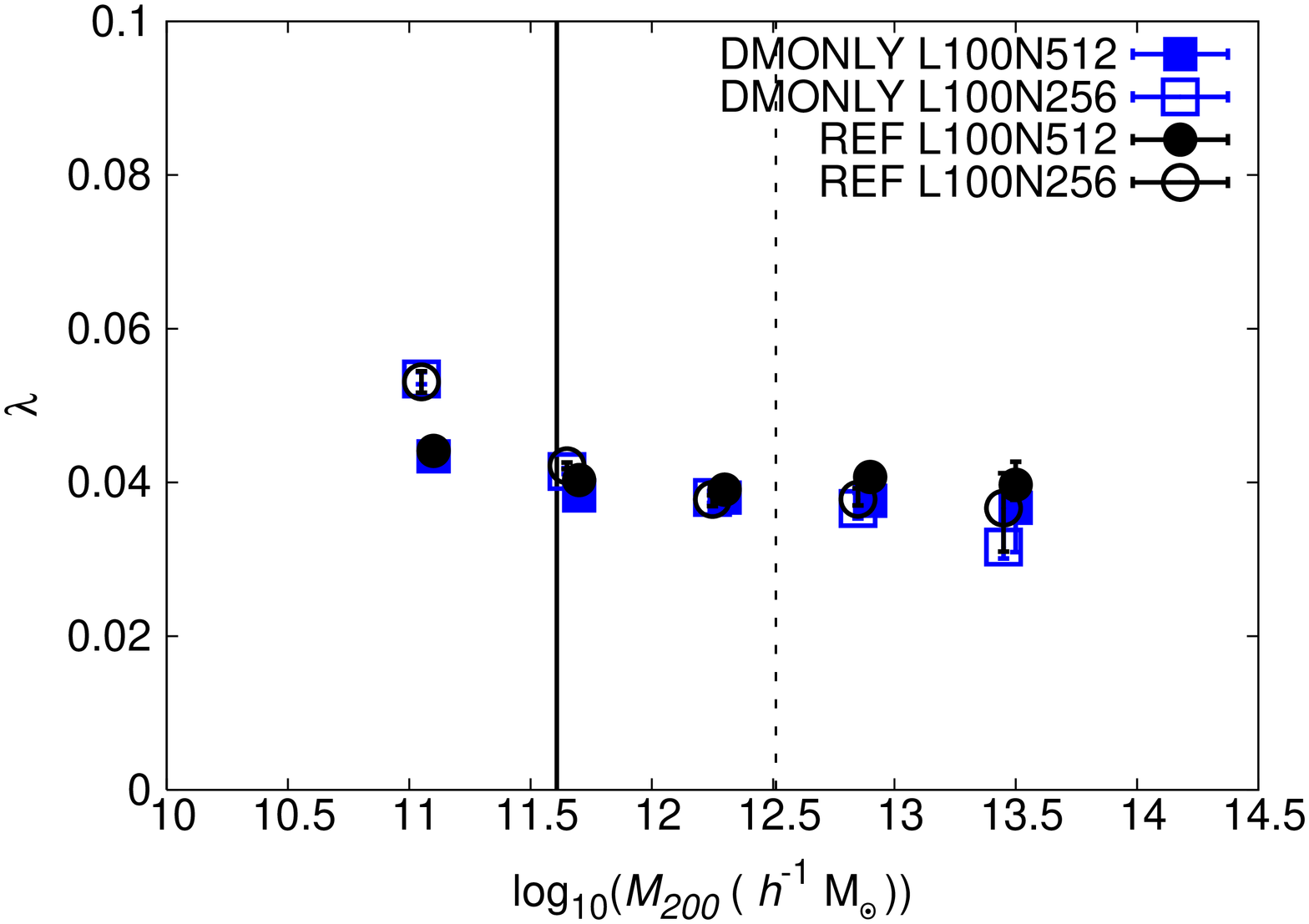}&
\includegraphics[width=7cm,height=7cm,angle=-90,keepaspectratio]{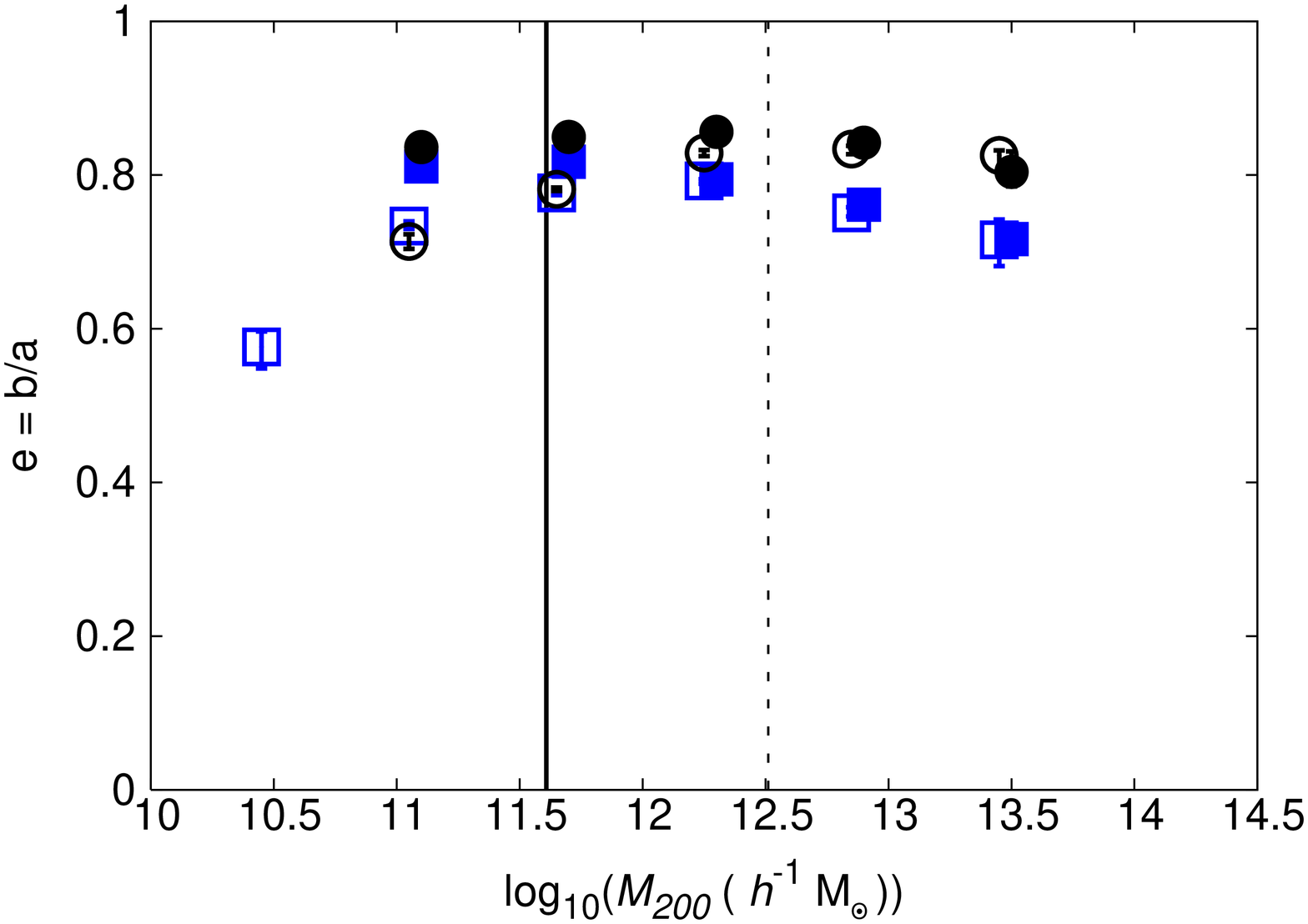}\\
\includegraphics[width=7cm,height=7cm,angle=-90,keepaspectratio]{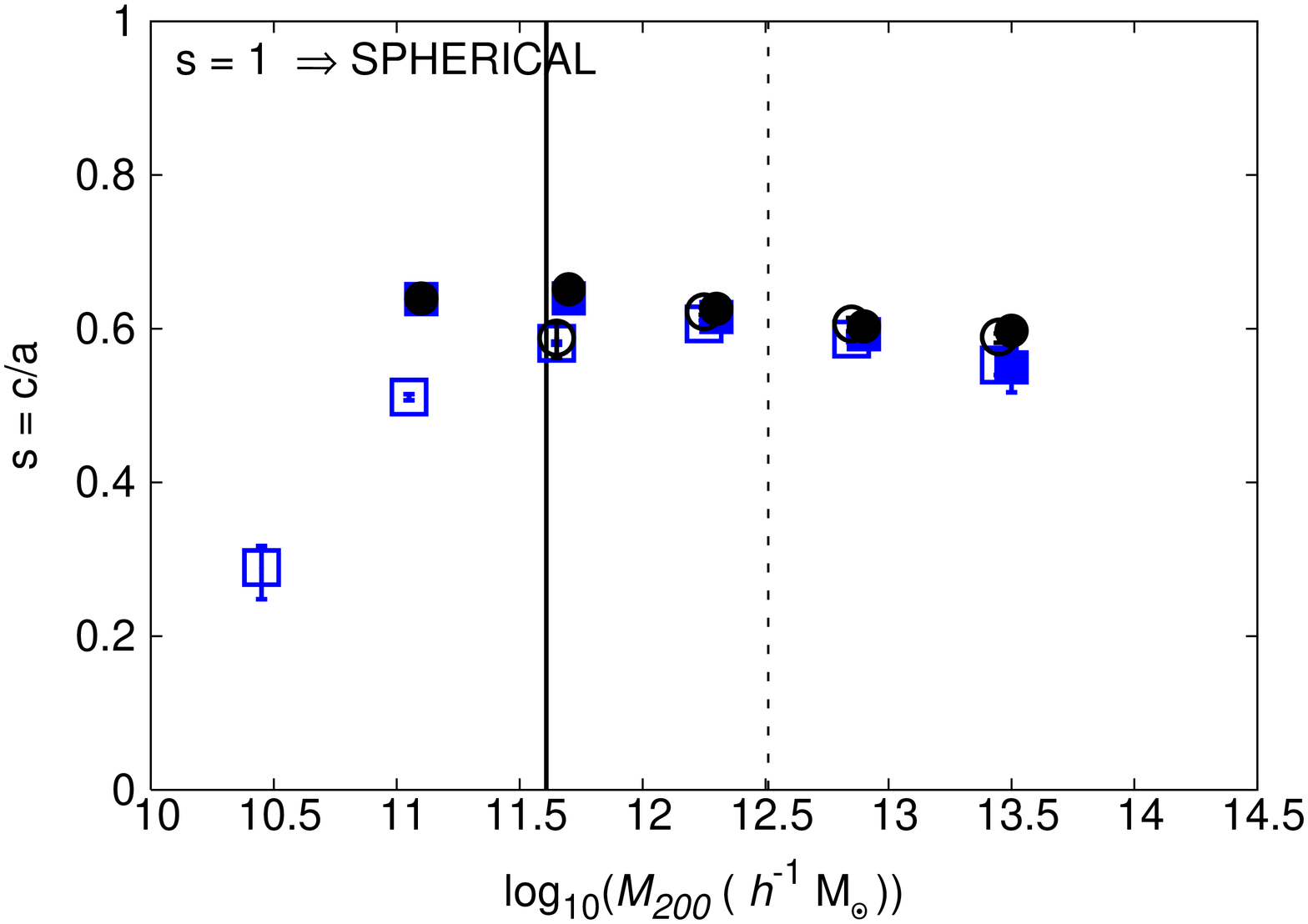} &
\includegraphics[width=7cm,height=7cm,angle=-90,keepaspectratio]{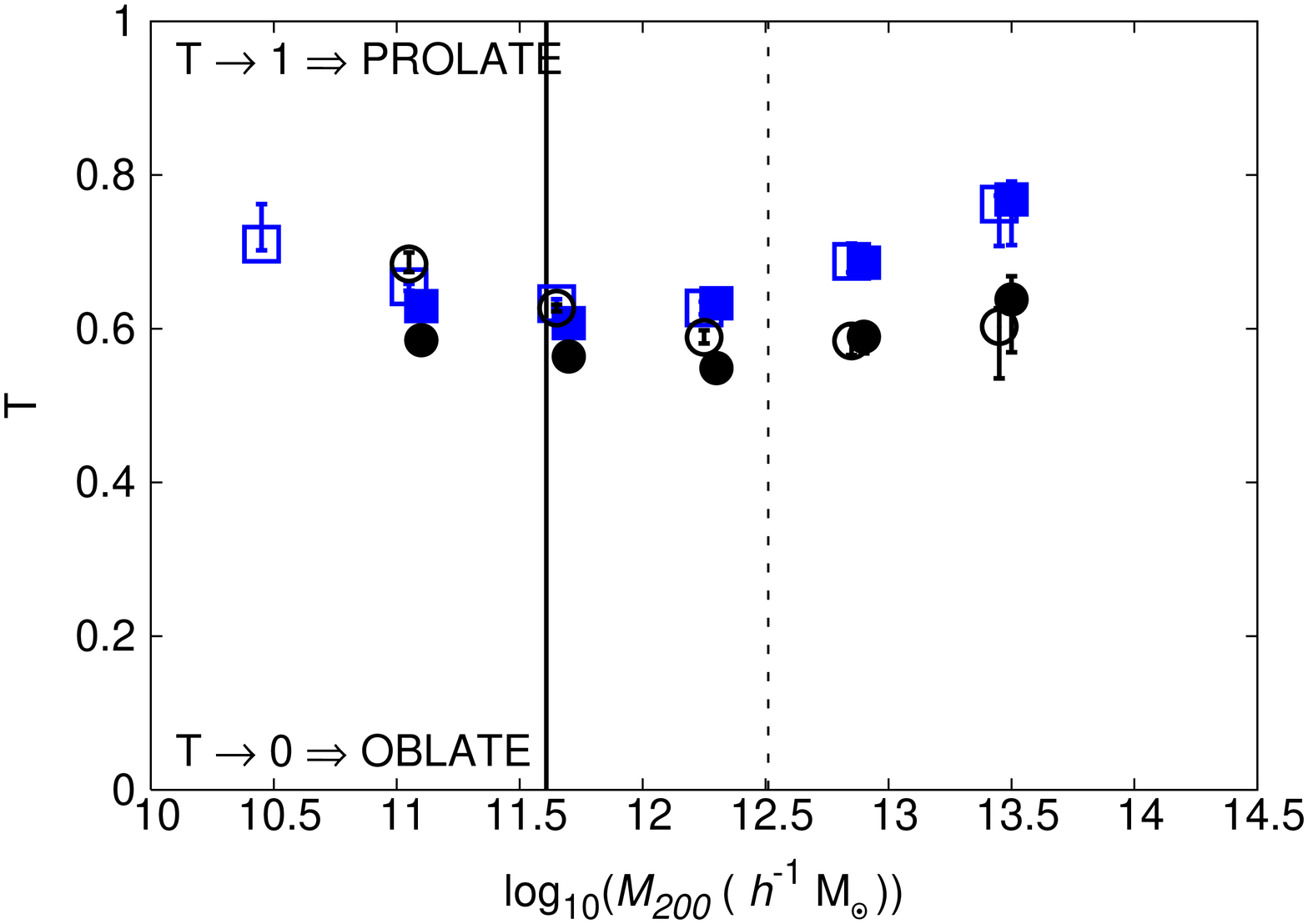}

\end{tabular} 
\end{center}
\caption[Resolution tests for REF halo triaxiality.]{\label{dmres}  As above but for halo properties computed within $r_{200}$. }
\end{figure*}

In order to quantify the effects of resolution, the shapes of haloes extracted
from a $512^3$ particle run with a maximum force softening length of 2 $h^{-1}$ kpc
(used in this analysis) are compared with those from a corresponding
lower-resolution run (containing  $256^3$ particles with a maximum softening length of
4 $h^{-1}$ kpc).  Resolution tests for the central regions ($0.25r_{200}$) of dark matter only haloes and the haloes from the weak stellar feedback run are shown in Fig. \ref{dmres_c}, while the resolution tests for the overall properties computed within $r_{200}$ are shown in Fig. \ref{dmres}.  These figures show the spin ($\lambda$), $e = b/a$, $s = c/a$ and triaxiality ($T$), respectively, as a function of halo mass.  In each of these plots the properties of haloes extracted from the $512^3$ particle simulations are shown with filled symbols and the results from the lower-resolution run with $256^3$ particles are shown with open symbols.  Dark matter only runs are shown as blue squares while the weak stellar feedback runs are shown as black circles.  The error bars are the one-sigma bootstrap resampled median distributions (1000 bootstrap samples have been used).  Vertical lines show the 1000-particle cuts that have been used in this analysis.  Clearly, the properties we discuss are well resolved beyond 1000 particles.

\end{appendix}

\label{lastpage}

\end{document}